\documentclass[prd,onecolumn]{revtex4}
\usepackage{dcolumn}
\usepackage{multirow}
\usepackage{graphicx}
\usepackage{amssymb}
\usepackage{bm}
\usepackage{hyperref}
\usepackage{epstopdf}
\usepackage{color}
\usepackage{mathrsfs}
\usepackage{amsmath,amssymb,amsthm}
\usepackage{rotating}
\usepackage{sverb, longtable}
\usepackage{subfigure}

\usepackage[graphicx]{realboxes}
\usepackage{adjustbox}
\begin{document}

\title{Testing modified gravity with 21 cm intensity mapping, HI galaxy, cosmic microwave background, optical galaxy, weak lensing, galaxy clustering, type Ia supernovae and gravitational wave surveys}
\author{Deng Wang}
\email{cstar@nao.cas.cn}
\affiliation{National Astronomical Observatories, Chinese Academy of Sciences, Beijing, 100012, China}
\begin{abstract}
In modern cosmology, an important task is investigating whether there exists a signal of modified gravity in the universe. Due to the limited resolutions and sensitivities of facilities, current observations can not detect any signal of modified gravity. As a consequence, it is urgent to predict the constraining power of future cosmological surveys on modified gravity. We constrain the Hu-Sawicki $f(R)$ gravity with eight future mainstream probes encompassing 21 cm intensity mapping, HI galaxy, cosmic microwave background, optical galaxy, weak lensing, galaxy clustering, type Ia supernovae and gravitational wave. We find that the HI galaxy survey SKA2 gives the strongest constraint $\sigma_{f_{R0}}=1.36\times10^{-8}$ among eight probes. The promising 21 cm intensity mapping survey SKA1-MID-B1 and optical galaxy survey Euclid also reach the order $\mathcal{O}$(-8). The fourth-generation CMB experiments SO and CORE produces the order $\mathcal{O}$(-6), while large scale structure surveys Euclid weak lensing, Euclid galaxy clustering and CSST weak lensing obtain the order $\mathcal{O}$(-5). Interestingly, CSST galaxy clustering gives the same order $\mathcal{O}$(-6) as SO and CORE, and the gravitational wave survey ET also obtain the order $\mathcal{O}$(-5). The combination of eight probes gives the tightest constraint $1.14\times10^{-8}$, which is just a little stronger than $1.34\times10^{-8}$ from the combination of SKA2 and SK1-MID-B1. This indicates that, to a large extent, future 21 cm intensity mapping and HI galaxy surveys can improve our understanding of modified gravity and energy budget in the cosmic pie.

\end{abstract}
\maketitle

\section{Introduction}
For the late-time universe, the standard cosmological model, $\Lambda$CDM, has achieved a great success in describing various aspects of the universe \cite{Planck:2018vyg}. In its cosmic pie, the universe consists of three ingredients, i.e., matter, dark matter (DM) and dark energy (DE). The matter including baryons, photons and neutrinos only occupies a small fraction of energy budget of our universe. The DM has a obvious clustering property and encodes important information of cosmic structure formation. Unlike DM, the DE which is responsible for the late-time cosmic acceleration, is homogeneously permeated in the universe at cosmological scales and hardly cluster. At least, the establishment of $\Lambda$CDM depends on two preconditions, i.e., general relativity (GR) and cosmological principle, which characterizes the geometry of the universe as homogeneous and isotropic at large scales. For the early-time universe, there is also a standard model called inflation, which explains the origin of the universe and generates the initial conditions of large scale structures. It indicates that the universe undergoes a quasi-exponential expansion at very early times. To sum up, current standard cosmological paradigm should be ``inflation+$\Lambda$CDM'', where variants of both parts predict possible new physics based on current and future cosmological surveys.   

In $\Lambda$CDM, DE is ascribed to the cosmological constant term, $\Lambda$, which gives a constant vacuum energy density. However, this model faces at least two intractable problems, i.e., the fine-tuning and cosmological constant problems \cite{Weinberg}. In order to solve these two problems and explain the late-time cosmic expansion, one class of important scenarios is modified gravity (MG) \cite{Koyama:2015vza,Clifton:2011jh}. When modifying GR, one may encounter various kinds of difficulties. A good starting point is making modifications based on the famous Lovelock's theorem, which indicates that Einstein's field equations are the sole second-order local equations of motion for a metric derived from the action in four dimension. This means that one can propose a MG model by using one or more of the following options: (i) extra dimensional spacetime; (ii) extra degrees of freedom; (iii) higher derivatives; (iv) non-locality. Once a MG theory is proposed, one shall investigate its stability and theoretical consistency such as satisfying the solar system constraints (see also for details \cite{Koyama:2015vza}).

MG can be constrained at different scales and regimes. The first results from Laser Interferometer Gravitational Wave Observatory (LIGO) give the first results on gravity in the strong field regime \cite{LIGOScientific:2016lio,LIGOScientific:2018dkp}. In the transition from weak to strong field regime, the pulsar timing array allows us to constrain many fundamental parameters which depict deviations from GR \cite{Boitier:2020xfx}. A number of classical tests of gravity ranging from laboratory experiments to solar system constraints have been performed in weak field regime \cite{Will:2014kxa}. Some astrophysical observations can also improve the constraints in a novel way \cite{Berti:2015itd}. The above all probes can present strong constraints on MG at small scales, however, the constraints become weaker at large scales. At current stage, the observations of large scale structure of the universe such as cosmic microwave background (CMB) \cite{Planck:2018vyg}, weak lensing (WL) \cite{Hildebrandt,Hamana,DES:2021wwk}, galaxy clustering (GC) and galaxy-galaxy lensing surveys \cite{DES:2021wwk,DES:2021bpo} just can provide limited constraining power for typical MG parameters independently \cite{Wang:2020zfv}. As a consequence, so far, combining them together to give a relatively tight constraint on MG is the best choice \cite{Wang:2020dsc,Wang:2020hqq,Wang:2021kuw}. Moreover, the geometrical probes such as baryon acoustic oscillations (BAO) \cite{Blake03,Seo03} and Type Ia supernovae (SNe Ia) \cite{Riess98,Perlmutter99,Wang:2018ahw} can also help improve the constraints and break the degeneracy between cosmological parameters.  

For a long time, there is a lack of forecasting constraints on a specific MG model from 21 cm intensity mapping (IM) and neutral hydrogen (HI) galaxy redshift experiments. Meanwhile, many authors just focus on the combined constraints on MG from current probes and do not know the possible errors on MG parameters from future combined main stream surveys. Based on these two motivations, in this study, we aim at performing a comprehensive forecasting constraint on one of the simplest extensions of GR, $f(R)$ gravity \cite{Buchdahl:1983zz}, by using future mainstream surveys such as 21 cm IM, neutral hydrogen (HI) galaxy redshift, CMB, optical galaxy redshift, WL, GC, SNe Ia and gravitational wave (GW). By implementing numerical analysis, we find that the HI galaxy redshift survey SKA2 gives the strongest constraint on the MG parameter among eight probes. Interestingly, the constraint from the combination of eight probes is just a little stronger than that from the combination of SKA2 and SK1-MID-B1, which implies that future HI surveys can help us explore new physics and improves our understanding of cosmic formation and evolution.   

This study is organized in the following manner. In the next section, we introduce the $f(R)$ gravity model to be constrained by forthcoming surveys. In Section III, we review the basic formula of each probe. In section IV, we exhibit the theoretical predictions of $f(R)$ gravity. In Section V, we present the experimental specifications of each probe and analysis methodology. In Section VI, we display the numerical results. The discussions and conclusions are presented in the final section.

\section{Modeling $f(R)$}
One of the most popular gravities is the so-called $f(R)$ gravity, where the action is generalized to be a function of the Ricci scalar curvature $R$. Since the equation of motion of this gravity model is fourth order, it usually belongs to the above third class of MG models, i.e., higher derivatives. Nonetheless, one can also make the equation of motion second order by introducing a scalar field. $f(R)$ gravity was firstly introduced in Ref.\cite{Buchdahl:1983zz} and more details can be found in recent reviews \cite{DeFelice:2010aj,Sotiriou:2008rp}. Its action reads as 
\begin{equation}
S=\int d^4x\sqrt{-g}\left[R+f(R)+\mathcal{L}_m\right], \label{1}
\end{equation}
where $g$, $f(R)$ and $\mathcal{L}_m$ are the trace of the metric, a function of $R$ and the standard matter Lagrangian, respectively. By taking a variation with respect to $g_{\mu\nu}$, we obtain the modified 
Einstein field equation as 
\begin{equation}
G_{\mu\nu}-\nabla_\mu\nabla_\nu f_R+(\Box f_R-\frac{f}{2})g_{\mu\nu}+f_RR_{\mu\nu}=8\pi GT_{\mu\nu}, \label{2}
\end{equation}
where $G_{\mu\nu}$ denotes the Einstein tensor, $f_R\equiv df/dR$ is an extra scalar degree of freedom, i.e., the so-called scalaron and $T_{\mu\nu}$ represents the energy-momentum tensor. For a spatially flat four-dimensional Friedmann-Robertson-Walker (FRW) universe, the equation governing the background dynamics in the framework of $f(R)$ gravity can be shown as 
\begin{equation}
H^2\frac{dR}{dN}f_{RR}-(H^2+H\frac{dH}{dN})f_R+\frac{f}{6}+H^2=\frac{8\pi G \rho_m}{3} ,              \label{3}
\end{equation} 
where $f\equiv f(R)$, $f_{RR}\equiv df_R/dR$, $N\equiv \mathrm{ln}\,a$, $a$ scale factor, $H$ Hubble parameter and $\rho_m$ matter energy density, respectively. 

Hereafter, we only consider the linear perturbations in $f(R)$ gravity. For sub-horizon modes, $k\gtrsim aH$, where $k$ is comoving wavenumber, the modified linear growth of matter density perturbations in the quasi-static approximation \cite{Bean:2006up} is expressed as
\begin{equation}
\frac{\mathrm{d}^2\delta}{\mathrm{d}a^2}+\left(\frac{1}{H}\frac{\mathrm{d}H}{\mathrm{d}a}+\frac{3}{a}\right)\frac{\mathrm{d}\delta}{\mathrm{d}a}-\frac{3H_0^2\Omega_{m}\delta a^{-5}}{H^2(1+f_R)}\left(\frac{1-2A}{2-3A}\right)=0, \label{4}
\end{equation}
where $H_0$ is Hubble constant, $\Omega_{m}$ is current matter fraction and the function $A(k,a)$ reads as 
\begin{equation}
A(k,a) = \frac{-2f_{RR}}{f_R+1}\left(\frac{k}{a}\right)^2. \label{5}
\end{equation}
It is worth noting that the function $A(k,a)$ occurring in Eq.(\ref{4}) introduces a scale dependence of linear growth factor $\delta(k,a)$ in $f(R)$ gravity, when the growth factor is just a function of scale factor in GR. 

In principle, a pure $f(R)$ gravity model should at least satisfy the stability conditions and pass the local gravity test. In light of current astrophysical and cosmological observations, we also expect that a viable $f(R)$ model can reproduce the late-time cosmic acceleration or early-time inflation. In order to explore the constraining power of future cosmological experiments better such as 21 cm IM and HI galaxy, we take the viable Hu-Sawicki $f(R)$ gravity into account in this analysis, which is denoted as ``HS model'' hereafter, and it is expressed as 
\begin{equation}
f(R)=\frac{-2\Lambda R^n}{R^n+\mu^{2n}}, \label{6}
\end{equation}  
where $\mu$ and $n$ denote two free parameters. In the high curvature regime, $R\gg\mu^2$, this model can explain the late-time cosmic expansion well and it looks like
\begin{equation}
f(R)=-2\Lambda-\frac{f_{R0}}{n}\frac{R_0^{n+1}}{R^n}, \label{7}
\end{equation}  
where $f_{R0}=df/dR|_{z=0}=-2\Lambda\mu^2/R_0^2$ and $R_0$ represents current Ricci scalar curvature. In order to confront the HS model with future observations, one should theoretically work out the evolutional behaviors of both background and perturbation by substituting Eq.(\ref{7}) into Eqs.(\ref{3}-\ref{4}).

\section{Basic formula of each probe}
We introduce the basic formula of the above mentioned cosmological probes and give some simple theoretical predictions.

\subsection{21 cm intensity mapping}  
In theory, the best we can do is mapping out the whole three-dimensional structure of the universe. In the near future, the HI observations can be a good tracer to achieve this goal. HI can trace the underlying DM distribution and has a characteristic 21 cm emission line, which originates from the transition between the hyperfine levels of HI atoms and corresponds to the frequency $\nu=1420$ MHz in the rest frame. The fluctuations of brightness temperature of the redshifted 21 cm line traces the HI distribution and consequently the large scale structure of our universe. In the Rayleigh-Jeans limit, the observed brightness temperature can be dissected into a background part $\tilde{T_b}$ and a perturbation part $\delta T_b$, i.e., $T_b=\tilde{T_b}+\delta T_b$, where \cite{Hall:2012wd}
\begin{equation}
\tilde{T_b}(z)=\frac{3}{32\pi}\frac{(hc)^3A_{10}}{k_Bm_pE_{21}^2}\frac{\Omega_{\mathrm{HI}}(z)\rho_{c,0}}{H(z)(1+z)}, \label{8}
\end{equation}
and 
\begin{equation}
\delta T_b(\mathbf{k},z)=\tilde{T_b}(z)\delta_{\mathrm{HI}}(\mathbf{k}, z), \label{9}
\end{equation}
where $h$ is Planck constant, $c$ speed of light, $k_B$ Boltzmann constant, $m_p$ proton mass, $z$ redshift, $\mathbf{k}$ comoving wave vector,
$E_21$ the energy of 21 cm transition in the rest frame,
$A_{10}=2.869\times10^{15}$ s$^{-1}$ the spontaneous emission coefficient, $\rho_{c,0}$ critical density, $\Omega_{\mathrm{HI}}(z)$ HI fraction and $H(z)$ the Hubble expansion rate at a given redshift. In the late-time universe, most HI is believed to be localized dense gas clouds in galaxies, where it is screened by ionizing photons. As a consequence, one can naturally treat HI as a biased tracer of DM distribution similar to galaxies. This fact allows us to express the HI density contrast as $\delta_{\mathrm{HI}}=b_{\mathrm{HI}}\diamond\delta$, where $\delta$ is matter density perturbation and the symbol $\diamond$ represents the convolution indicating the possibility of time- and scale-dependent biasing.

In the literature, there are two methods to calculate the theoretical predictions, i.e., HI angular power spectrum (APS) \cite{Hall:2012wd} and HI power spectrum (PS) \cite{Bull:2014rha}. In this study, we focus on constraining the HS model using the HI APS method.

For the HI APS method, APS between two redshift windows $i$ and $j$ is written as \cite{Hall:2012wd,Challinor:2011bk}
\begin{equation}
C_\ell^{ij}=4\pi\int \mathrm{d} \, \mathrm{ln}k P_\mathcal{R}(k)\Delta^i_\ell(k)\Delta^j_\ell(k), \label{10}
\end{equation}
where $P_\mathcal{R}(k)$ denotes the power spectrum of dimensionless primordial curvature perturbation $\mathcal{R}$, while $\Delta^i_\ell(k)=\Delta^i_\ell(\mathbf{k})/\mathcal{R}(\mathbf{k})$. Integrating over a window function $W(z)$, the source of 21 cm APS is expressed as \cite{Challinor:2011bk}
\begin{equation}
\Delta^i_\ell(\mathbf{k})=\int \mathrm{d}zW_i(z)\Delta^i_\ell(\mathbf{k},z), \label{11}
\end{equation}
where $\Delta^i_\ell(\mathbf{k,z})$ denotes the spherical harmonic expansion of brightness temperature perturbation. Specifically, in this study, we adopt a Gaussian window function.

\subsection{Galaxy redshift survey}
Up to now, there are two main ways to explore the universe for a 21 cm experiment, i.e., the above 21 cm IM and corresponding HI galaxy redshift survey. A galaxy redshift survey enables us to measure the cosmic expansion history via the measurement of BAO as well as the growth history of large scale structure via the measurement of RSD. As well as known, galaxies are biased traces of density fields. One can constrain the MPS or correlation function by identifying individual galaxies and confirming their redshifts. Optical galaxy redshift surveys have achieve great success in cosmology during the last several years, future HI galaxy surveys such as Baryon acoustic oscillations from Integrated Neutral Gas Observations (BINGO) \cite{Costa:2021jsk} and Square Kilometre Array (SKA) \cite{SKA} can also give strong constraining power on cosmic geometry and growth.

The main observable for a optical galaxy survey is the galaxy power spectrum (GPS),
\begin{equation}
P(\mathbf{k},z)=\left[b(z)+f\mu^2\right]^2e^{-\frac{\left[k\sigma_{\mathrm{NL}}(z,\mu)\right]^2}{2}}P(k,z),\label{14}
\end{equation} 
where the first term describes the Kaiser effect \cite{Kaiser1987}, the second term is responsible for the '``Finger of God'' effect due to uncorrelated velocities at small scales, which washes out the radial structure below the nonlinear velocity dispersion scale $\sigma_{\mathrm{NL}}$,  $b(z)$ denotes the galaxy bias as a function of redshift, $f$ is the linear growth rate, $\mu=\hat{k}\cdot\hat{z}$.   and 
\begin{equation}
\sigma_{\mathrm{NL}}(z,\mu)=\sigma_{\mathrm{NL}}D(z)\sqrt{1+f\mu^2(2+f)},\label{15}
\end{equation} 
where $D(z)$ is the linear growth factor and $f\equiv \mathrm{d \,log} \,D/\mathrm{d \,log}\,a$.

The quality of a galaxy redshift survey is subject to complex systematics. In particular, stars are a dominated contaminant in large optical galaxy surveys. Bright stars can effectively mask galaxies behind them, when one distinguishes stars from galaxies by their color. This will leads to a very complicated angular selection function on the sky. Interestingly, this problem is not serious in the radio wavelength. Although there are other contaminants such as some non-galaxy point sources and diffuse galactic synchrotron emission affecting the source-finding process and final galaxy catalogue. Another important systematic effect is source evolution. For instance, the luminosity function of the tracer population usually changes with redshift, which affects the detected number of galaxies. For a specific tracer, this effect limits the effective redshift range of a survey and make its selection function more complicated. Furthermore, this effect is generally characterized by the galaxy bias $b(z)$, the effect of which on BAO is varying the shot noise by varying the
effective galaxy number density $n(z)$. More details about these effects can be found in Ref.\cite{Raccanelli:2015qqa}. Notice that Eqs.(\ref{14}) and (\ref{15}) are also applied into the HI galaxy redshift survey. 

\subsection{Cosmic microwave background}
During the past three decades, the CMB experiments have made great progresses in improving sensitivity and resolution \cite{Planck:2018nkj}. CMB observations can measure many aspects of formation and evolution of the universe such as matter components, topology and large scale structure effects.
Recently, the Planck 2018 final release \cite{Planck:2018vyg} with improved measurements of large scale polarization has given high precision constraints on cosmological parameters. 

The CMB temperature field in the universe can be shown as
\begin{equation}
T(\theta, \phi)=T_0\left[1+\Theta(\theta, \phi)\right],\label{16}
\end{equation}
where $T_0$ is background temperature, $\Theta$ denotes fluctuations of the temperature field $T$, and $\theta$ and $\phi$ are polar and azimuthal angles in spherical coordinates, respectively. By expanding the fluctuations into spherical harmonic functions, the CMB anisotropy spectrum can be expressed as \cite{Dodelson03} 
\begin{equation}
C_\ell^{TT}=\frac{2}{\pi}\int^\infty_0 \mathrm{d}k \, k^2 P(k)\left[\frac{\Theta_\ell(k)}{\delta(k)}\right]^2, \label{17}
\end{equation}
where $\Theta_\ell$ denotes the expanded perturbation temperature field at a given multipole $\ell$. Besides the temperature spectrum, CMB observations also include E-modes ($C_\ell^{EE}$) and B-modes ($C_\ell^{BB}$) polarization information, which can also be used for exploring cosmic neutrino background, DM particle mass, DE and inflation. Specific formula of CMB polarization $C_\ell^{EE}$ and $C_\ell^{BB}$ and more details can be found in Ref.\cite{Dodelson03}.

\subsection{Weak lensing}
WL depicts the distortion of images of distant galaxies caused by the intervening matter between observer and source along the line of sight. Based on the fact that weak lensing is very sensitive to the background geometry of the universe and matter distribution of large scale structure, one can study the cosmic expansion history and structure growth by measuring the gravitational distortions from source galaxies as a function of redshift, i.e., the so-called lensing tomography technique.

The cosmic shear correlation function in real space reads as  
\begin{equation}
\xi_{\pm}^{ij}(\theta)=\frac{(1+m_i)(1+m_j)}{2\pi}\int \mathrm{d} \ell \, \ell J_{0/4}(\ell\theta)C_{\gamma\gamma}^{ij}(\ell),\label{18}
\end{equation}
where $m_i$ and $m_j$ are multiplicative factors, which explain the shear calibration bias, 1 for each tomographic bin, $J_0$ and $J_4$ correspond to the first kind of Bessel functions, and $C_{\gamma\gamma}^{ij}(\ell)$ denotes the observed cosmic shear PS. In general, $C_{\gamma\gamma}^{ij}(\ell)$ can be divided into the following four parts,
\begin{equation}
C_{\gamma\gamma}^{ij}(\ell)=C_{GG}^{ij}(\ell)+C_{GI}^{ij}(\ell)+C_{IG}^{ij}(\ell)+C_{II}^{ij}(\ell)+C_N^{ij}(\ell), \label{19}
\end{equation}  
where the first term $C_{GG}^{ij}(\ell)$ is true shear PS and three medium terms $C_{GI}^{ij}(\ell)$, $C_{IG}^{ij}(\ell)$, $C_{II}^{ij}(\ell)$ are contaminants relative to true spectrum due to intrinsic alignment (IA), and the final term $C_N^{ij}(\ell)$ is noise PS. 

Taking the Limber approximation and the flat-sky assumption, the true shear PS is expressed as 
\begin{equation}
C_{GG}^{ij}(\ell)=\int_0^{\chi_H} \mathrm{d}\chi\frac{q_i(\chi)q_j(\chi)}{\chi^2}P\left(\frac{\ell}{\chi},\chi\right), \label{20}
\end{equation}
where $\chi$ and $\chi_H$ denote the comoving distance for the lens and comoving horizon. Subsequently, the lensing kernel $q_i(\chi)$, also called the lensing weighting function, is in the $i$-th tomographic bin reads as
\begin{equation}
q_i(\chi)=\frac{3\chi\Omega_mH_0^2}{2a(\chi)c^2}\int_{\chi}^{\chi_H}\mathrm{d} \tilde{\chi}\frac{\tilde{n}_i(\tilde{\chi})(\tilde{\chi}-\chi)}{\tilde{\chi}}, \label{21}
\end{equation}
where $\tilde{\chi}$ is the comoving distance of source galaxies and $\tilde{n}_i(\tilde{\chi})$ denotes the normalized source galaxy distribution in the $i$-th bin.

For the contamination parts, two types ``GI'' and ``IG'' represent the correlations between the intrinsic ellipticity (tidally torqued) of a foreground galaxy and the gravitational shear of a background galaxy. Their contribution is shown as
\begin{equation}
C_{GI}^{ij}(\ell)+C_{IG}^{ij}(\ell)=\int_{0}^{\chi_H}\mathrm{d}\chi\frac{q_i(\chi)\tilde{n}_j(\chi)+q_j(\chi)\tilde{n}_i(\chi)}{\chi^2}P_{GI}\left(\frac{\ell}{\chi},\chi\right). \label{22}
\end{equation}
The ``II'' type describes the correlations of intrinsic eplliticities between neighboring galaxies due to local tidal gravitational field and it reads as
\begin{equation}
C_{II}^{ij}(\ell)=\int_{0}^{\chi_H}\mathrm{d}\tilde{\chi}\frac{\tilde{n}_i(\tilde{\chi})\tilde{n}_j(\tilde{\chi})}{\tilde{\chi}^2}P_{II}\left(\frac{\ell}{\tilde{\chi}},\tilde{\chi}\right). \label{23}
\end{equation}
$P_{II}$ and $P_{GI}$ are IA power spectra. By assuming the amplitudes of IA power spectra are linearly related to local density fields, we have
\begin{equation}
P_{II}(k,z)=F^2(z)P(k,z), \label{24}
\end{equation}
\begin{equation}
P_{GI}(k,z)=F(z)P(k,z), \label{25}
\end{equation}
where the function $F(z)$ reads as \cite{Hirata:2004gc} 
\begin{equation}
F(z)=-\frac{A_{IA}U_1\rho_{c,0}\Omega_{m}}{D(z)}\left(\frac{1+z}{1+z_0}\right)^{\eta_{IA}}\left(\frac{L_i}{L_0}\right)^{\alpha_{IA}},\label{26}
\end{equation}
where $U_1$ is a normalized constant ($U_1=5\times10^{-14}h^{-2}\mathrm{M}_\odot\,\mathrm{Mpc}^3$ \cite{Brown:2000gt}), $h=H_0/100$ km$s^{-1}$Mpc$^{-1}$ the pivot redshift $z_0=0.6$, $L_0$ is luminosity, and $A_{IA}$, $\eta_{IA}$ and $\alpha_{IA}$ are typical free parameters in this nonlinear model, respectively. Furthermore, for simplicity, we set the baseline values $A_{IA}=-1$ and $\eta_{IA}=0$. Moreover, we choose $\alpha_{IA}=0$, since the average luminosity variation can be neglected across different bins.

\subsection{Galaxy clustering}
As we know, galaxies are not randomly distributed in the universe. There are major concentrations of galaxies we refer to as clusters, nearly empty areas that we refer to as voids, and more complicated structures such as filaments and sheets. The homogeneity of matter distribution in space can be traced via the galaxy distribution, which is modulated by the galaxy bias depending on redshift and scales as well as the number of galaxies. The overabundance of pairs at angular separation $\theta$ in a random distribution, $\omega(\theta)$, is one of the most elegant ways to measure GC. It quantifies the scale dependence and strength of GC, and consequently affects the underlying matter clustering. Using a simple scale-independent linear bias model, the galaxy-galaxy auto correlation function in the $i$-th tomographic bin is shown as 
\begin{equation}
\omega_i(\theta)=\frac{1}{2\pi}\int \mathrm{d}\ell J_0(\ell\theta)C^{ii}_{gg}(\ell), \label{27}
\end{equation} 
where the auto APS of GC $C^{ii}_{gg}(\ell)$ reads as
\begin{equation}
C^{ii}_{gg}(\ell)=(b_i)^2\mathrm{d}\chi\int_{0}^{\chi_H}\left[\frac{\bar{n}_i(\chi)}{\chi}\right]^2P\left(\frac{\ell}{\chi},\chi\right), \label{28}
\end{equation}
where $b_i$ and $\bar{n}_i(\chi)$ are the linear bias and the normalized lens galaxy distribution in the $i$-th tomographic bin, respectively.

\subsection{Type Ia supernovae}
As is well known, the absolute magnitudes of all SNe Ia are believed to be the same, because all the SNe Ia almost explode at the same mass ($M\approx-19.3\pm0.3$). As a consequence, SNe Ia can act as a powerful distance indicator in theory. The observations of SNe Ia provide an excellent way to explore the background evolution of the universe, particularly, the equation of state of DE and Hubble parameter.

The observational cosmological quantity of SNe Ia is the luminosity distance,
\begin{equation}
D_L(z)=\frac{c(1+z)}{H_0\sqrt{|\Omega_k|}}\mathrm{sinn}\left[\sqrt{|\Omega_k|}\int^z_0\frac{\mathrm{d}z'}{E(z')}\right], \label{29}
\end{equation}
where the dimensionless Hubble parameter $E(z)\equiv H(z)/H_0$, $\Omega_k$ is cosmic curvature, and $\mathrm{sinn}(x)=\mathrm{sin}(x), \, x, \, \mathrm{sinh}(x)$ for $k=1,\,0,\,-1$. Note that, in this study, we consider a flat universe $\Omega_k=0$. Furthermore, the theoretical distance modulus can be written as
\begin{equation}
\mu(z)=5\log_{10}D_L(z)+\mu_0, \label{30}
\end{equation}
where $\mu_0=42.39-5\log_{10}h$.

\subsection{Gravitational wave}
GWs are ripples in the fabric of spacetime generated by the acceleration of astrophysical objects. Similar to SNe Ia, GW standard sirens can act as a promising probe to explore the nature of DE or MG. The key cosmological quantity for gravitational sirens is also the luminosity distance (see Eq.(\ref{29})). Because GW standard sirens are self-calibrating, the luminosity distance $D_L(z)$ of a source can be directly inferred from the
observed GW signal, without the help of a cosmic distance ladder. We will describe the details about how to use GW sirens to implement the numerical analysis in the following sections.

\begin{figure}[htbp]
	\centering
	\includegraphics[scale=0.55]{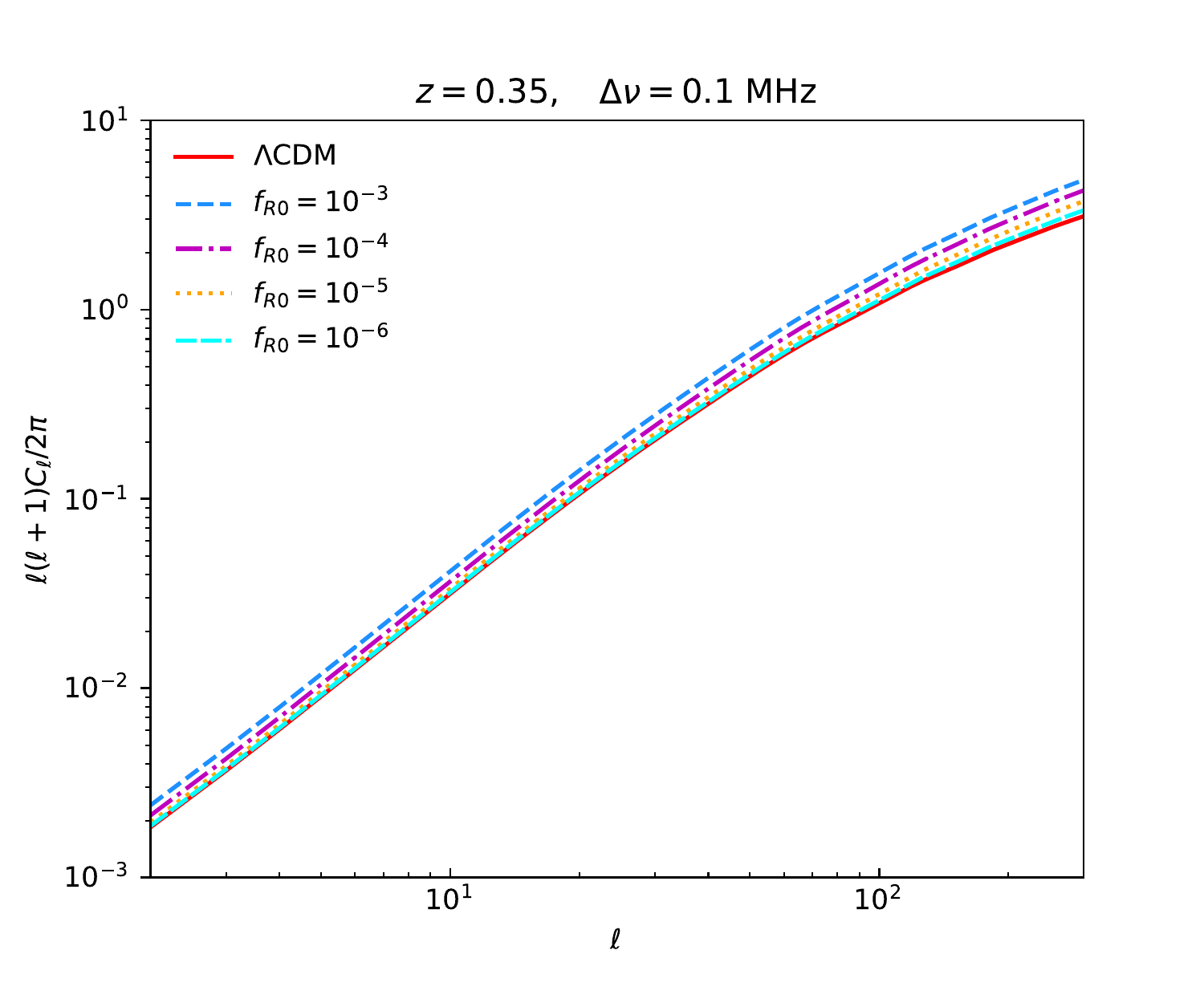}
	\includegraphics[scale=0.55]{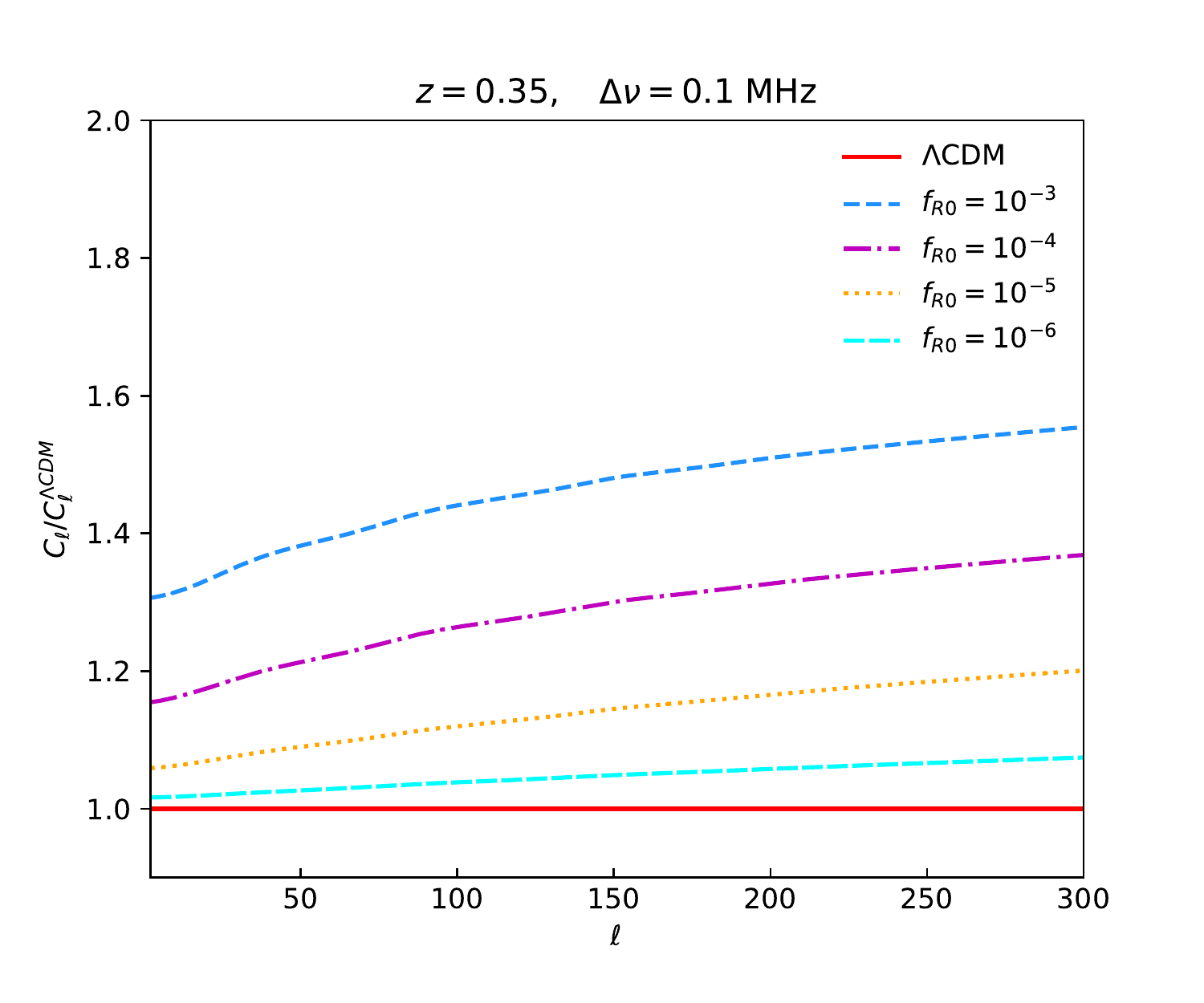}
	\includegraphics[scale=0.55]{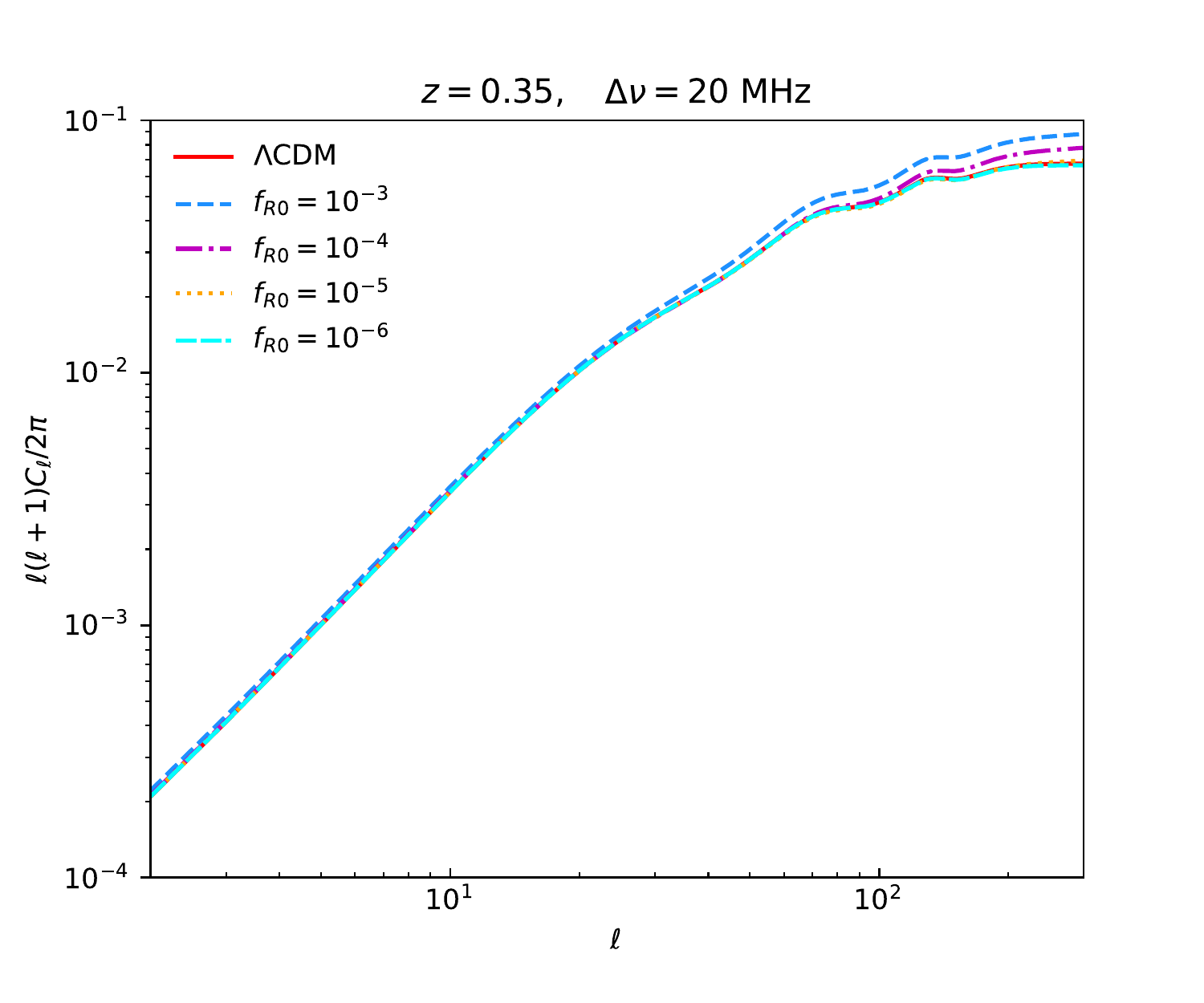}
	\includegraphics[scale=0.55]{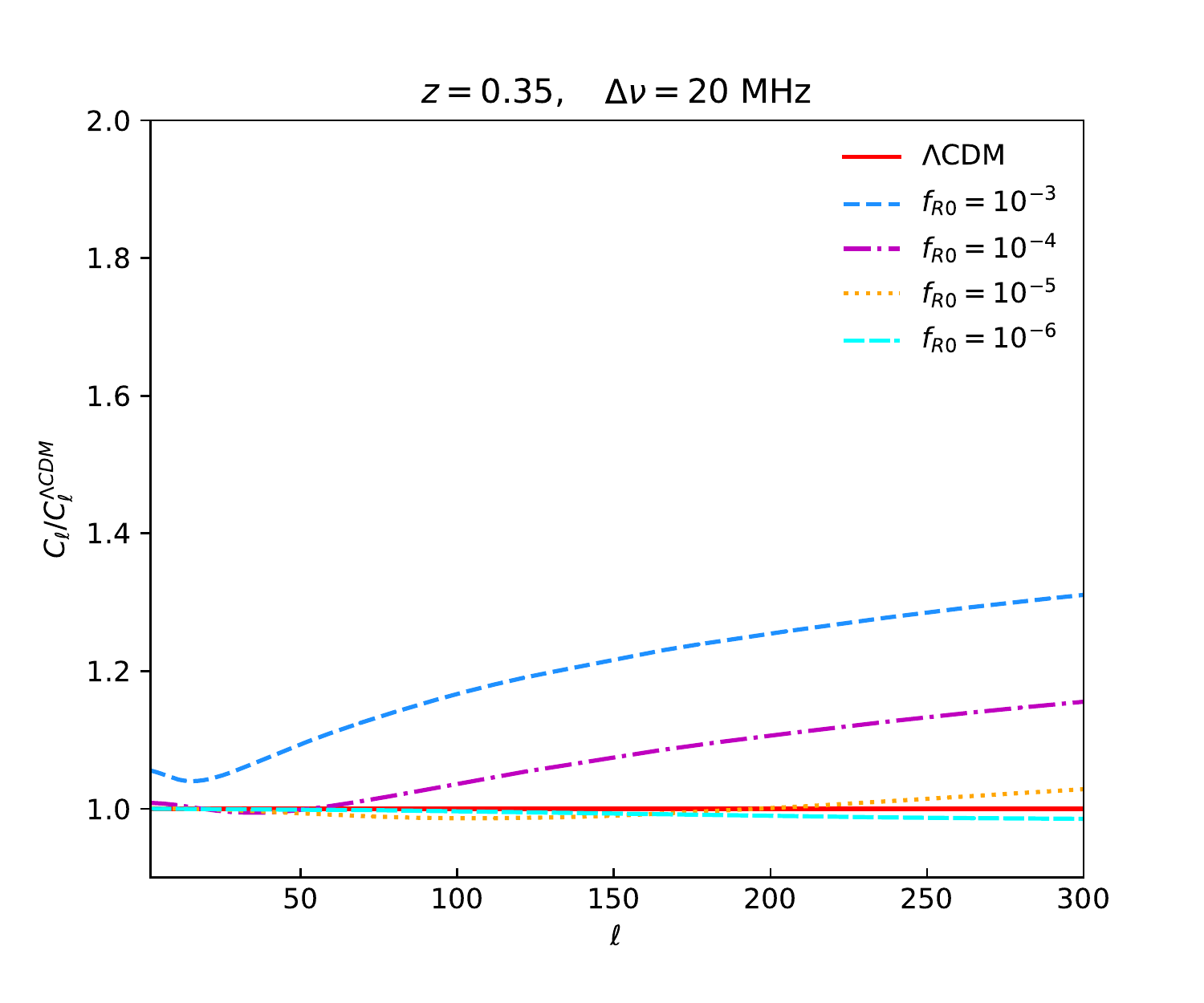}
	\caption{The dimensionless 21 cm auto APS ({\it left}) in the HS $f(R)$ gravity and their ratios ({\it right}) relative to the $\Lambda$CDM model are shown at $z=0.35$ for a narrow window $\Delta\nu=0.1$ MHz ({\it top}) and a broad window $\Delta\nu=20$ MHz ({\it bottom}), respectively. The red solid, blue short-dashed, magenta dash-dotted, orange dotted and cyan long-dashed lines denote $\Lambda$CDM, $f_{R0}=10^{-3},\,10^{-4}, \, 10^{-5}$ and $10^{-6}$, respectively.} 
	\label{f1}
\end{figure} 

\begin{figure}[htbp]
	\centering
	\includegraphics[scale=0.55]{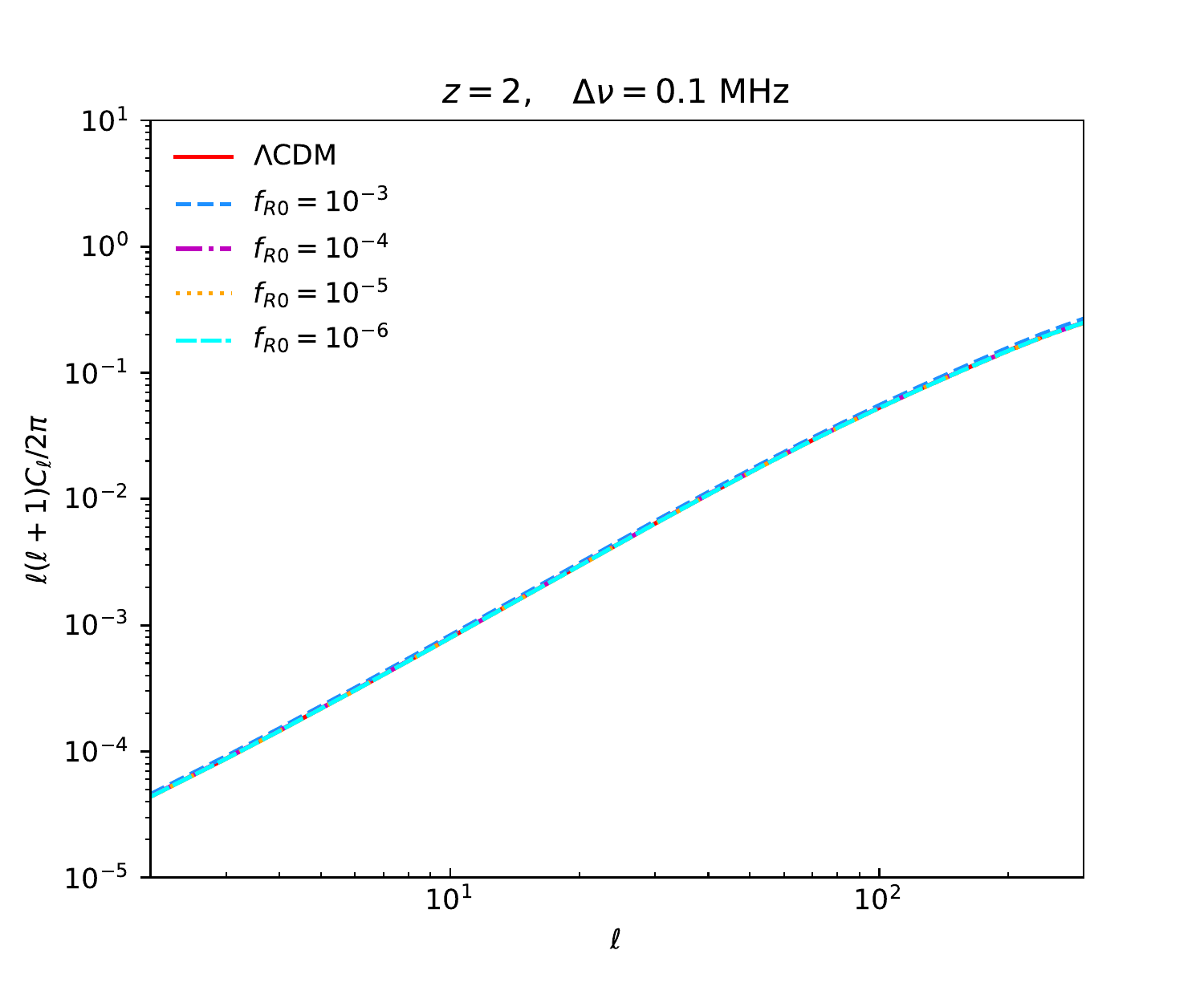}
	\includegraphics[scale=0.55]{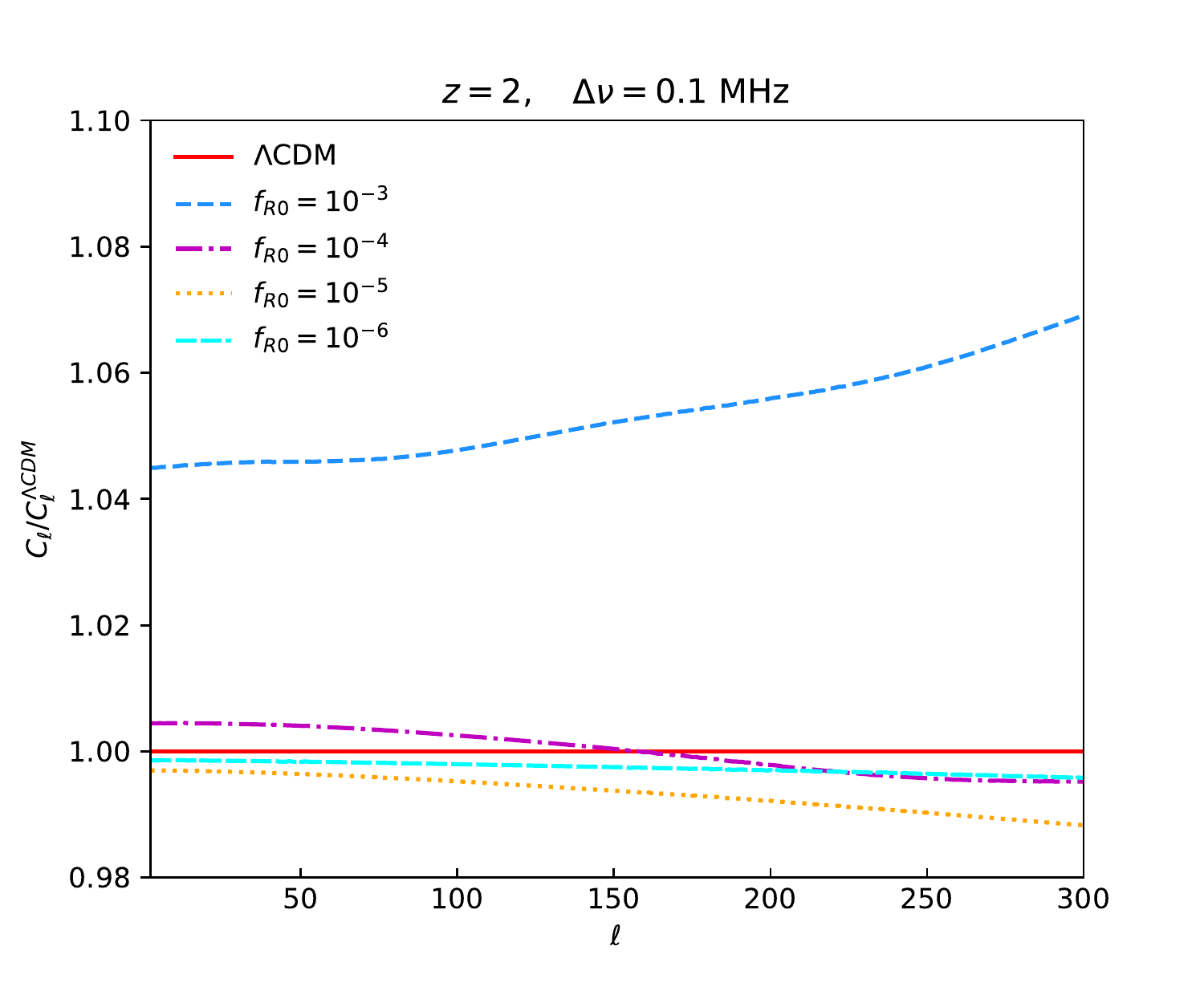}
	\includegraphics[scale=0.55]{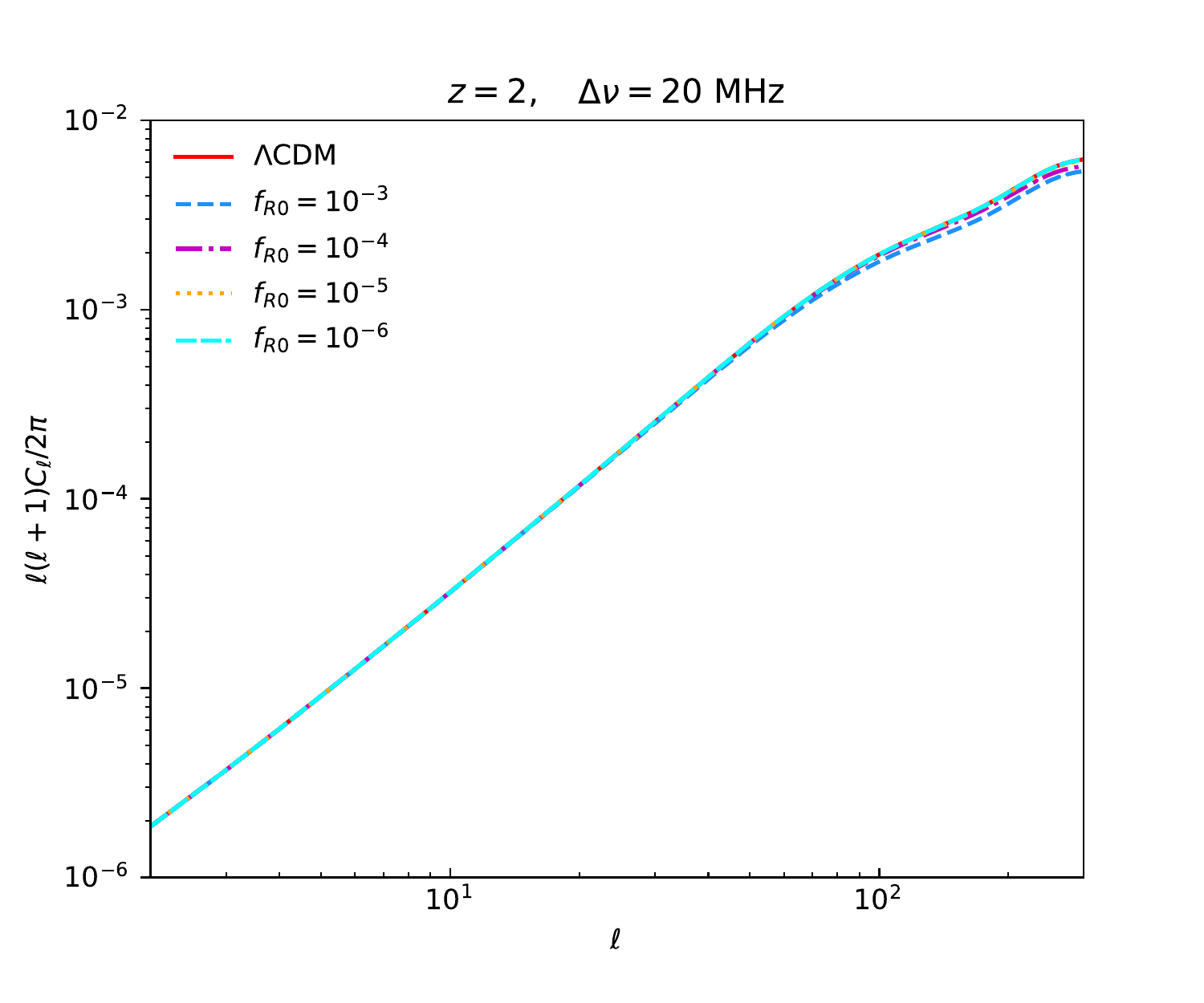}
	\includegraphics[scale=0.55]{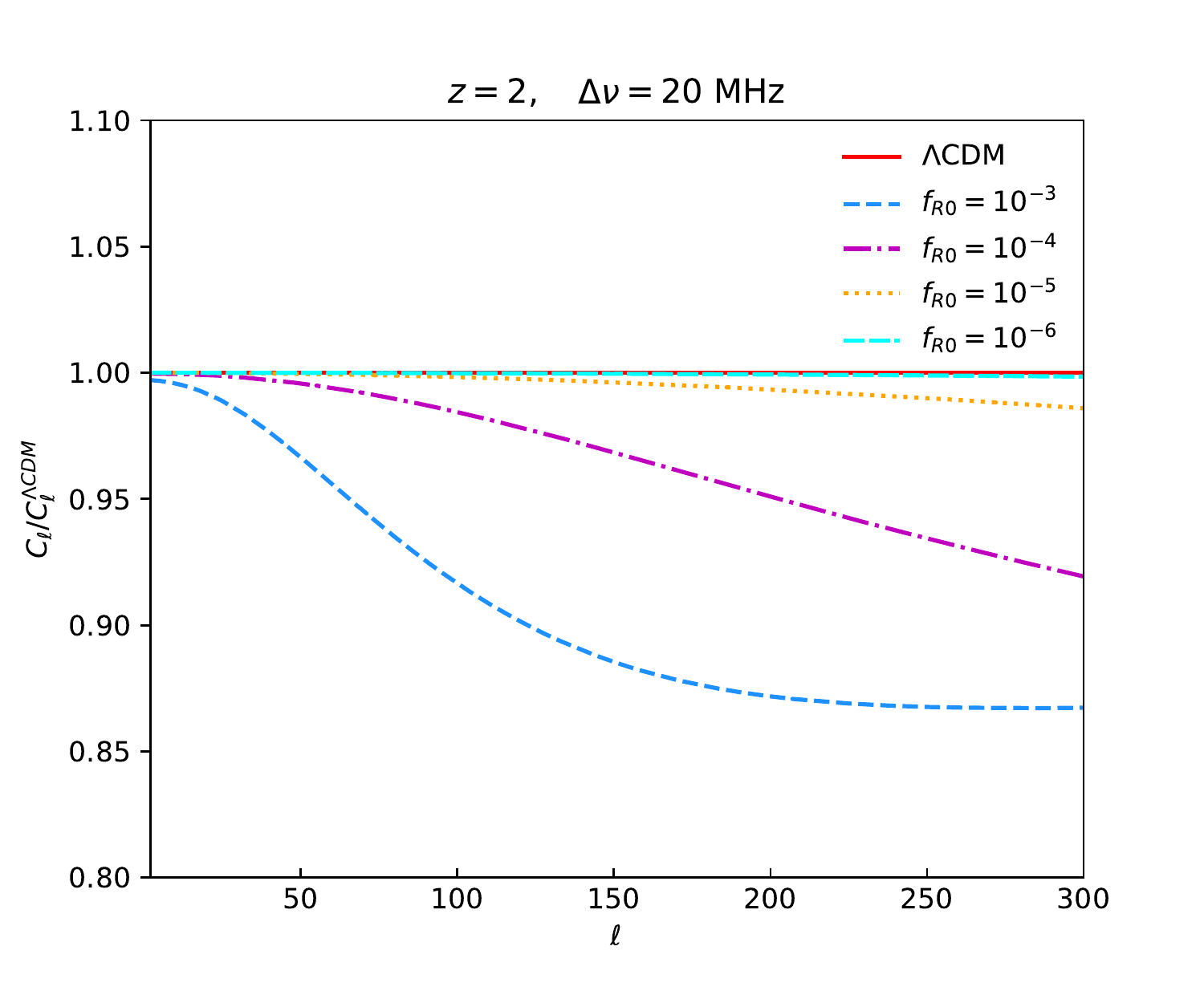}
	\caption{The dimensionless 21 cm auto APS ({\it left}) in the HS $f(R)$ gravity and their ratios ({\it right}) relative to the $\Lambda$CDM model are shown at $z=2$ for a narrow window $\Delta\nu=0.1$ MHz ({\it top}) and a broad window $\Delta\nu=20$ MHz ({\it bottom}), respectively. The red solid, blue short-dashed, magenta dash-dotted, orange dotted and cyan long-dashed lines denote $\Lambda$CDM, $f_{R0}=10^{-3},\,10^{-4}, \, 10^{-5}$ and $10^{-6}$, respectively. }
	\label{f2}
\end{figure} 

\begin{figure}[htbp]
	\centering
	\includegraphics[scale=0.55]{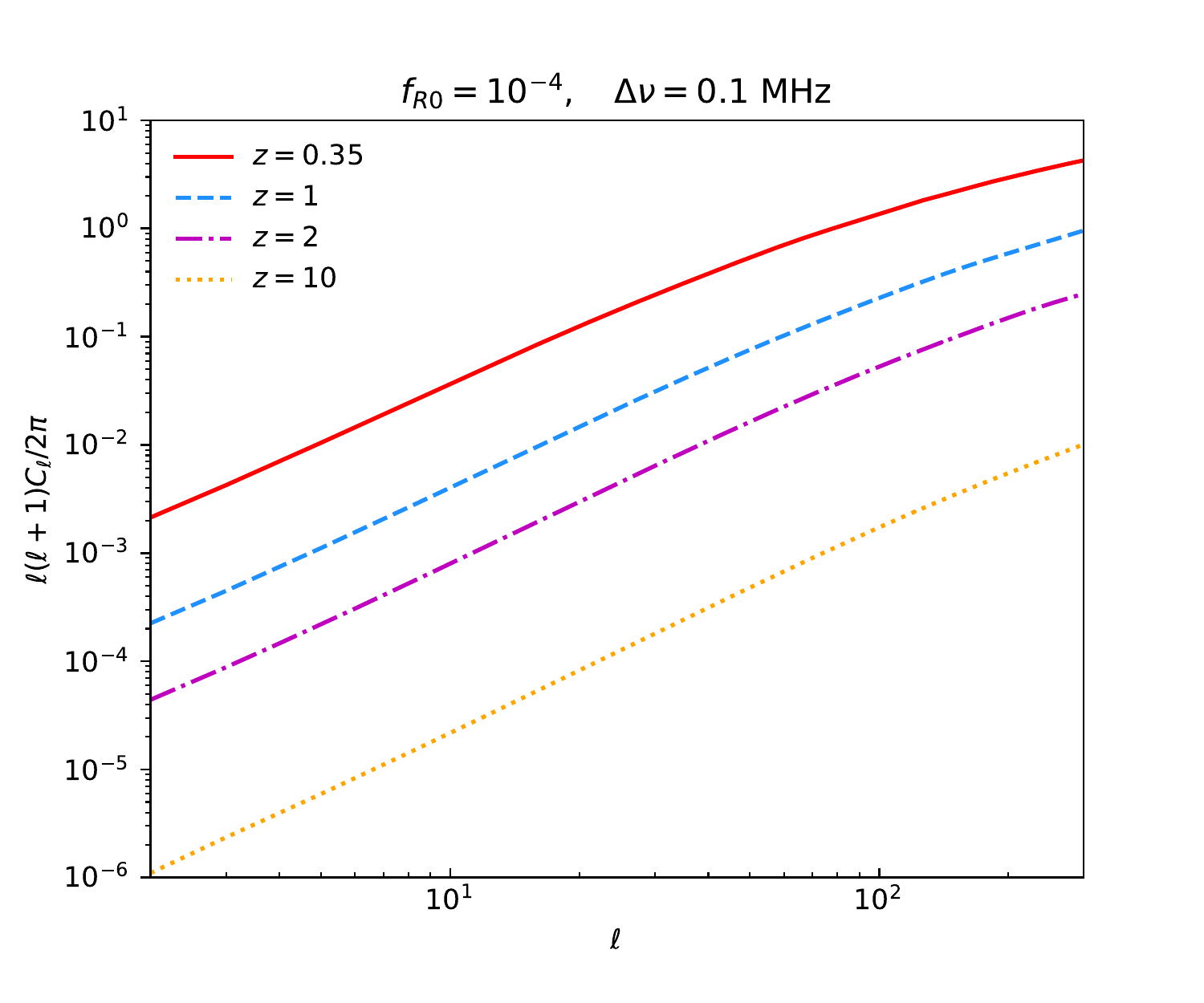}
	\includegraphics[scale=0.55]{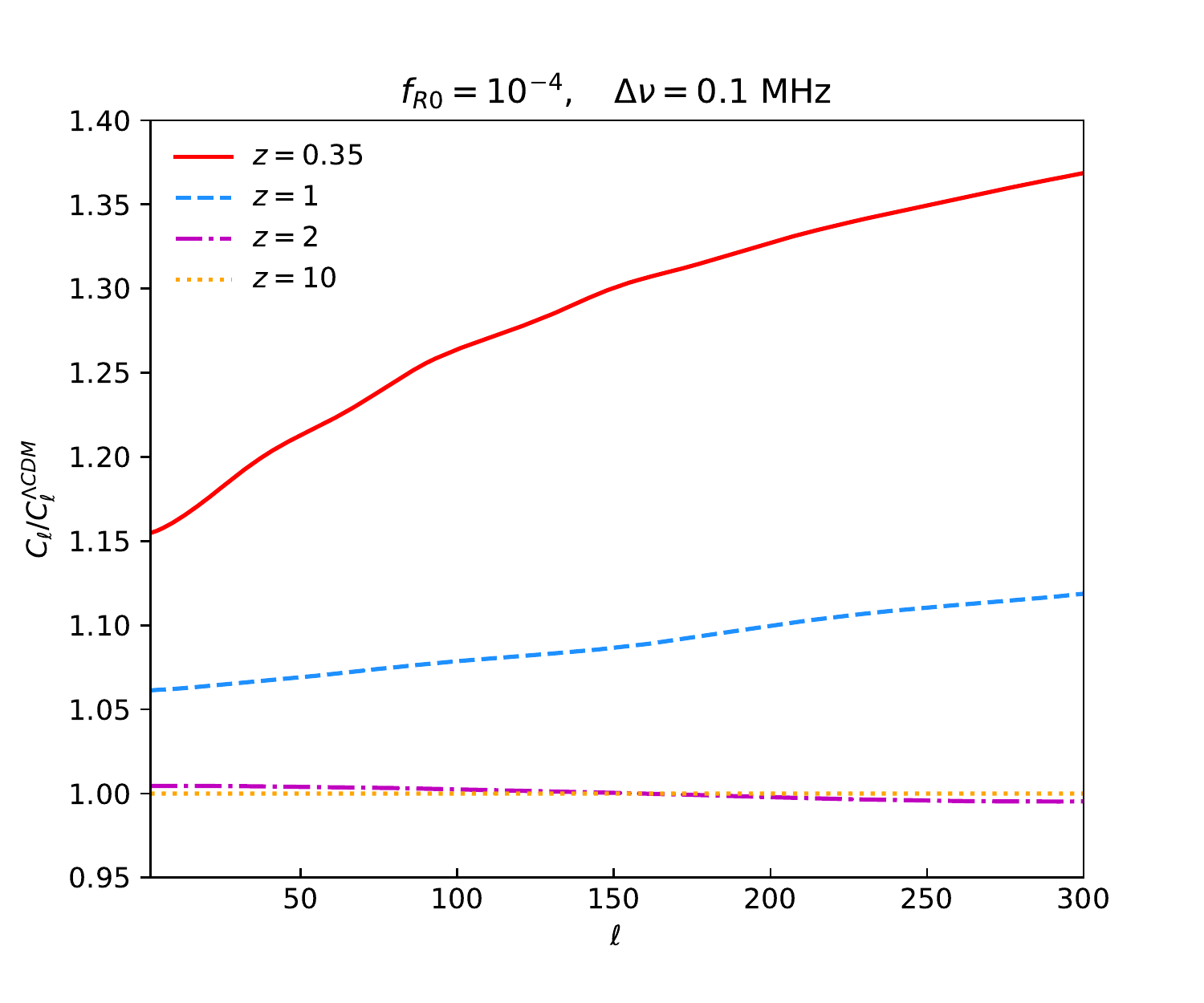}
	\includegraphics[scale=0.55]{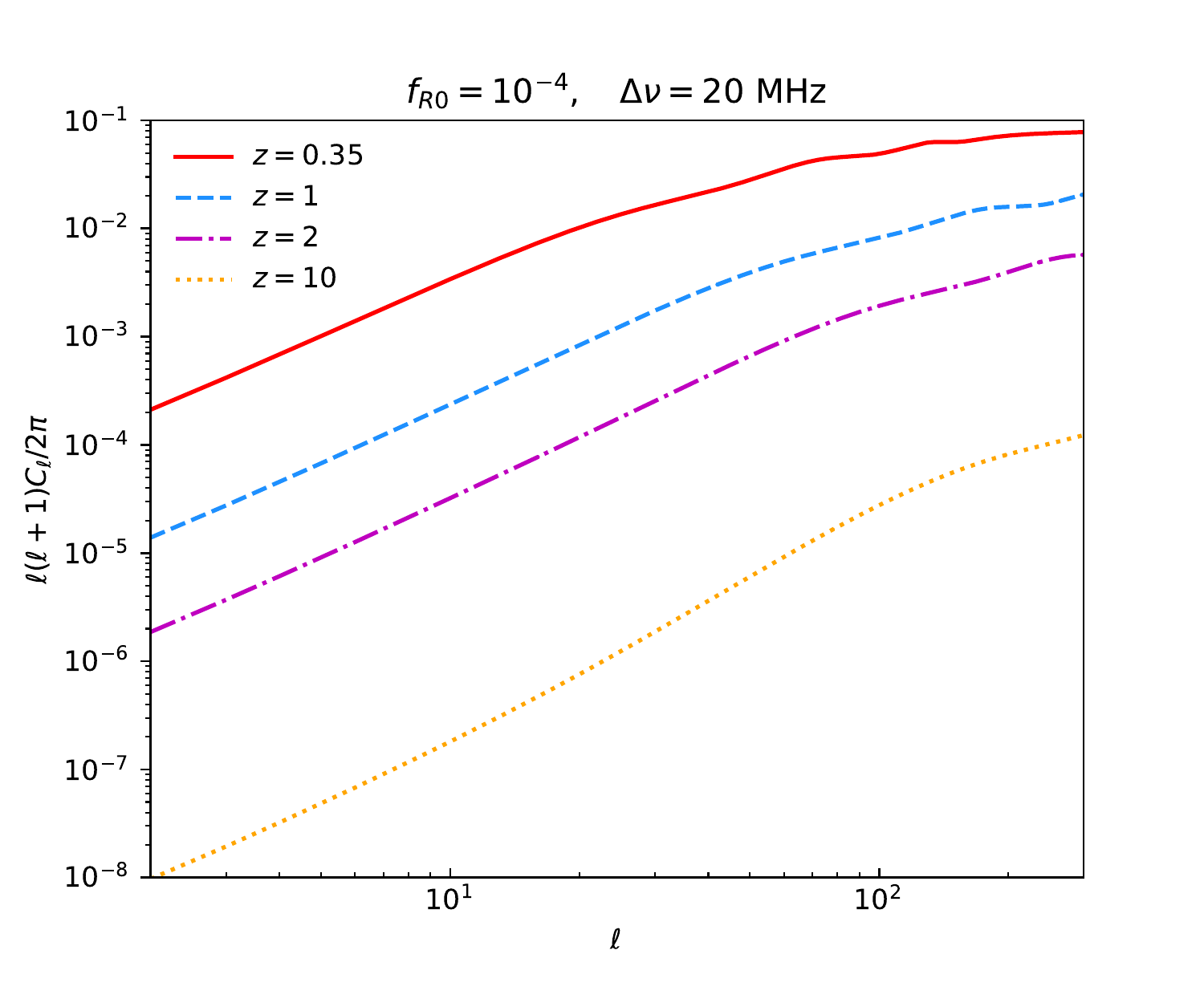}
    \includegraphics[scale=0.55]{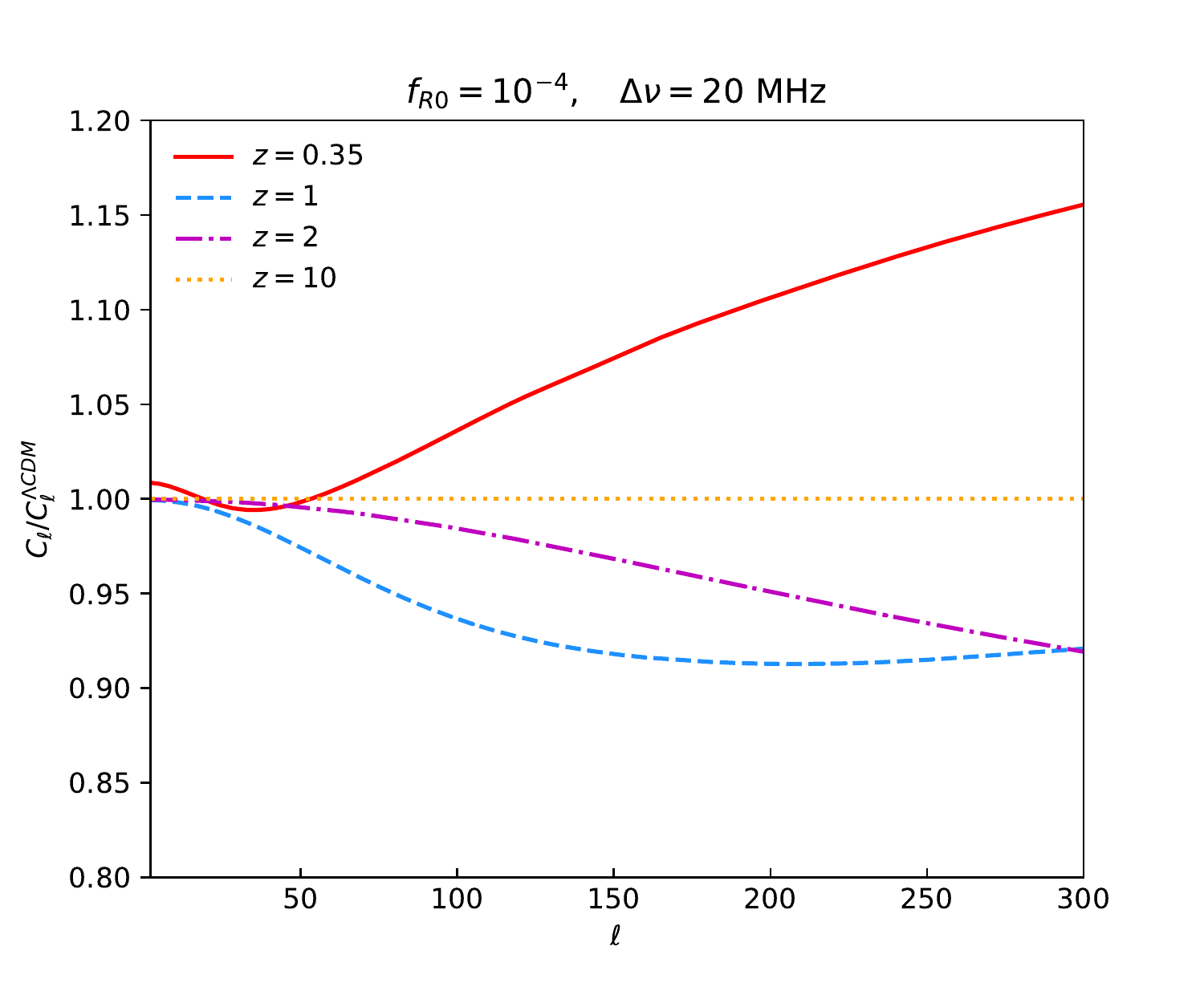}
	\caption{The dimensionless 21 cm auto APS ({\it left}) for the case of $f_{R0}=10^{-4}$ in the HS $f(R)$ gravity and their ratios ({\it right}) relative to the $\Lambda$CDM model are shown at different redshifts for a narrow window $\Delta\nu=0.1$ MHz ({\it top}) and a broad window $\Delta\nu=20$ MHz ({\it bottom}), respectively. The red solid, blue dashed, magenta dash-dotted and orange dotted lines denote $z=0.35, \, 1, \, 2$ and 10, respectively.}
	\label{f3}
\end{figure}

\begin{figure}[htbp]
	\centering
	\includegraphics[scale=0.55]{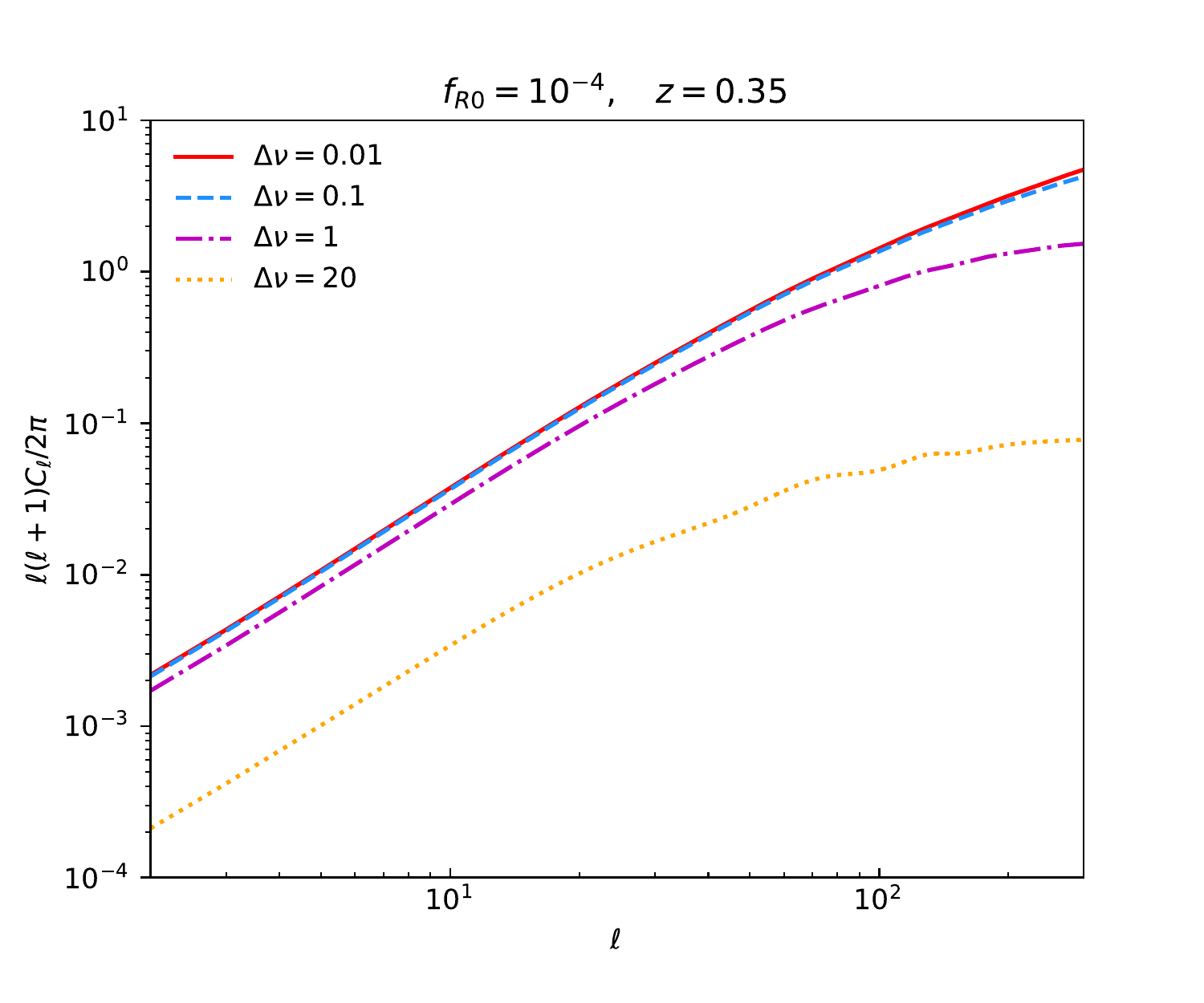}
	\includegraphics[scale=0.55]{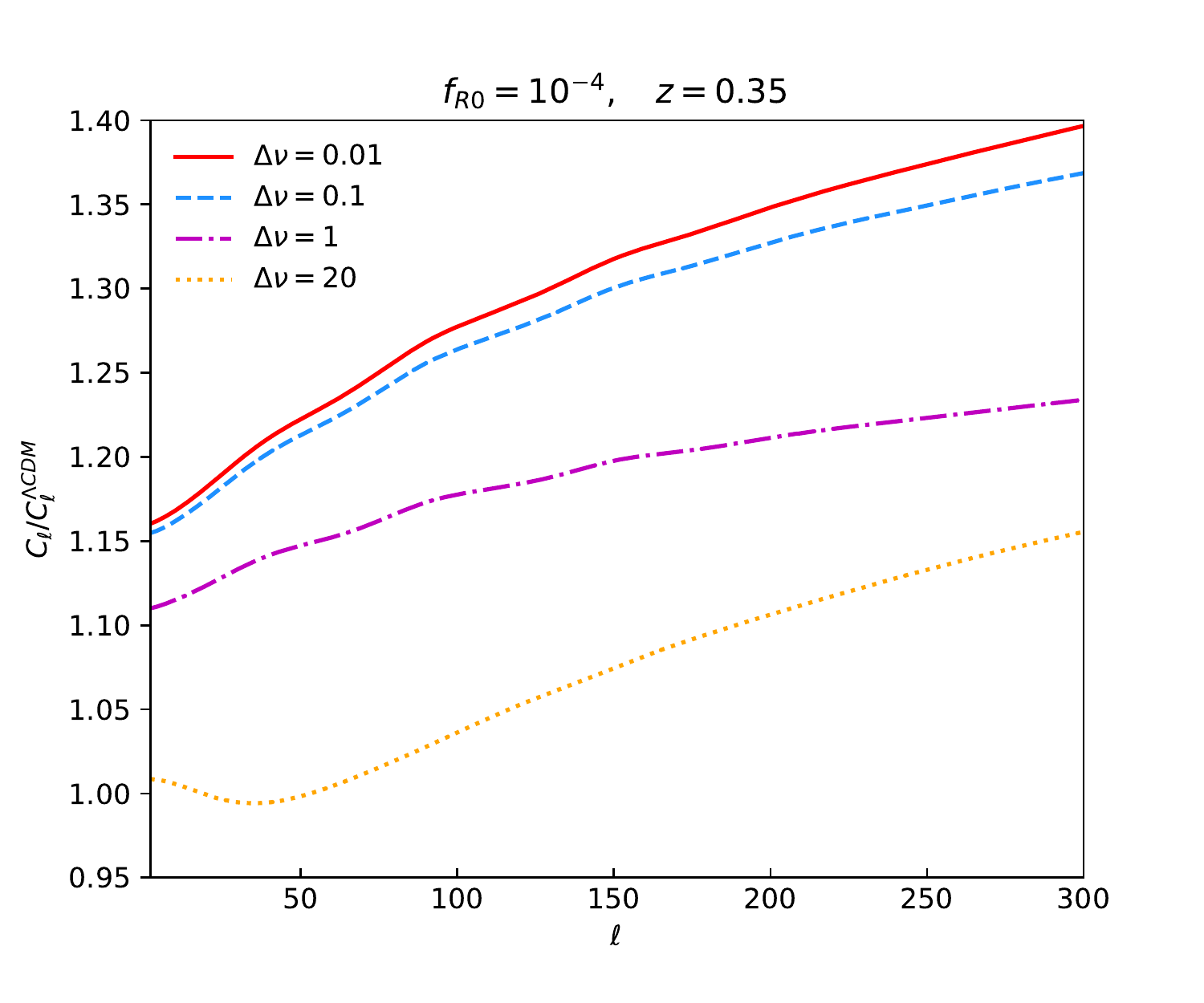}
	\includegraphics[scale=0.55]{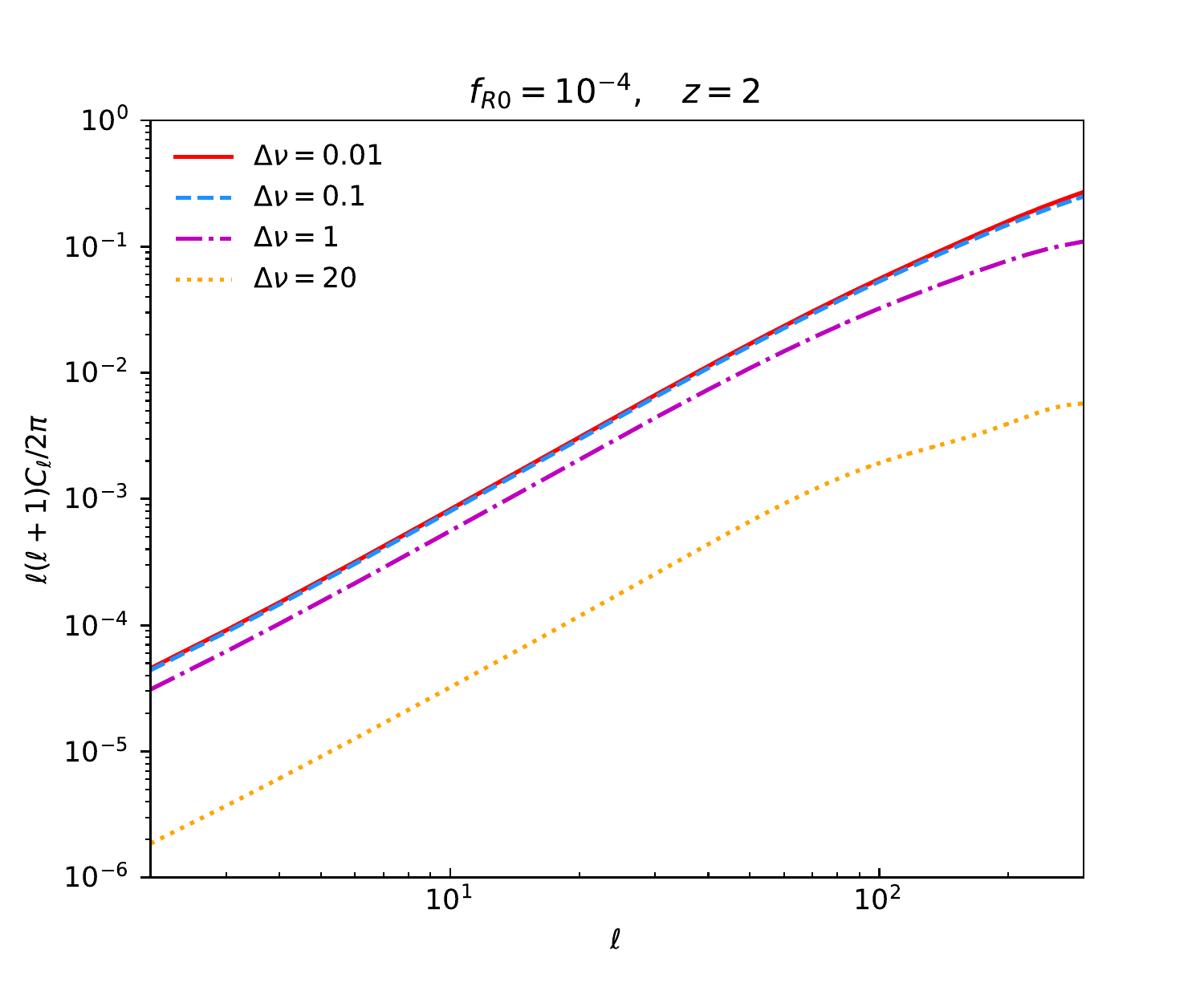}
    \includegraphics[scale=0.55]{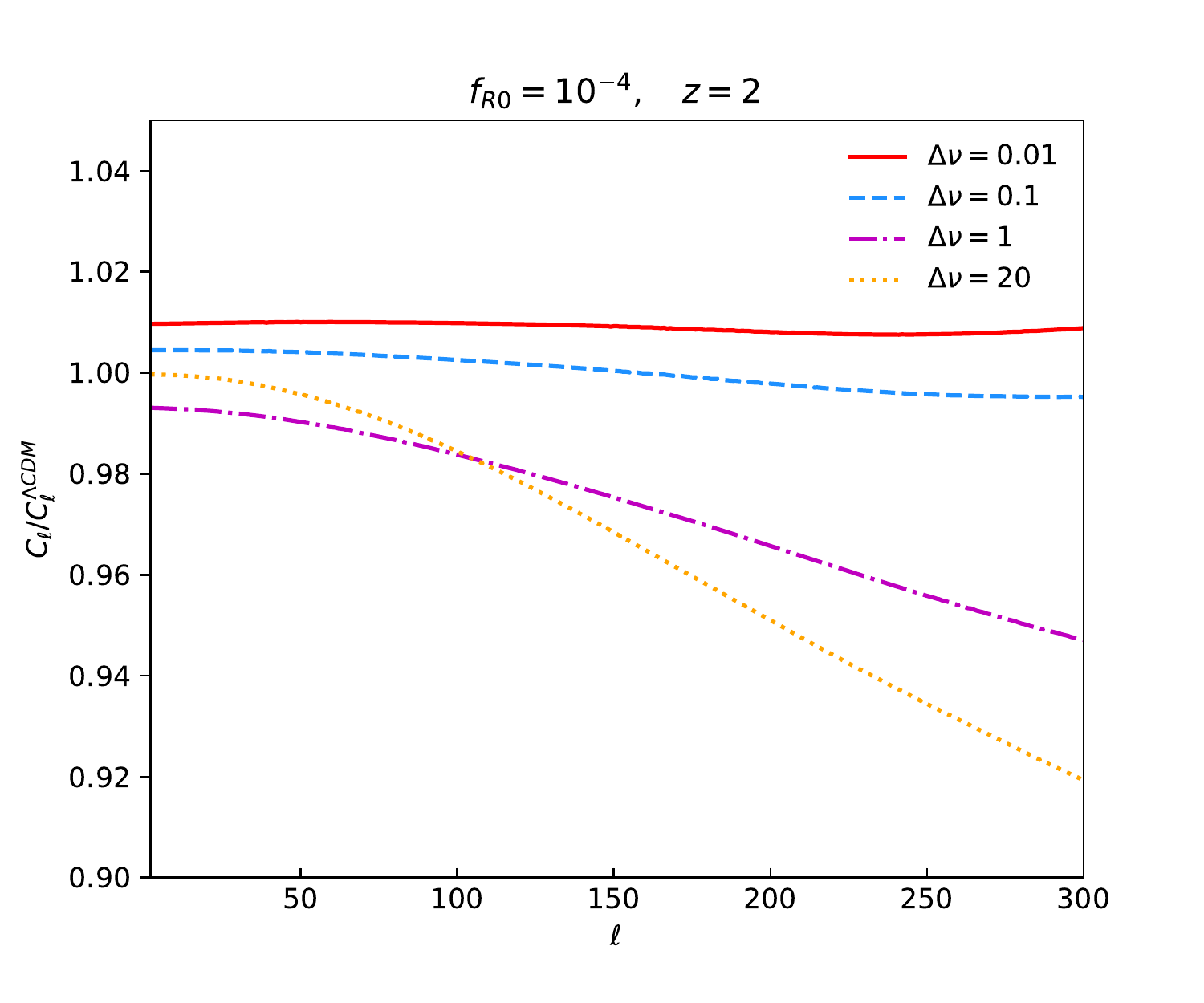}
	\caption{The dimensionless 21 cm auto APS ({\it left}) for the case of $f_{R0}=10^{-4}$ in the HS $f(R)$ gravity and their ratios ({\it right}) relative to the $\Lambda$CDM model are shown at $z=0.35$ ({\it top}) and $z=2$ ({\it bottom})  for different window widths, respectively. The red solid, blue dashed, magenta dash-dotted and orange dotted lines denote $\Delta\nu=0.01, \, 0.1, \, 1,$ and 20 MHz, respectively. }
	\label{f4}
\end{figure}

\begin{figure}[htbp]
	\centering
	\includegraphics[scale=0.55]{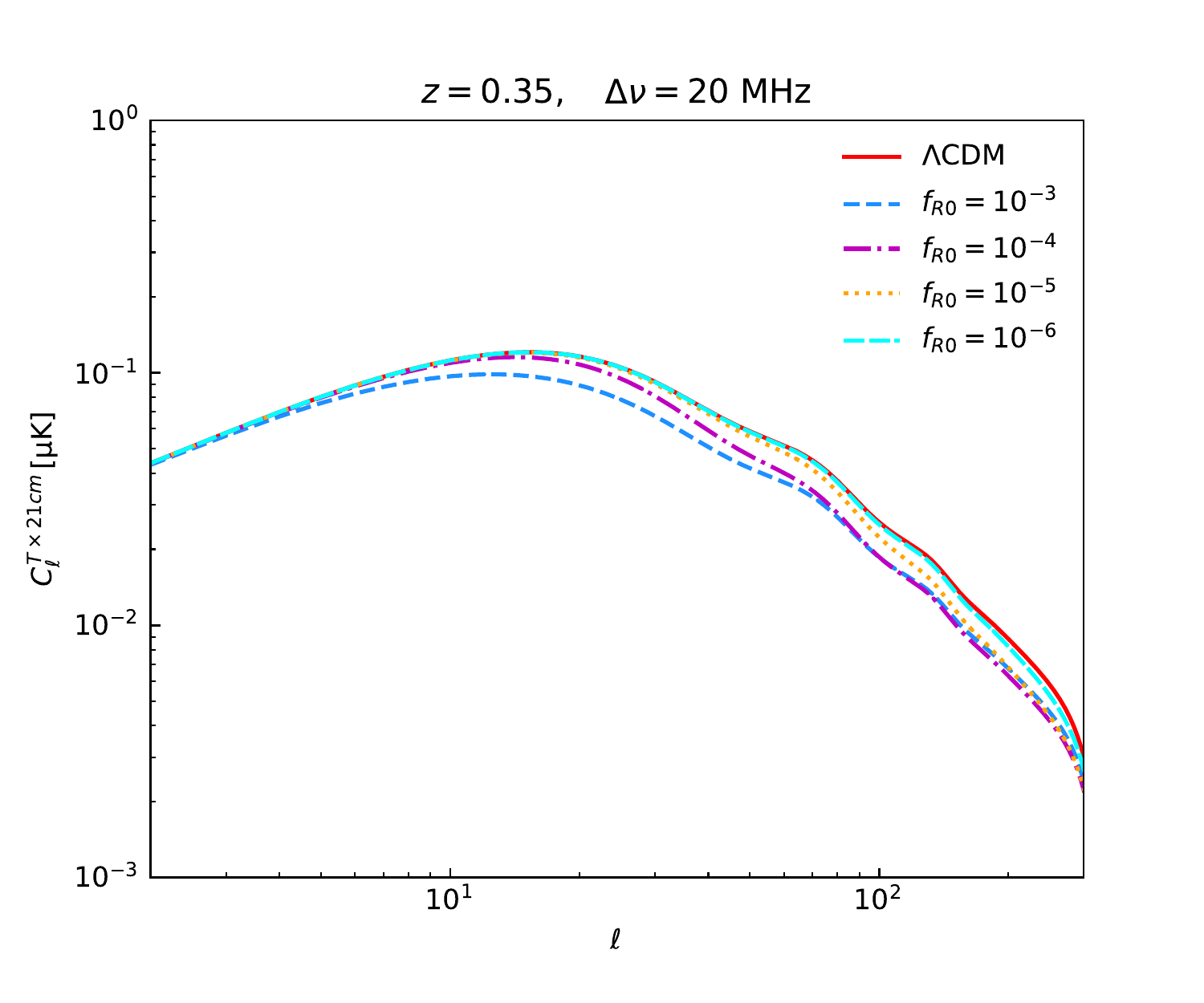}
	\includegraphics[scale=0.55]{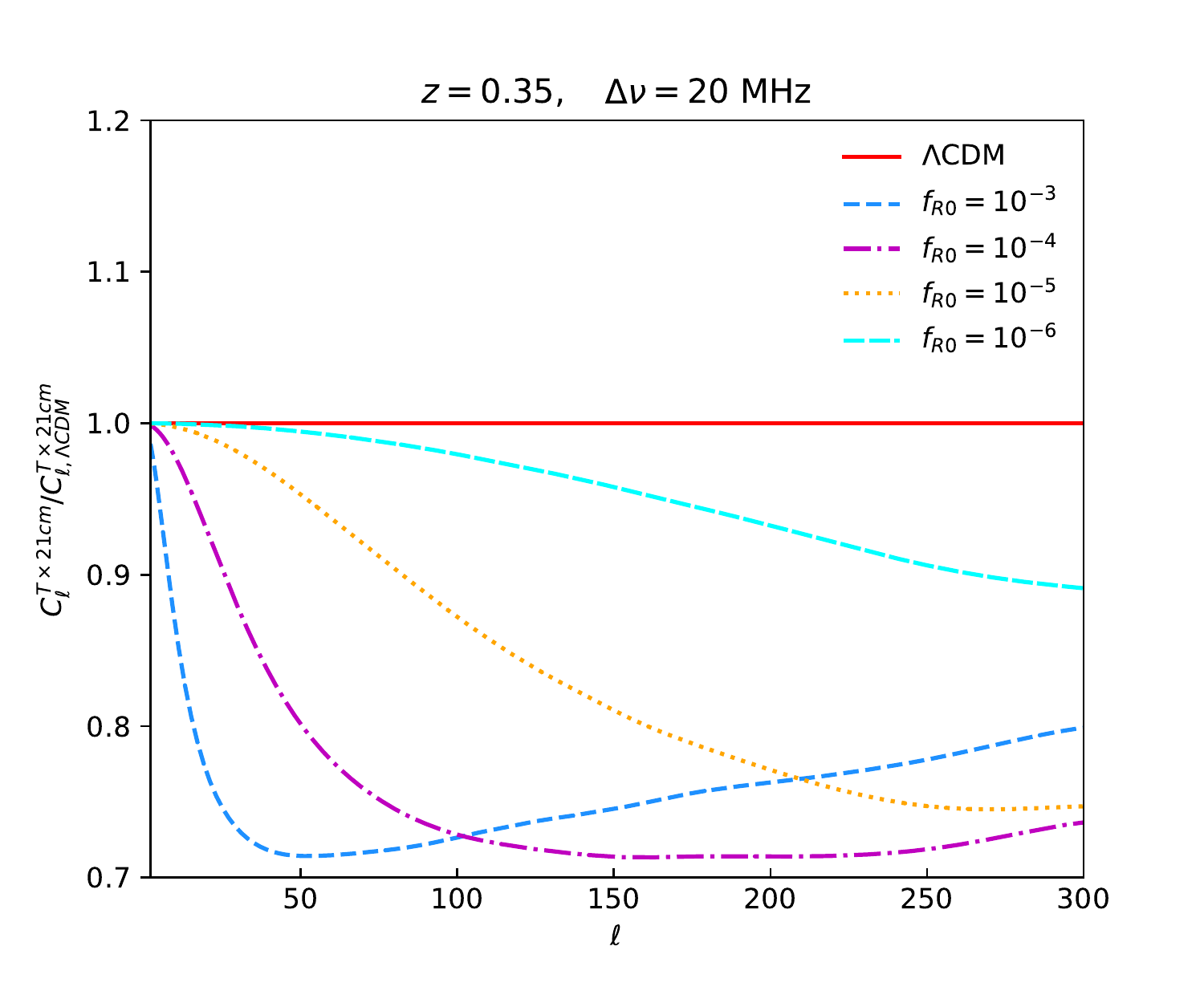}
	\includegraphics[scale=0.55]{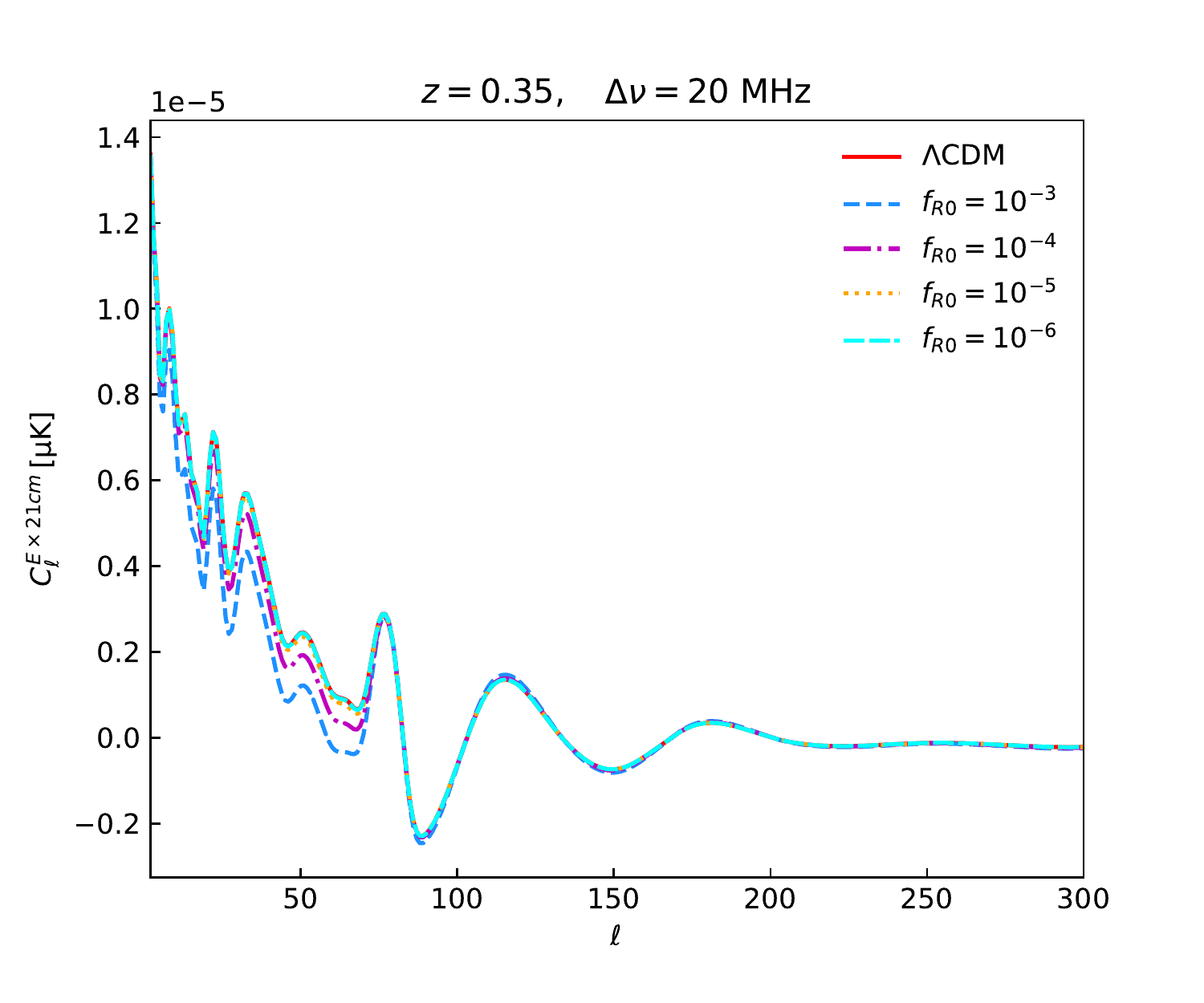}
    \includegraphics[scale=0.55]{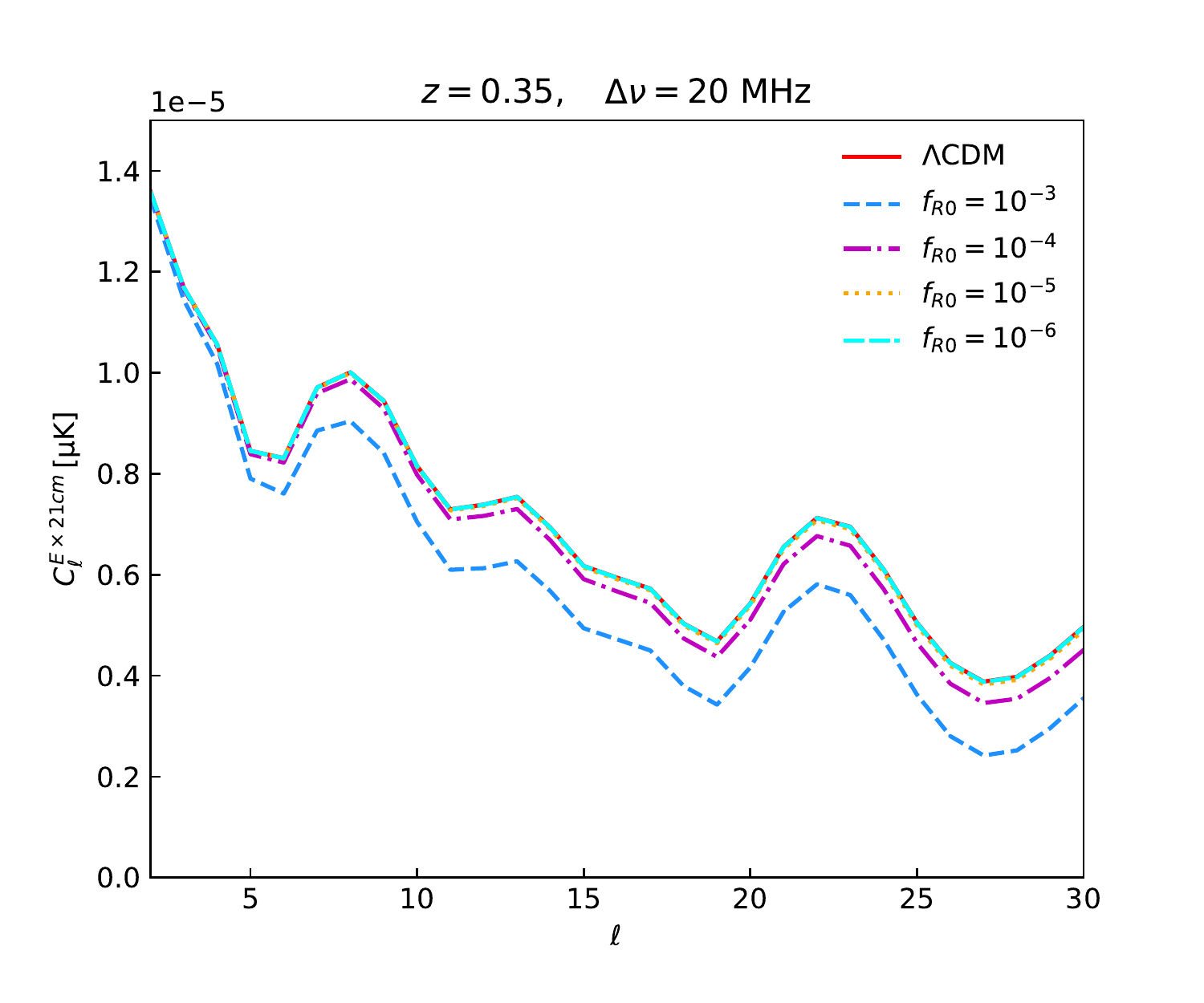}
    \includegraphics[scale=0.55]{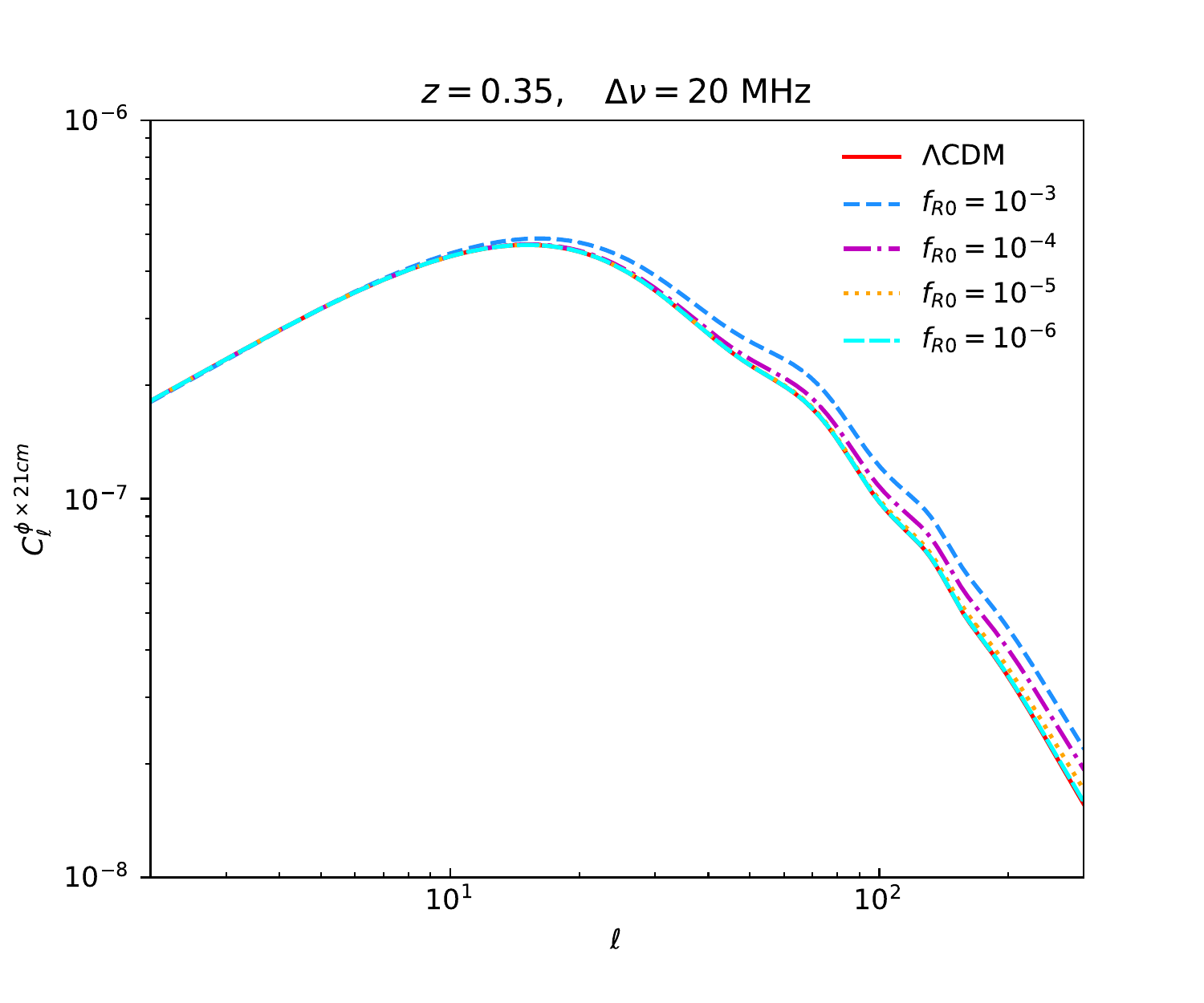}
    \includegraphics[scale=0.55]{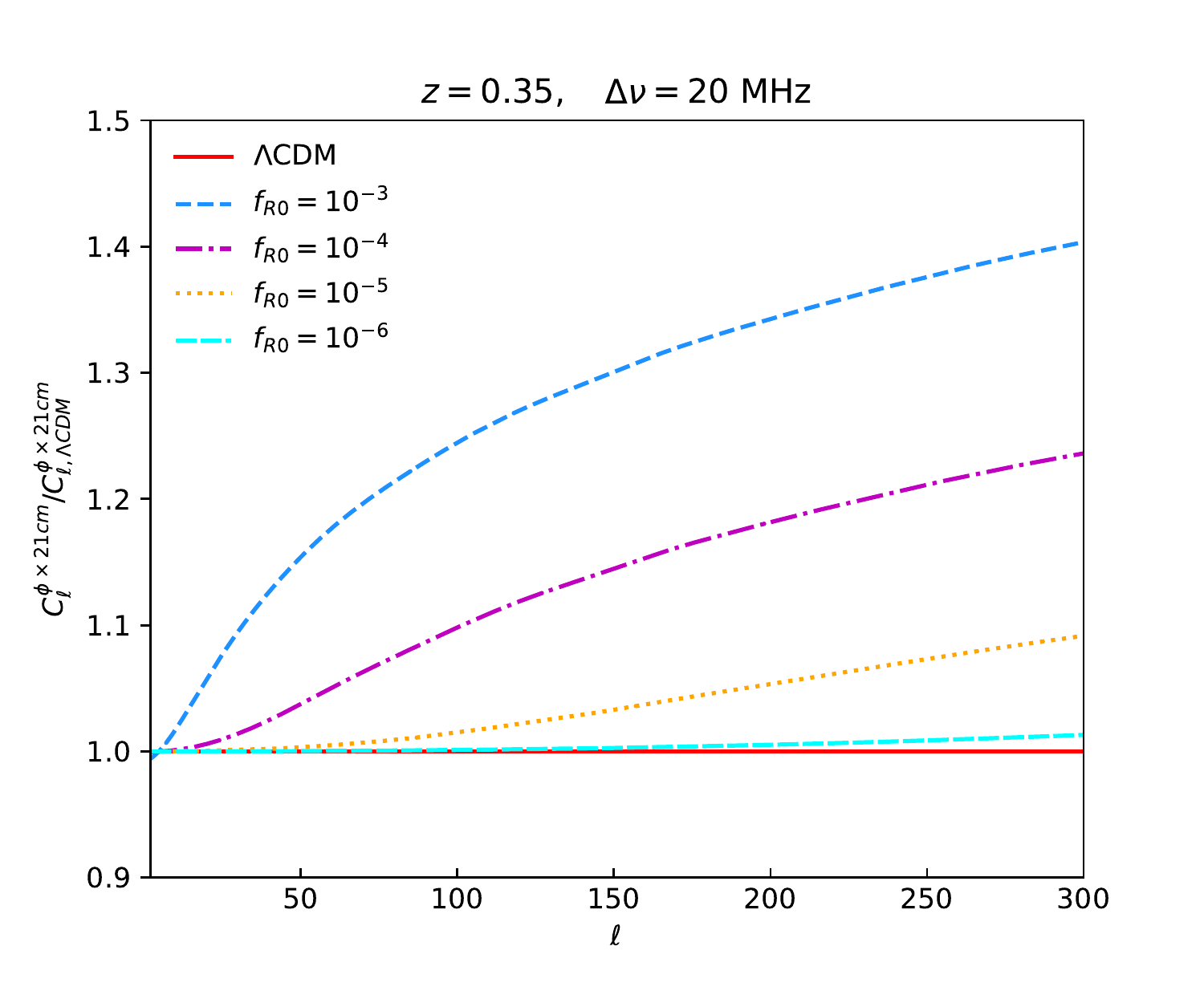}
	\caption{{\it Top panels.} The cross APS ({\it left}) between the CMB temperature and 21 cm radiation in the HS $f(R)$ gravity and their ratios ({\it right}) relative to the $\Lambda$CDM model are shown. {\it Medium panels.} The cross APS ({\it left}) between the CMB E-mode polarization and 21 cm radiation in the HS $f(R)$ gravity and their behaviors at large angular scales ({\it right}) are shown. {\it Bottom panels.} The cross APS ({\it left}) between the CMB lensing potential and 21 cm radiation in the HS $f(R)$ gravity and their ratios ({\it right}) relative to $\Lambda$CDM are shown. The red solid, blue short-dashed, magenta dash-dotted, orange dotted and cyan long-dashed lines denote $\Lambda$CDM, $f_{R0}=10^{-3},\,10^{-4}, \, 10^{-5}$ and $10^{-6}$, respectively. Here we take $z=0.35$ and a broad window $\Delta\nu=20$ MHz for the 21 cm radiation. Note that for all the cross APS, the unit should be $\mathrm{\mu K}$, since we take the dimensionless 21 cm APS in the numerical calculations.}
	\label{f5}
\end{figure}

\begin{figure}[htbp]
	\centering
	\includegraphics[scale=0.55]{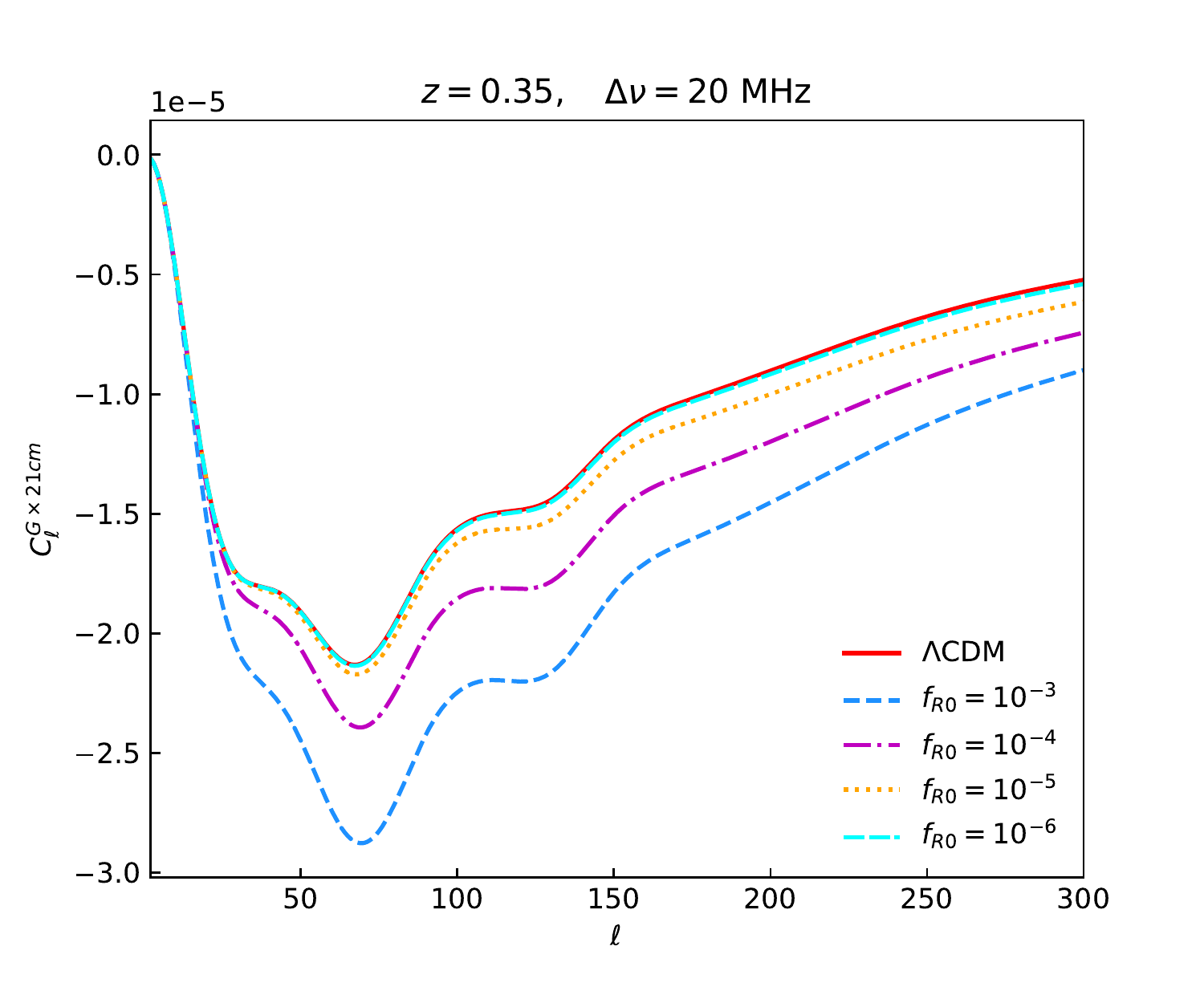}
	\includegraphics[scale=0.55]{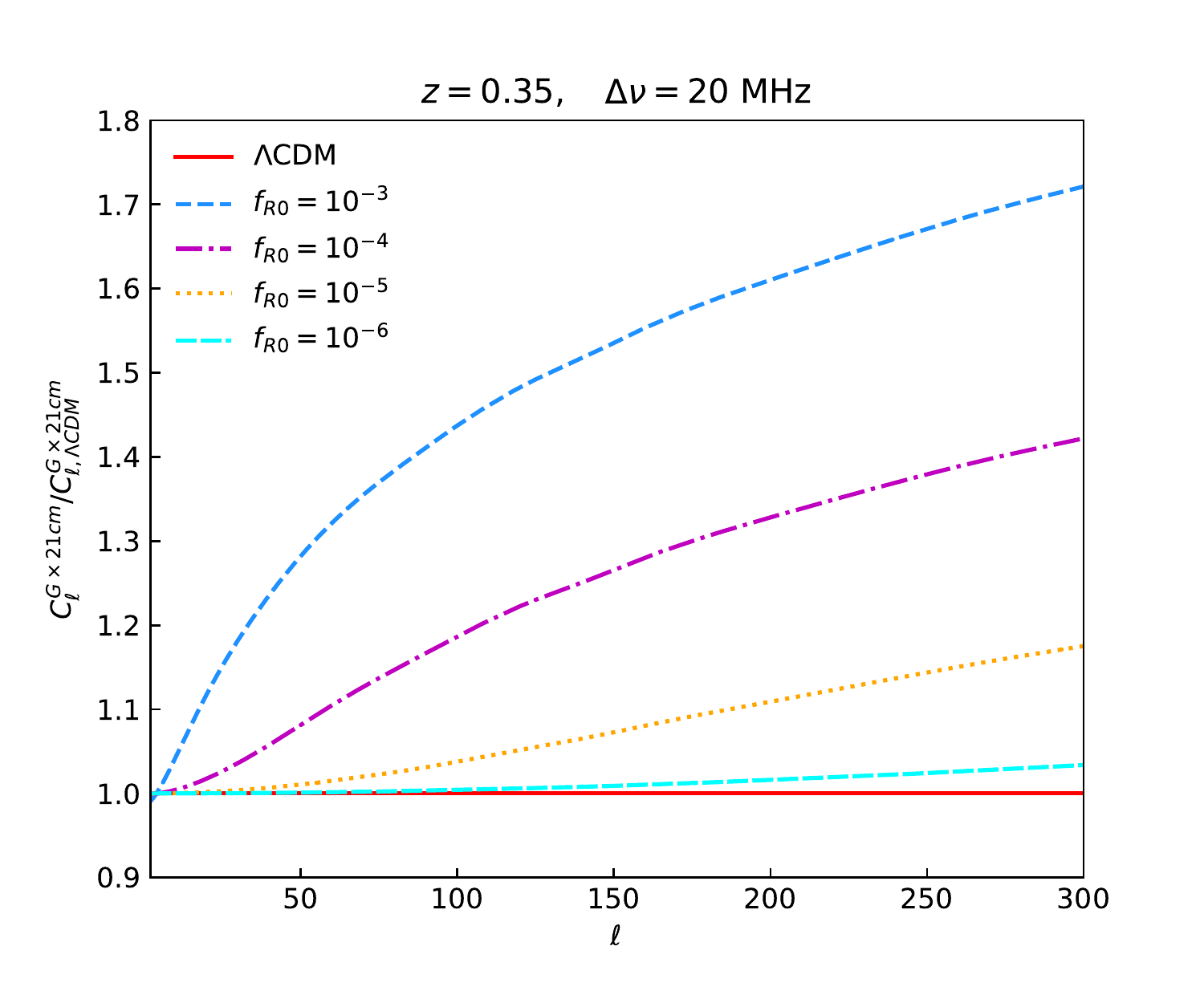}
	\includegraphics[scale=0.545]{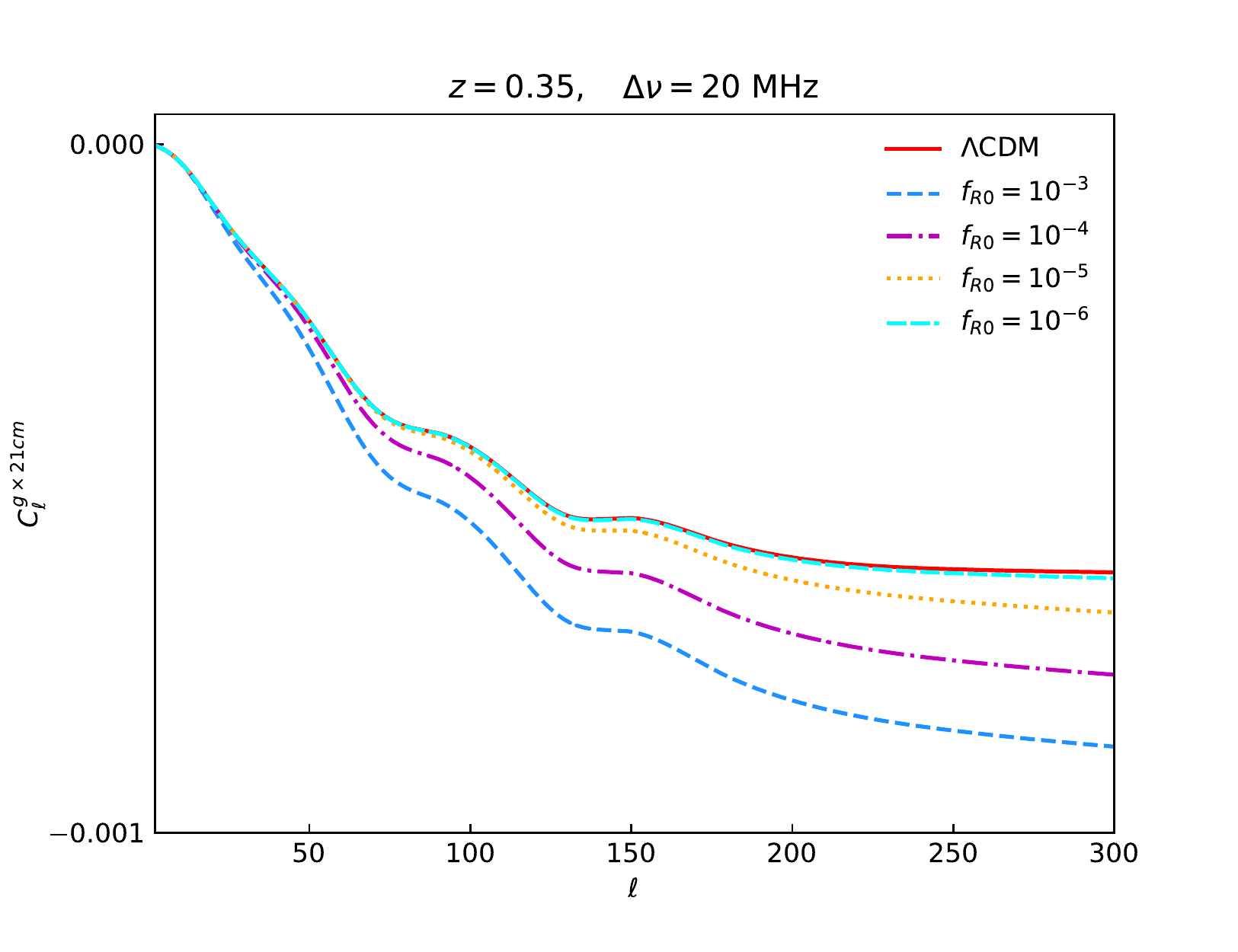}
	\includegraphics[scale=0.55]{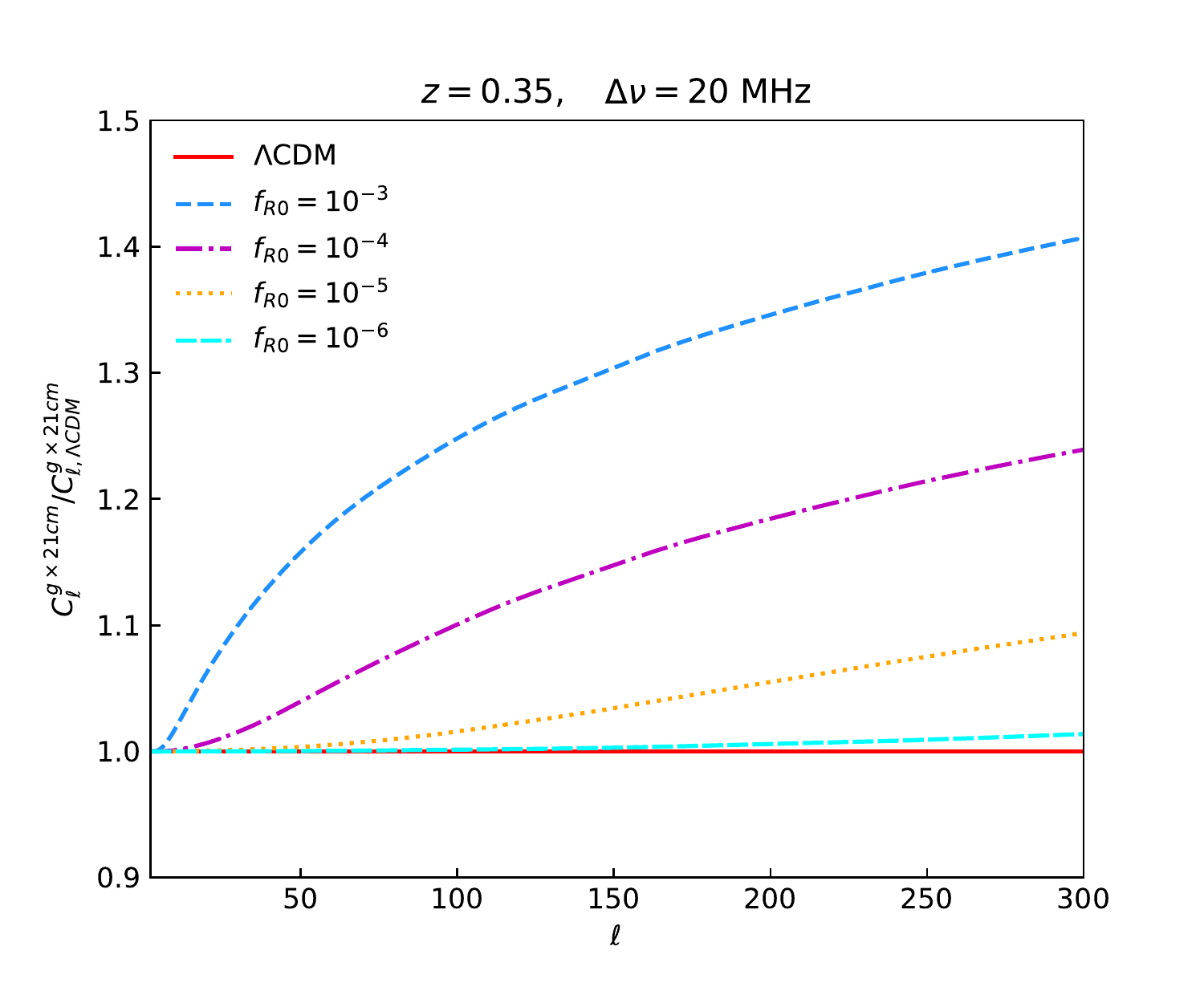}
	\caption{{\it Top panels.} The cross APS ({\it left}) between the cosmic shear and 21 cm radiation in the HS $f(R)$ gravity and their ratios ({\it right}) relative to the $\Lambda$CDM model are shown. {\it Bottom panels.} The cross APS ({\it left}) between the galaxy clustering and 21 cm radiation in the HS $f(R)$ gravity and their ratios ({\it right}) relative to $\Lambda$CDM are shown. The red solid, blue short-dashed, magenta dash-dotted, orange dotted and cyan long-dashed lines denote $\Lambda$CDM, $f_{R0}=10^{-3},\,10^{-4}, \, 10^{-5}$ and $10^{-6}$, respectively. Here we take $z=0.35$ and a broad window $\Delta\nu=20$ MHz for the 21 cm radiation.}
	\label{f6}
\end{figure}

\begin{figure}[htbp]
	\centering
	\includegraphics[scale=0.55]{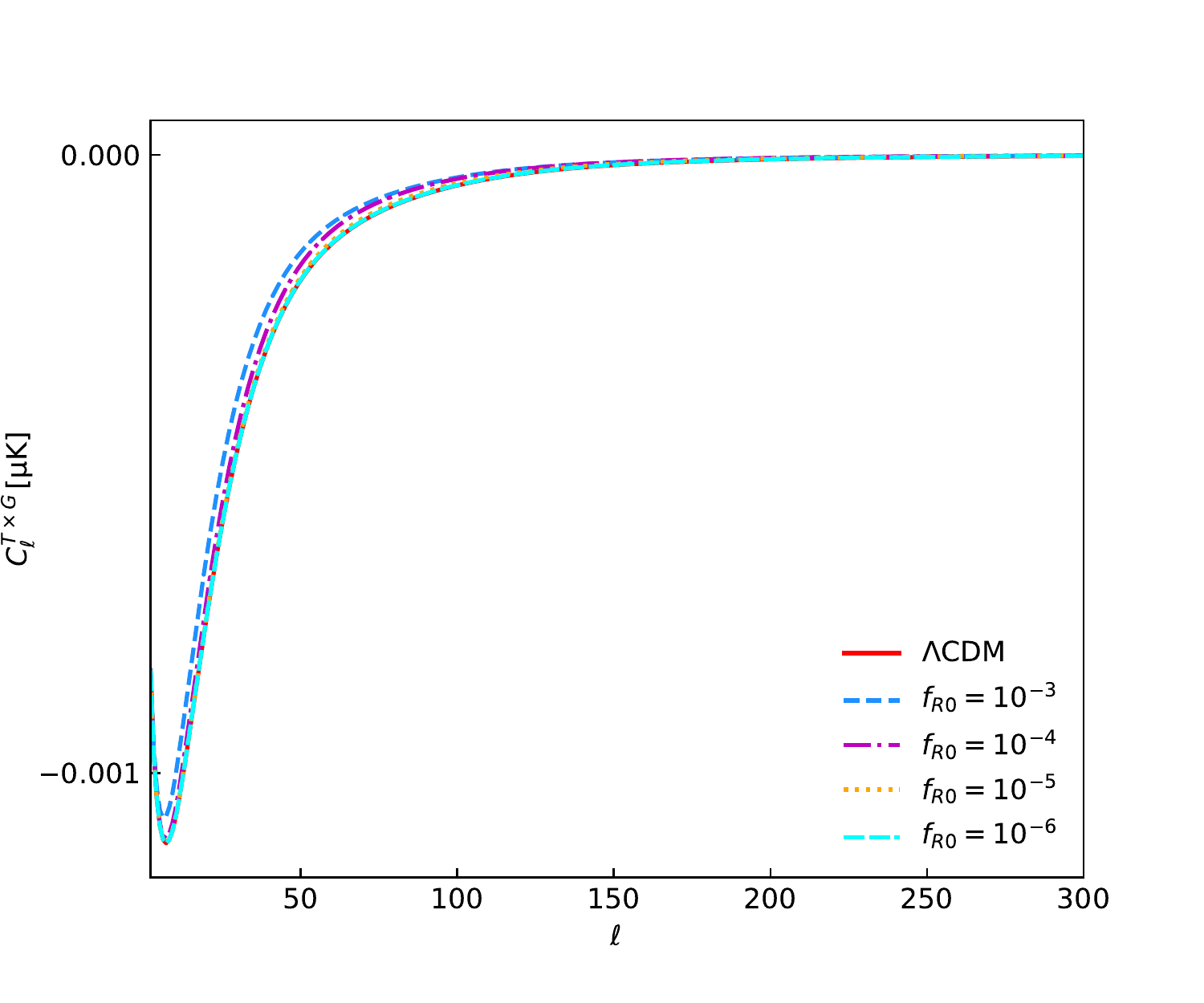}
	\includegraphics[scale=0.55]{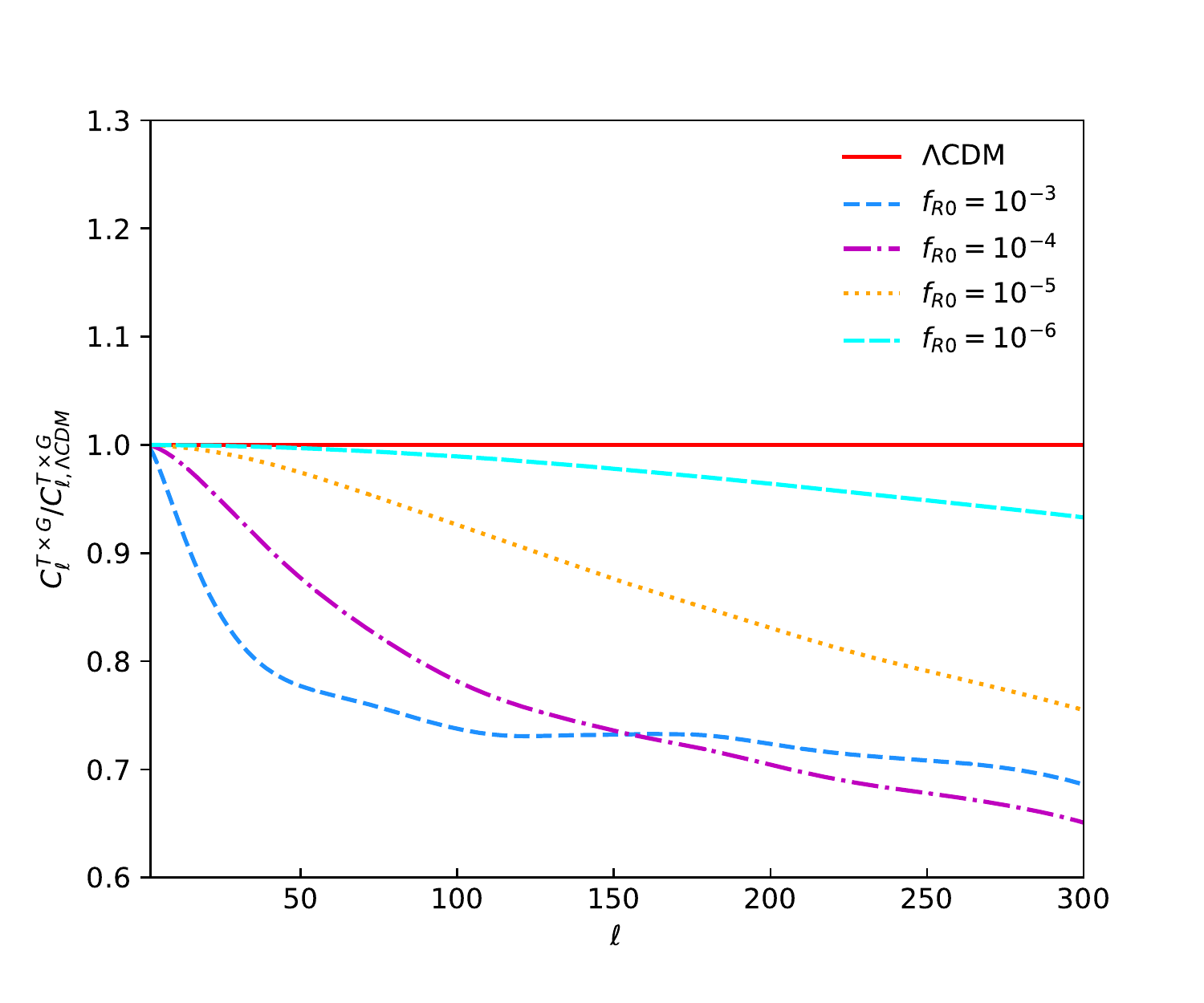}
	\includegraphics[scale=0.55]{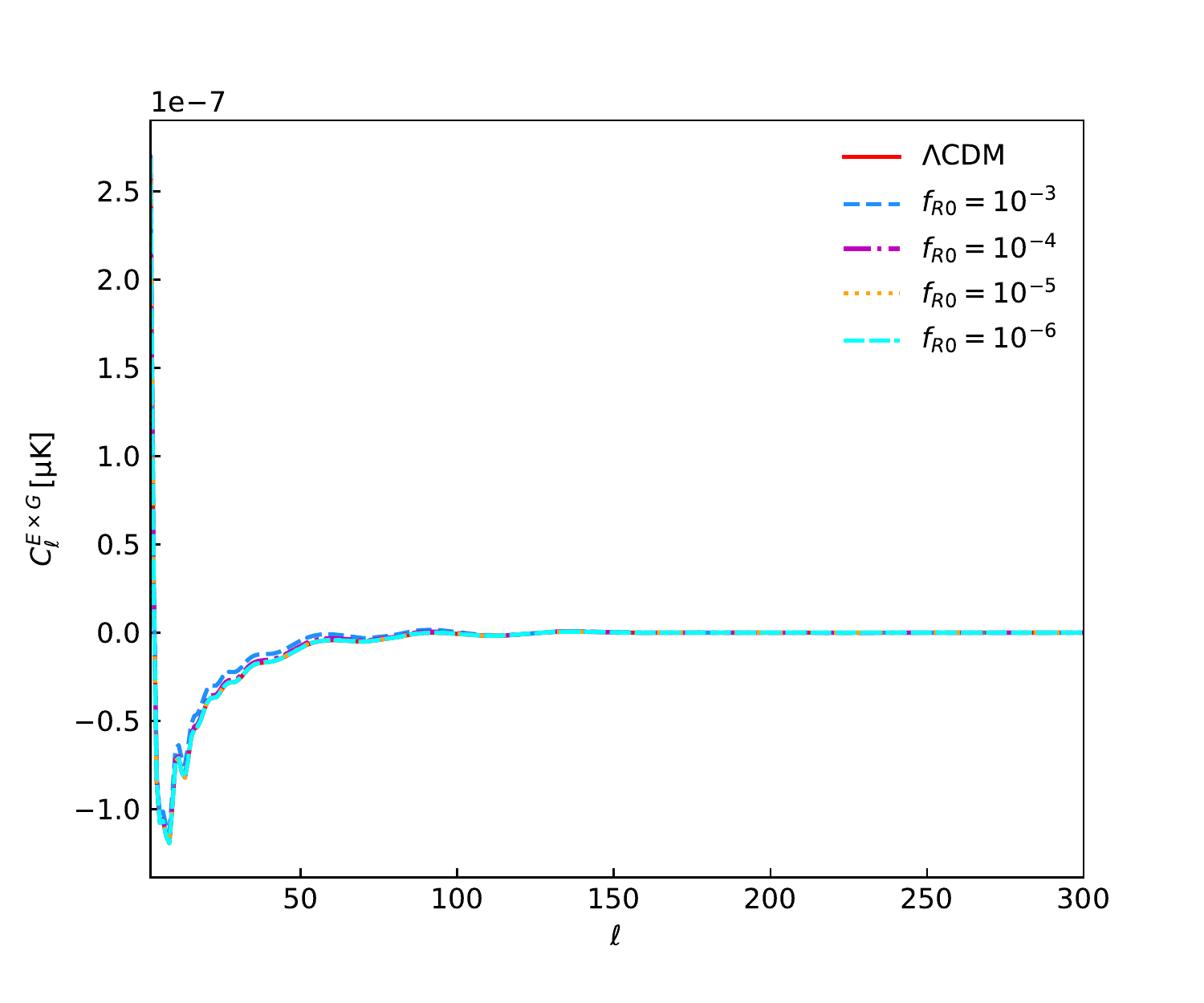}
    \includegraphics[scale=0.55]{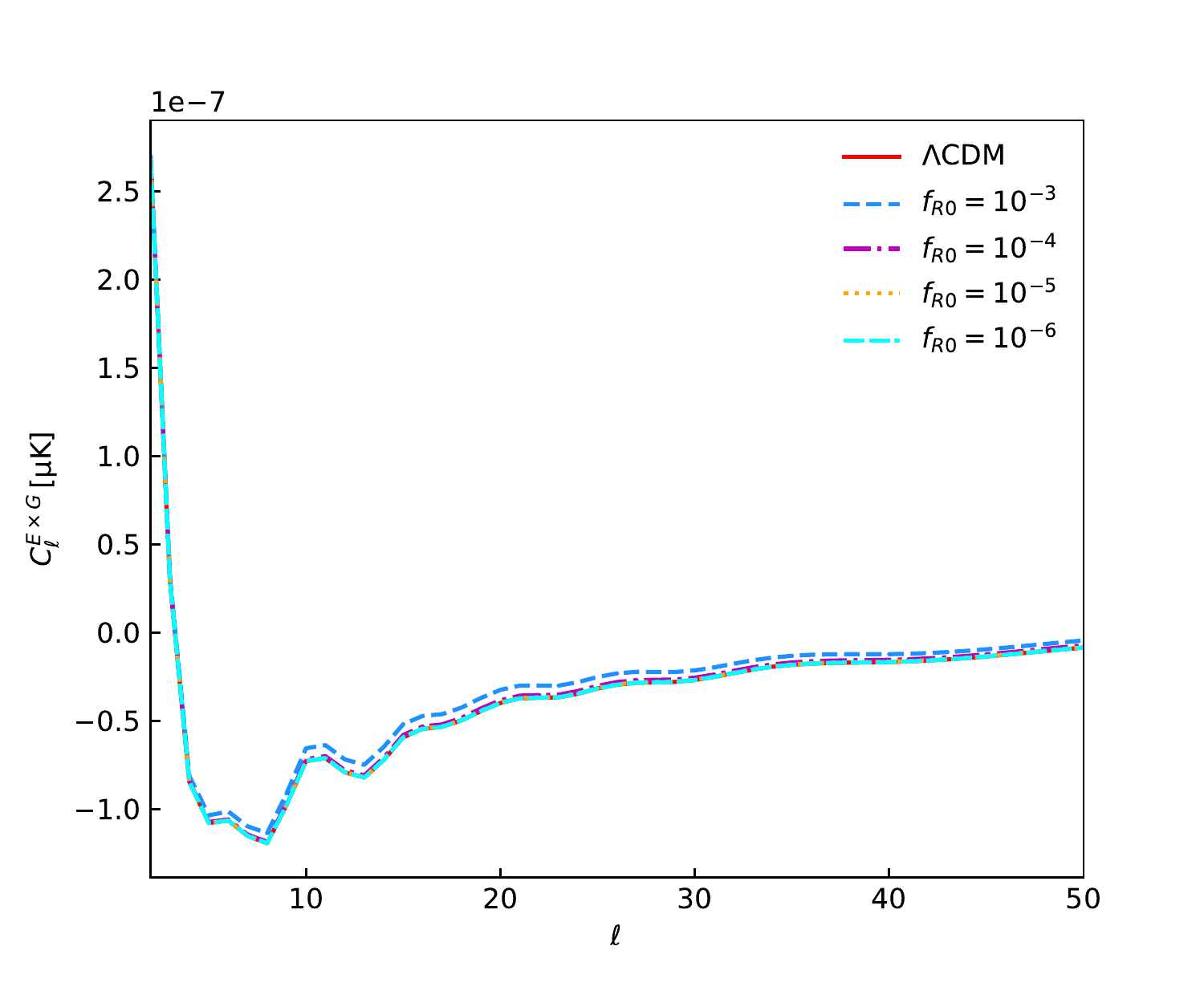}
    \includegraphics[scale=0.55]{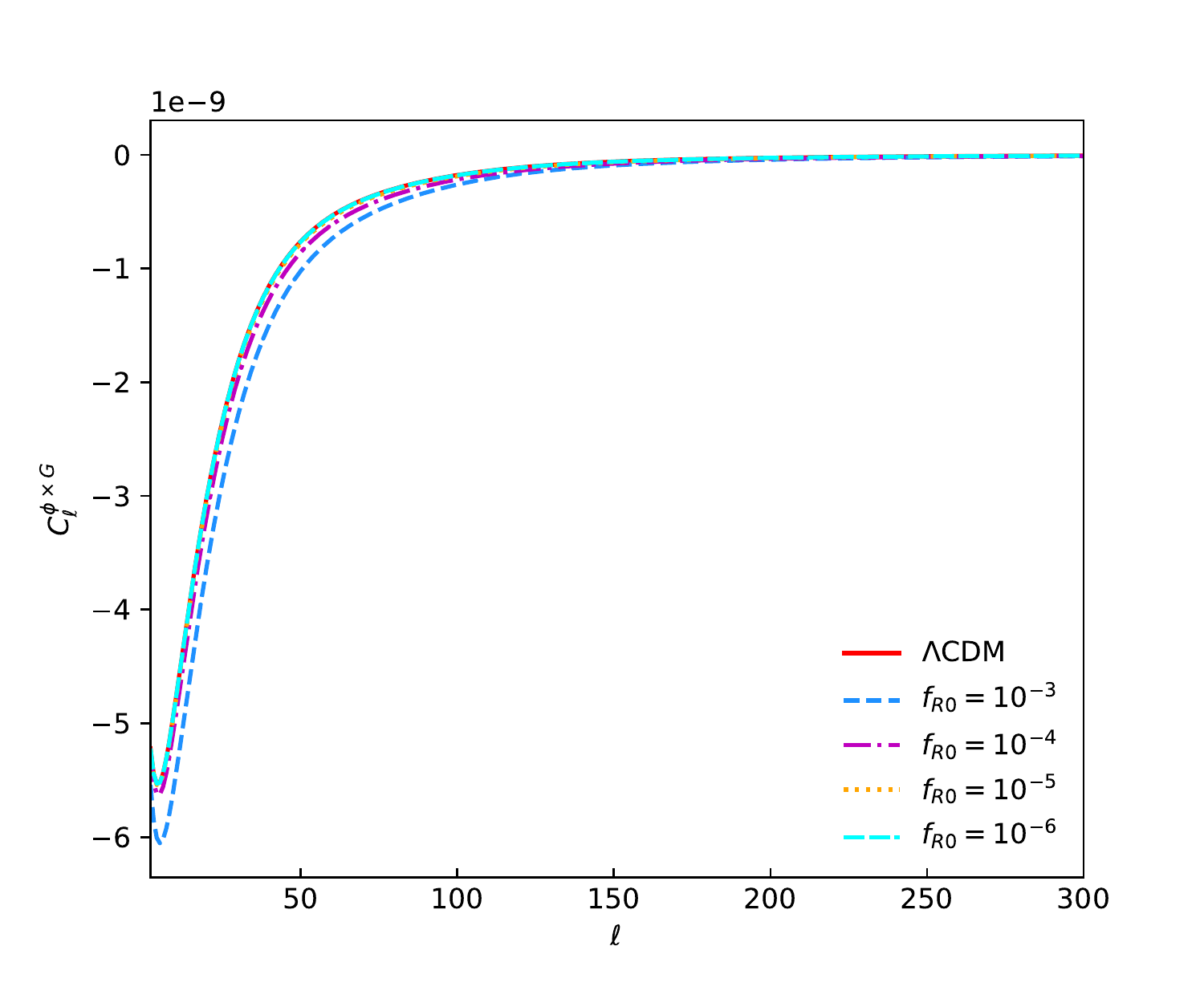}
    \includegraphics[scale=0.55]{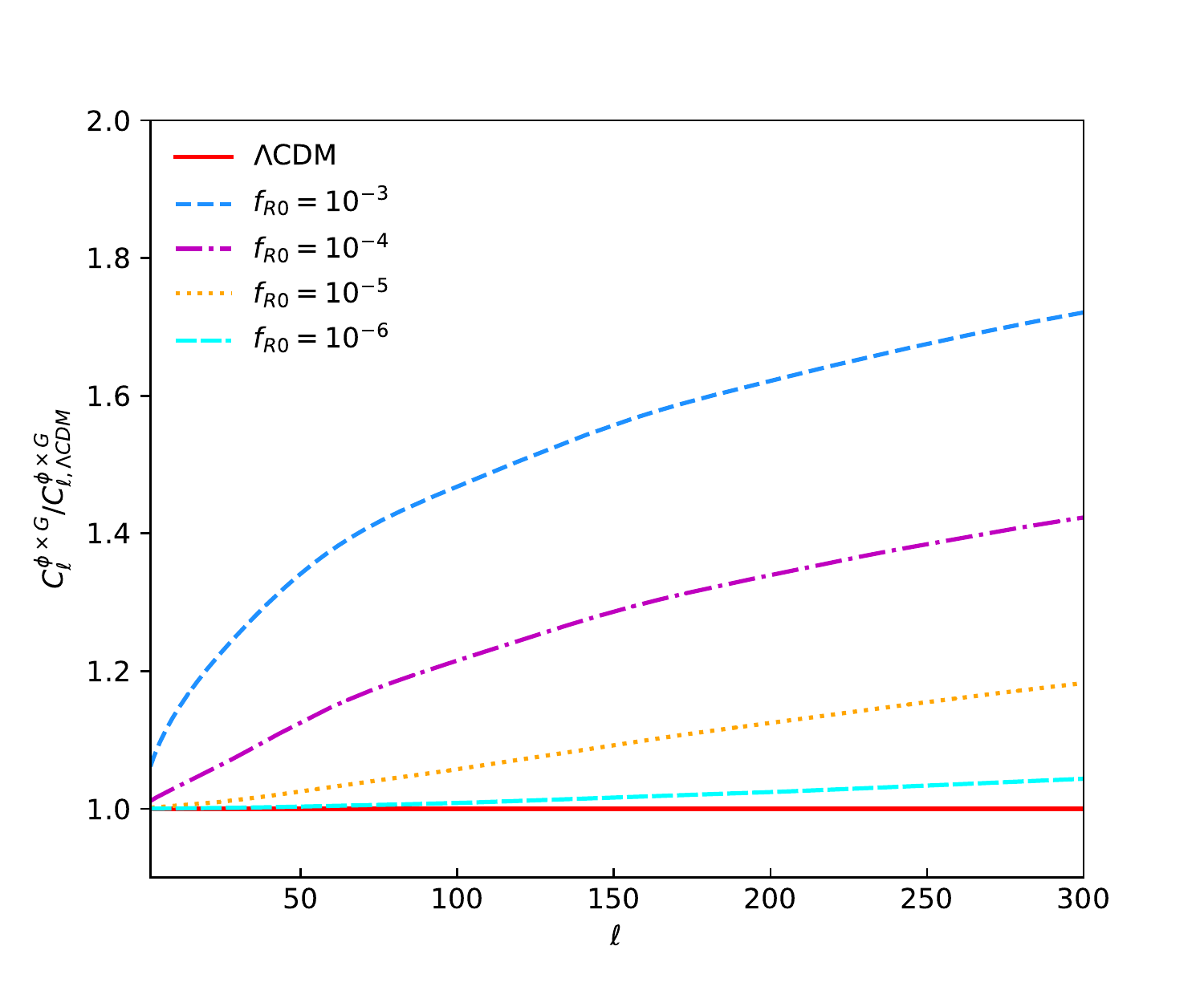}
	\caption{{\it Top panels.} The cross APS ({\it left}) between the CMB temperature and cosmic shear in the HS $f(R)$ gravity and their ratios ({\it right}) relative to the $\Lambda$CDM model are shown. {\it Medium panels.} The cross APS ({\it left}) between the CMB E-mode polarization and cosmic shear in the HS $f(R)$ gravity and their behaviors at large angular scales ({\it right}) are shown. {\it Bottom panels.} The cross APS ({\it left}) between the CMB lensing potential and cosmic shear in the HS $f(R)$ gravity and their ratios ({\it right}) relative to $\Lambda$CDM are shown. The red solid, blue short-dashed, magenta dash-dotted, orange dotted and cyan long-dashed lines denote $\Lambda$CDM, $f_{R0}=10^{-3},\,10^{-4}, \, 10^{-5}$ and $10^{-6}$, respectively. }
	\label{f7}
\end{figure}

\begin{figure}[htbp]
	\centering
	\includegraphics[scale=0.55]{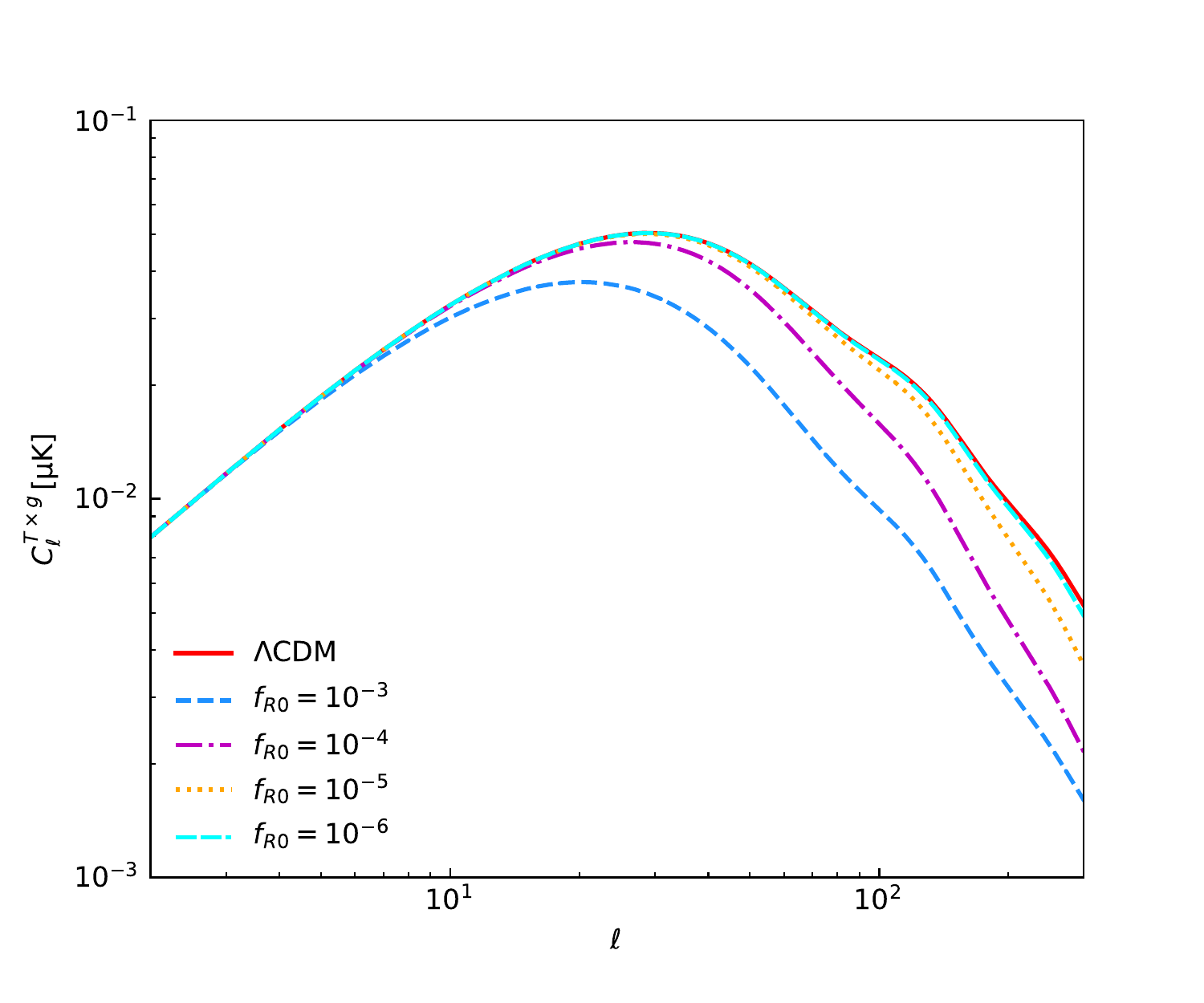}
	\includegraphics[scale=0.55]{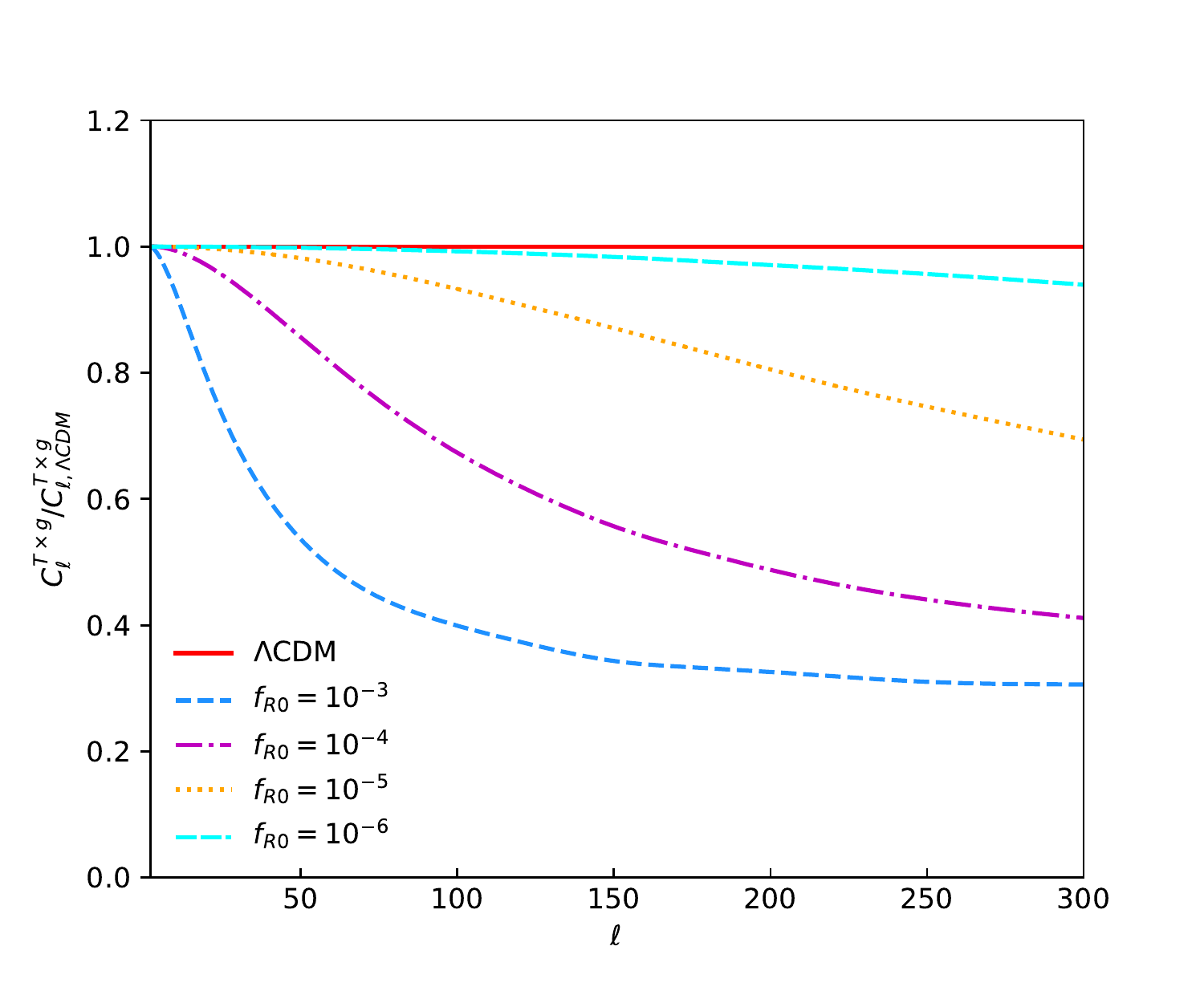}
	\includegraphics[scale=0.55]{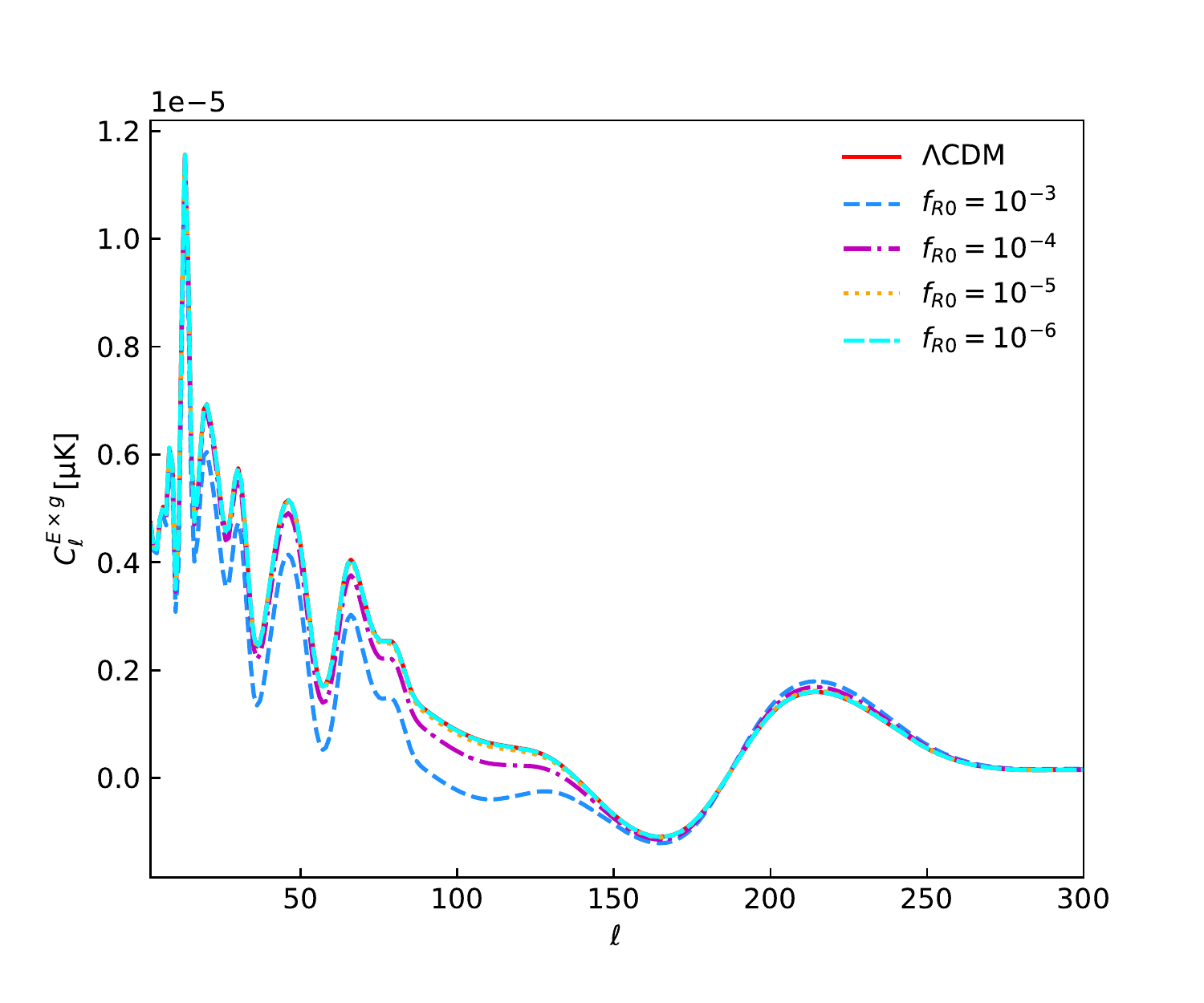}
	\includegraphics[scale=0.55]{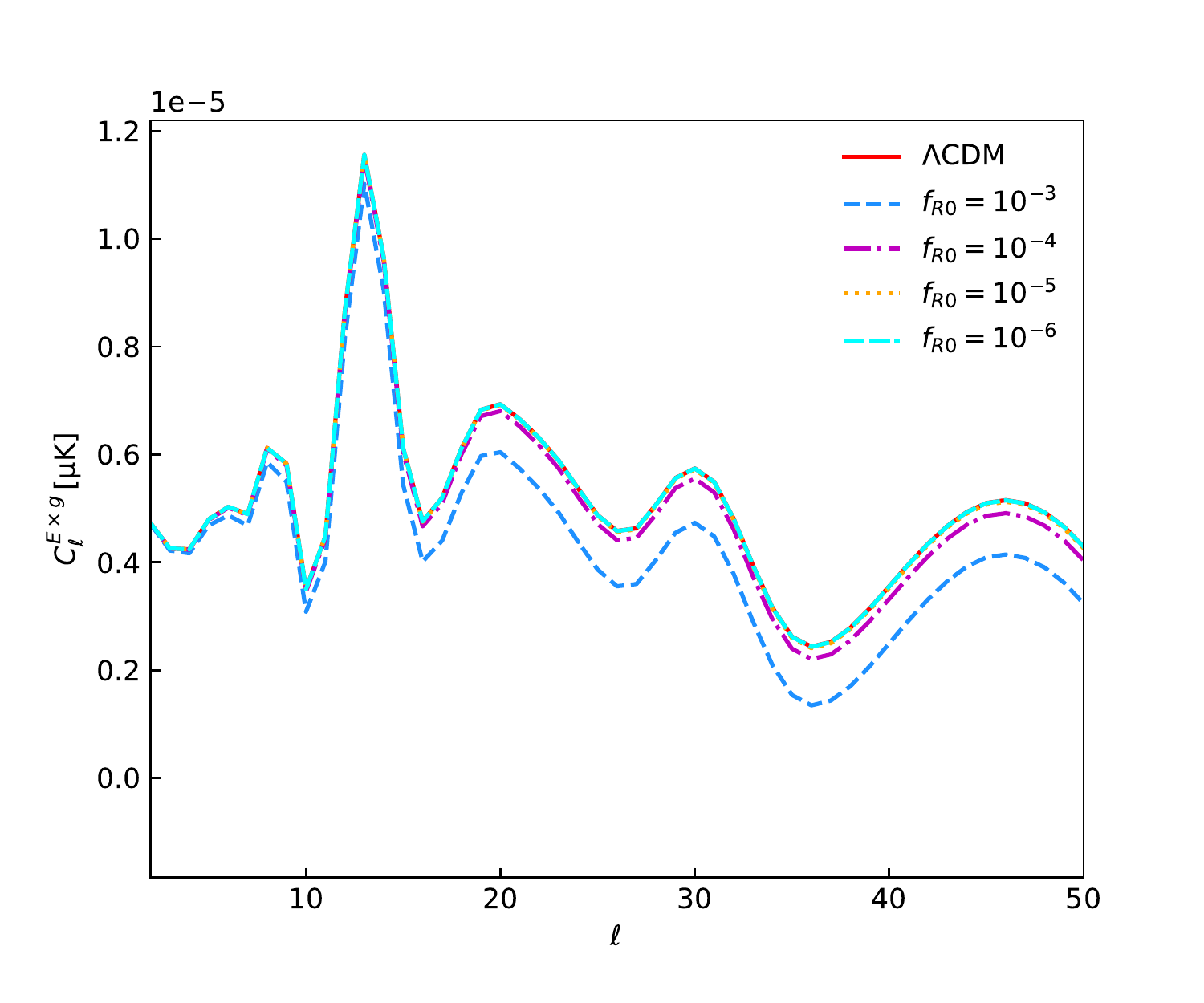}
	\includegraphics[scale=0.55]{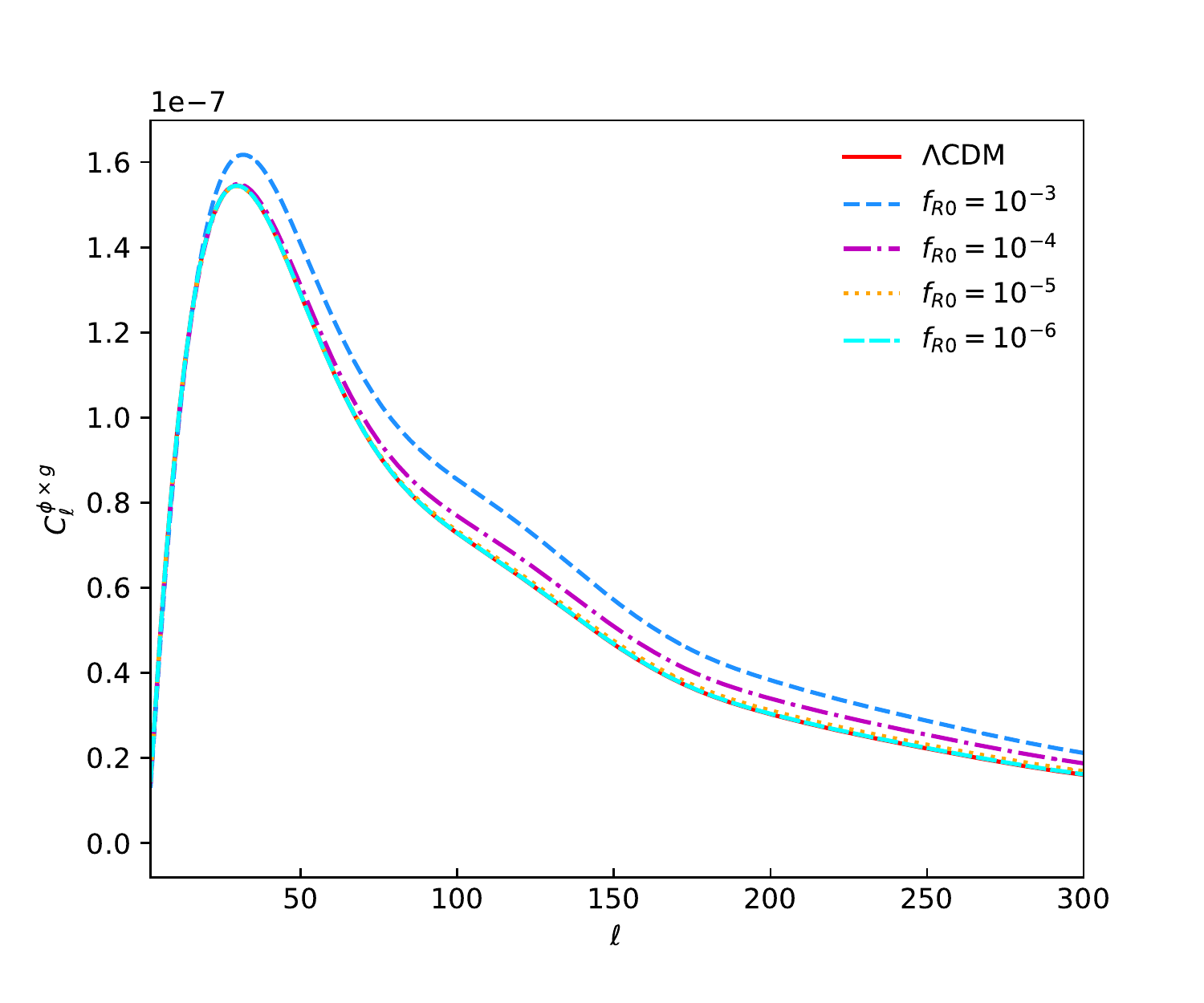}
	\includegraphics[scale=0.55]{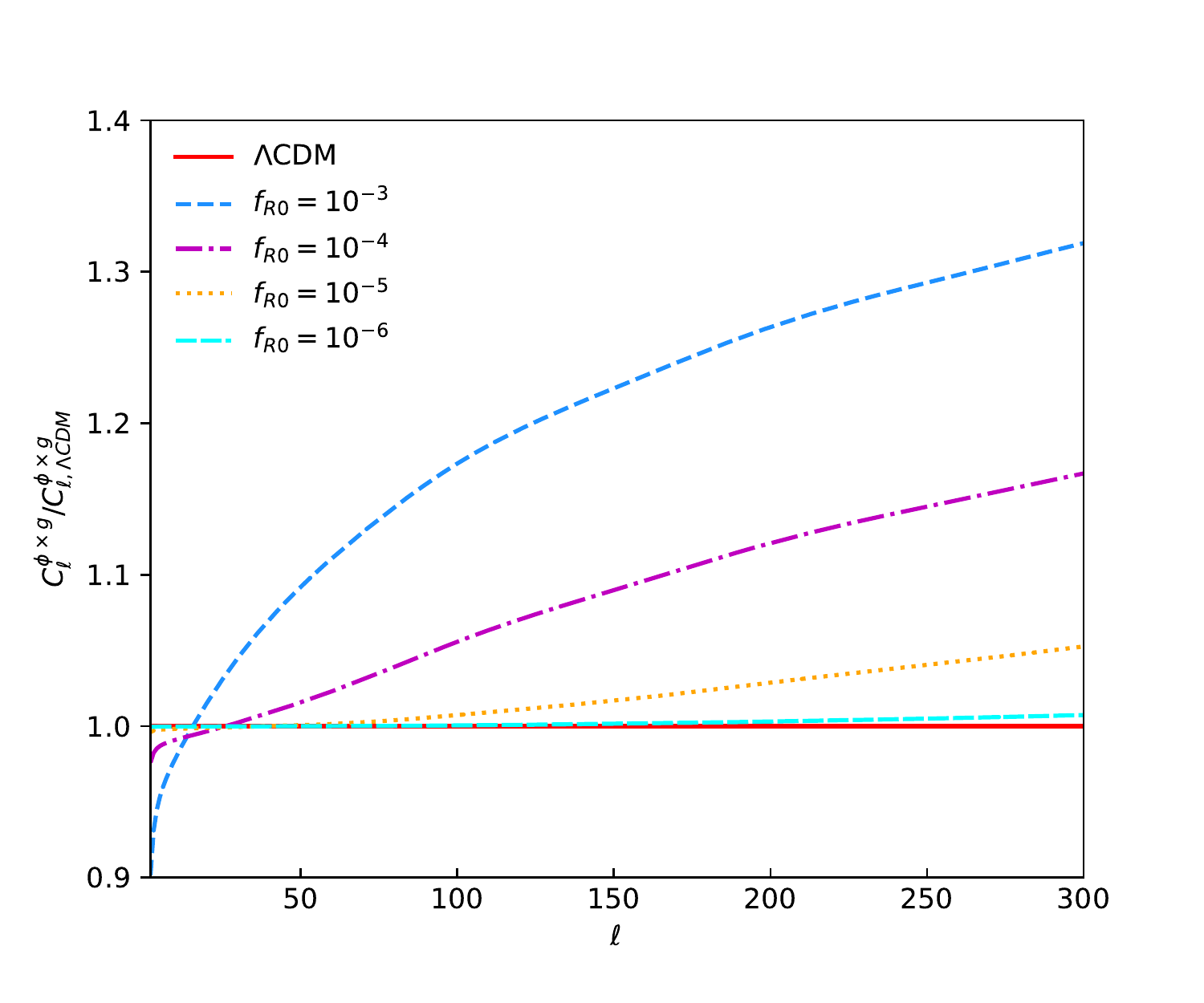}
	\caption{{\it Top panels.} The cross APS ({\it left}) between the CMB temperature and galaxy clustering in the HS $f(R)$ gravity and their ratios ({\it right}) relative to the $\Lambda$CDM model are shown. {\it Medium panels.} The cross APS ({\it left}) between the CMB E-mode polarization and galaxy clustering in the HS $f(R)$ gravity and their behaviors at large angular scales ({\it right}) are shown. {\it Bottom panels.} The cross APS ({\it left}) between the CMB lensing potential and galaxy clustering in the HS $f(R)$ gravity and their ratios ({\it right}) relative to $\Lambda$CDM are shown. The red solid, blue short-dashed, magenta dash-dotted, orange dotted and cyan long-dashed lines denote $\Lambda$CDM, $f_{R0}=10^{-3},\,10^{-4}, \, 10^{-5}$ and $10^{-6}$, respectively.  }
	\label{f8}
\end{figure}

\section{Theoretical predictions}
In this section, we will study the theoretical predictions of HS $f(R)$ gravity. Specifically, for the HS model, we modify carefully the public available package $\mathtt{CAMB}$ \cite{Lewis13} at the background and perturbation levels. This can let us study the predictions of 21 cm IM, CMB, WL, GC, SNe Ia and GW in the HS $f(R)$ gravity very conveniently. To implement the calculations, we take the following Planck 2018 fiducial cosmology \cite{Planck:2018vyg},
\begin{equation}
\left\{\Omega_bh^2=0.02237, \; \Omega_mh^2=0.12, \; \mathrm{ln}(10^{10}A_s)=3.044, \; \tau=0.0544, \; n_s=0.9649, \;\Omega_k=0, \; H_0=67.36\right\},  \label{a1}
\end{equation} 
where $\Omega_b$ denotes the baryon fraction, $A_s$ the amplitude of primordial PS at the pivot scale $k_p=0.05$ Mpc$^{-1}$, $\tau$ the optical depth due to reionization and $n_s$ the spectral index of the scalar spectrum. 

First of all, we investigate the behaviors of dimensionless 21 cm auto APS in the HS $f(R)$ gravity at different redshifts for different bin widths. In the top left panel of Fig.\ref{f1}, for a narrow frequency window $\Delta\nu=0.1$ MHz at $z=0.35$, we find that the 21 cm spectrum exhibit an overall increase at all scales with increasing values of $f_{R0}$. To see this property clearly, we show the ratios between the 21 cm spectrum in HS and that in LCDM in the top right panel of Fig.\ref{f1}, and find that the ratios not only increase monotonically at all angular scales for a given model but also increase rapidly with increasing values of $f_{R0}$. For example, choosing $f_{R0}=10^{-3},\,10^{-4}, \, 10^{-5}$ and $10^{-6}$, the increase can reach about $50\%, \, 30\%, \, 20\%$, and $5\%$ of $C_\ell^{\mathrm{\Lambda CDM}}$ at multipole $\ell=300$, respectively. Notice that for 21 cm IM, we just consider its contribution at large scales. However, in the bottom panels of Fig.\ref{f1}, for a wide window $\Delta\nu=20$ MHz, we find that the values of 21 APS are lower than those for $\Delta\nu=0.1$ MHz by about 1 order of magnitude, that the ratios which decrease at largest scales and increase at relatively small scales are not monotonic, and that the increase of 21 cm spectrum relative to $\Lambda$CDM becomes obviously small at $\ell=300$. It is interesting that one can significantly observe the BAO signals at around $\ell=100$ \cite{Lewis13}. In Fig.\ref{f2}, at a high redshift $z=2$, the values of 21 APS are lower than those at $z=0.35$ by at least 1 order of magnitude and they are closer to the $\Lambda$CDM case. This is because the number of HI-dominated galaxies is not large enough and 21 cm radiation is not strong enough at $z=2$. We find that there is a degeneracy between a broad window and a high redshift, which can significantly decrease the power of 21 cm radiation, in affecting the values of 21 cm APS. More specifically, in the top right panel of Fig.\ref{f2}, we find that all the ratios except for $f_{R0}=10^{-3}$ lie very close to $\Lambda$CDM and decrease with increasing $\ell$, that when $f_{R0}=10^{-3}$, the ratio departs from $\Lambda$CDM obviously, and that, very interestingly, there is a cross point at around $\ell=150$ for the case of $f_{R0}=10^{-4}$. Nonetheless, if considering  $\Delta\nu=20$ MHz, the inconsistent monotonicity between the ratios in four HS $f(R)$ models and the cross point disappear (see the bottom right panel of Fig.\ref{f2}). In light of the above analysis, we are interested in studying carefully the redshift dependence of 21 cm APS. In Fig.\ref{f3}, choosing $f_{R0}=10^{-4}$, we observe that higher redshifts actually lead to lower 21 cm spectrum \cite{Lewis13}. For a narrow window $\Delta\nu=0.1$ MHz, the ratios at $z=2$ and $10$ are very close to each other, which means the 21 cm APS of $f_{R0}=10^{-4}$ gravity is very similar to $\Lambda$CDM from the beginning of reionization (here we roughly take $z=10$) to $z=2$. However, this similar behaviors also disappear for a broad window $\Delta\nu=20$ MHz (see the right panels of Fig.\ref{f3}). As for the frequency dependence of 21 cm APS, in Fig.\ref{f4}, we find that the smaller the frequency window width becomes, the less increase the ratios of 21 cm APS have at a given redshift for a given $f(R)$ model.

In the second place, in the framework of HS $f(R)$ gravity, we study the cross APS between the CMB temperature ($C_\ell^{T\times21cm}$), E-mode polarization ($C_\ell^{E\times21cm}$) and lensing potential ($C_\ell^{\phi\times21cm}$) and 21 cm radiation at $z=0.35$ for a broad window $\Delta\nu=20$ MHz. Similar to the 21 cm auto APS, we focus on the large scale range $\ell\in[2,300]$. In the top left panel of Fig.\ref{f5}, we find that $C_\ell^{T\times21cm}$ is lower than $C_\ell^{TT}$ \cite{Planck:2018vyg} by about 4 orders of magnitude, that the case of $f_{R0}=10^{-3}$ starts to deviate from $\Lambda$CDM when $\ell\geqslant5$ and the departure reaches its maximum at around $\ell=50$, and that although the cases of $f_{R0}=10^{-4}$ and $f_{R0}=10^{-5}$ starts to deviate later than $f_{R0}=10^{-3}$, they will catch up with $f_{R0}=10^{-3}$ before $\ell=300$. The related tendency can also be found in the top right panel of Fig.\ref{f5}. We find that $C_\ell^{T\times21cm}$ in the cases of $f_{R0}=10^{-3},\,10^{-4}, \, 10^{-5}$ and $10^{-6}$ are lower than that in $\Lambda$CDM by $20\%$, $26\%$, $25\%$ and $10\%$ at $\ell=300$. It's interesting that $f_{R0}=10^{-3}$ does not lead to the largest departure from $\Lambda$CDM at all scales. In the medium panels of Fig.\ref{f5}, $C_\ell^{E\times21cm}$ first exhibits interesting oscillatory behaviors for different models until $\ell=75$, and then tend to be same and very small at small scales. We find that $C_\ell^{E\times21cm}$ is lower than $C_\ell^{T\times21cm}$ by at least 3 orders of magnitude at large scales, and that $f_{R0}=10^{-3}$ produces the largest deviation from $\Lambda$CDM at large scales. This means that a cross correlation between E-mode polarization and 21 cm radiation is sensitive to the HS $f(R)$ gravity at large scales. 
In the bottom panels of Fig.\ref{f5}, $C_\ell^{\phi\times21cm}$ is basically lower than $C_\ell^{T\times21cm}$ by about 5 orders of magnitude, although they have similar behaviors in logarithmic space. Different from $C_\ell^{T\times21cm}$, $C_\ell^{\phi\times21cm}$ for different $f(R)$ models are all larger than $\Lambda$CDM, and all their ratios increase monotonically with decreasing angular scales.

In the third place, we study the cross APS between the WL ($C_\ell^{G\times21cm}$) and GC ($C_\ell^{g\times21cm}$) and 21 cm radiation at $z=0.35$ for a broad window $\Delta\nu=20$ MHz. Here we still focus on the multipole range $\ell\in[2,300]$. In order to calculate $C_\ell^{G\times21cm}$, we take a small single-bin Euclid-like WL survey by assuming the central redshift $z_c=0.7$ with a bin width $\Delta z=0.1$, $z_0=0.6374$, $\sigma_{\mathrm{pho}}=0.05$, mean internal ellipticity 0.22, $\tilde{\alpha}=2$ and $\tilde{\beta}=1.5$. To compute $C_\ell^{g\times21cm}$, we also take a Euclid-like GC survey by assuming the galaxy bias $b=1.304$, $z_c=0.7$, $\Delta z=0.1$, $z_0=0.6374$, $\sigma_{\mathrm{pho}}=0.05$, $\tilde{\alpha}=2$ and $\tilde{\beta}=1.5$. For the meanings of $z_0$, $\sigma_{\mathrm{pho}}$, $\tilde{\alpha}=2$ and $\tilde{\beta}=1.5$, we refer the readers to Section V. In Fig.\ref{f6}, we observe that both $C_\ell^{G\times21cm}$ and $C_\ell^{g\times21cm}$ are negative, that their ratios have a very similar scale-dependent increasing tendency, that the ratio of $C_\ell^{G\times21cm}$ in the case of $f_{R0}=10^{-3}$ produces a $73\%$ deviation from $\Lambda$CDM,
and that $f_{R0}=10^{-3}$ has the smallest spectrum at all scales for both WL and GC surveys. Interestingly, $C_\ell^{g\times21cm}$ shows a monotonic decreasing behavior for GC. However, for WL, $C_\ell^{g\times21cm}$ of all five models first reach their minima at around $\ell=75$ and then increase monotonically. It is worth noting that $C_\ell^{G\times21cm}$ is larger than $C_\ell^{g\times21cm}$ by at least 2 orders of magnitude and $C_\ell^{g\times21cm}$ is comparable to $C_\ell^{E\times21cm}$.
 
In the fourth place, we investigate the cross correlations between the CMB temperature ($C_\ell^{T\times G}$), E-mode polarization ($C_\ell^{E\times G}$) and lensing potential ($C_\ell^{\phi\times G}$) and WL. From Fig.\ref{f7}, we find that although they have different orders of magnitude, $C_\ell^{T\times G}$, $C_\ell^{E\times G}$ and $C_\ell^{\phi\times G}$ have similar global behaviors, i.e., first decreasing to minima at extremely large scales and then increasing rapidly to stable values that are very close to zero. This implies that they just produce cross correlations at large scales. It is interesting that the ratios of $C_\ell^{T\times G}$ in the cases of $f_{R0}=10^{-3}$ and $10^{-4}$ have a cross point during the process of monotonically decreasing. This behavior is different from $C_\ell^{T\times21cm}$ and leads to the largest departure from $\Lambda$CDM occurring in the case $f_{R0}=10^{-4}$. Note that the ratio of $C_\ell^{\phi\times G}$ in the case of $f_{R0}=10^{-3}$ also generates a $70\%$ deviation from $\Lambda$CDM.

Finally, we study the cross correlations between the CMB temperature ($C_\ell^{T\times g}$), E-mode polarization ($C_\ell^{E\times g}$) and lensing potential ($C_\ell^{\phi\times g}$) and GC. In Fig.\ref{f8}, we find that $C_\ell^{T\times g}$ exhibits a similar global behavior to $C_\ell^{T\times21cm}$, but $C_\ell^{T\times g}$ for different models have significantly larger separations when $\ell>100$. This indicates that the cross correlations between CMB temperature and GC at small scales may be a good tool to distinguish the HS $f(R)$ gravity from $\Lambda$CDM. Although $C_\ell^{E\times g}$ has a highly oscillatory behavior like $C_\ell^{E\times21cm}$, its oscillation will ends a little later at around $\ell=150$. $C_\ell^{\phi\times g}$ first reach rapidly their maximum at large scales and then decrease slowly. Very interestingly, all the ratios of $f(R)$ models
have cross points with $\Lambda$CDM, i.e., they vary from $<1$ region to $>1$ region with increasing tendencies.

\section{Experimental specifications and analysis methodology}
In this section, we specify the experimental configurations of each probe and describe the corresponding forecasting methodology. 

For a given survey, to precisely forecast its performance, the ultimate method is implementing a full simulation in supercomputers and containing a number of systematic and instrumental effects and then producing mock data and via the actual analysis pipeline. However, this approach needs to spend a great computational cost. Fisher Matrix is a much simpler method, which takes the expected properties of signal and noise in theoretical quantities for a given survey to derive a Gaussian approximation to the underlying likelihood for a set of parameters to be measured. So far, this method plays a key role in describing the ability of a given experiment to constrain parameters of interest.

\begin{figure}[htbp]
	\centering
	\includegraphics[scale=0.55]{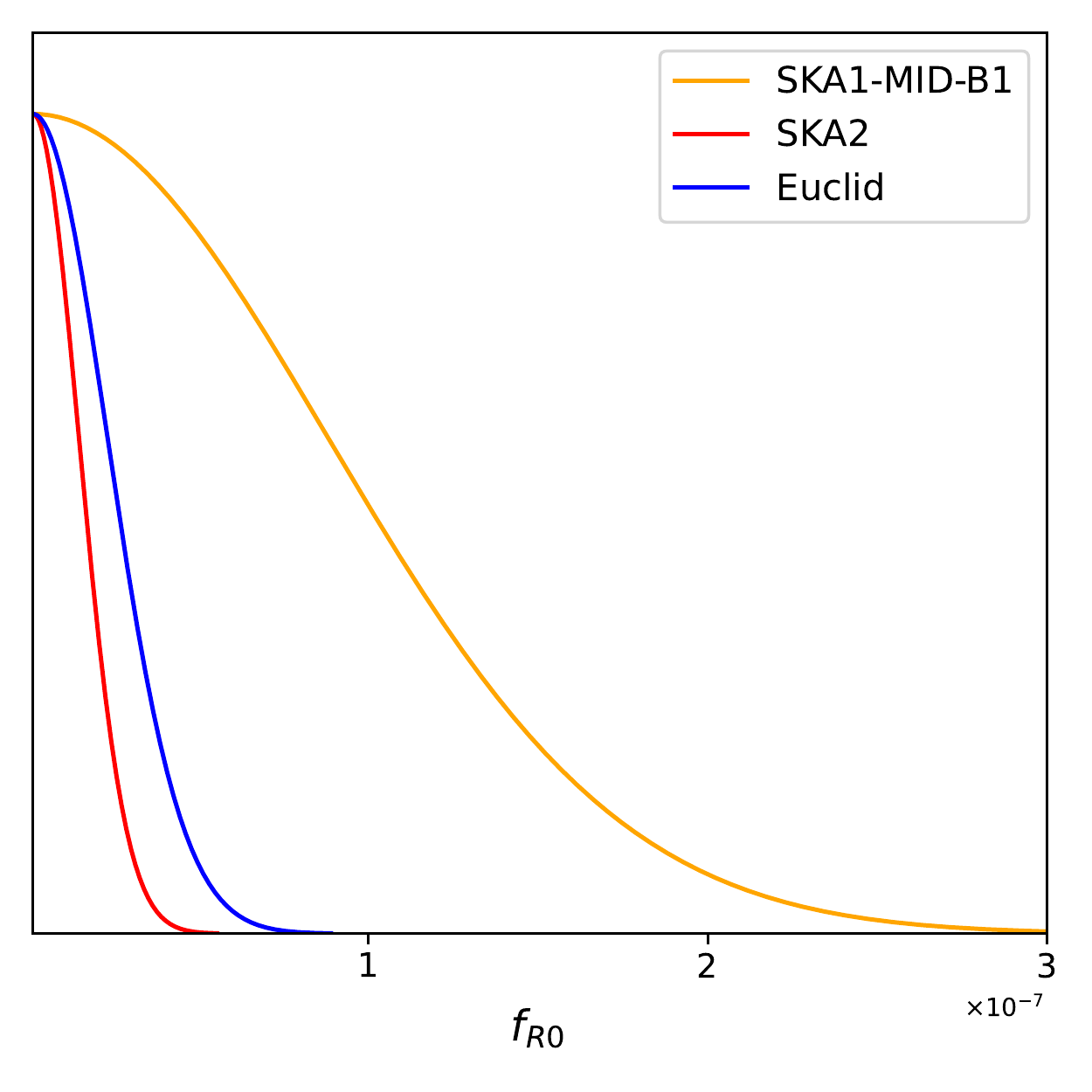}
	\includegraphics[scale=0.55]{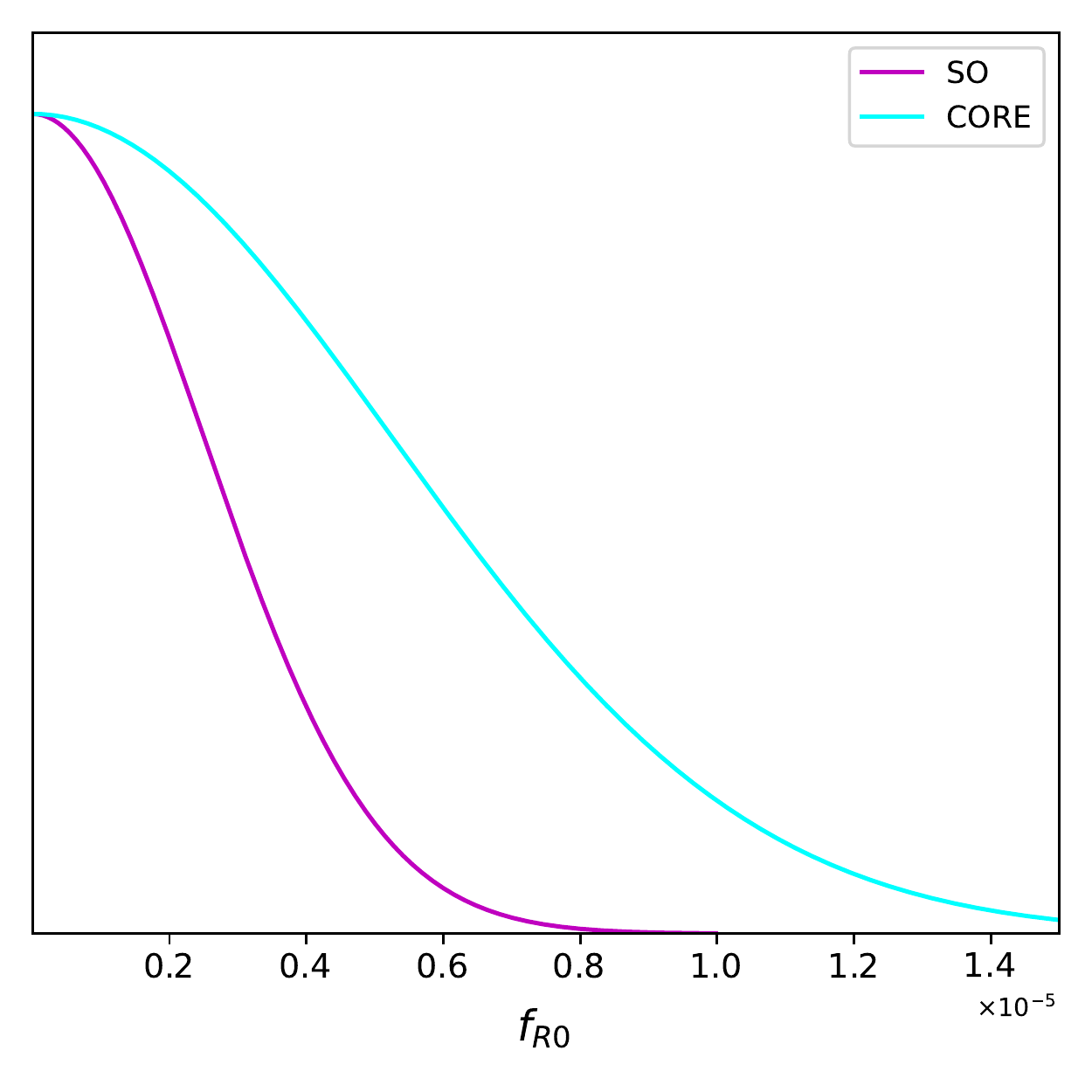}
	\includegraphics[scale=0.55]{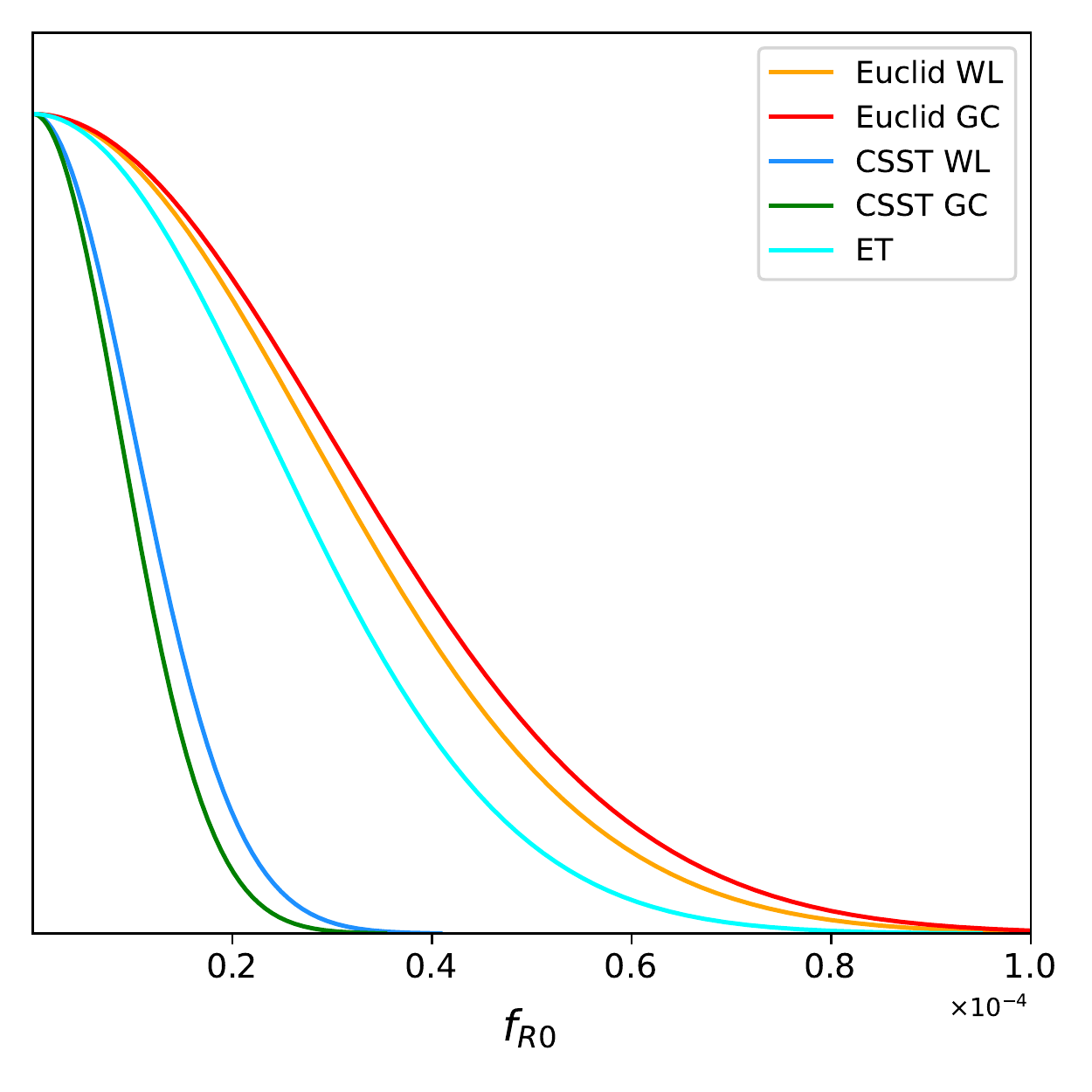}
	\includegraphics[scale=0.55]{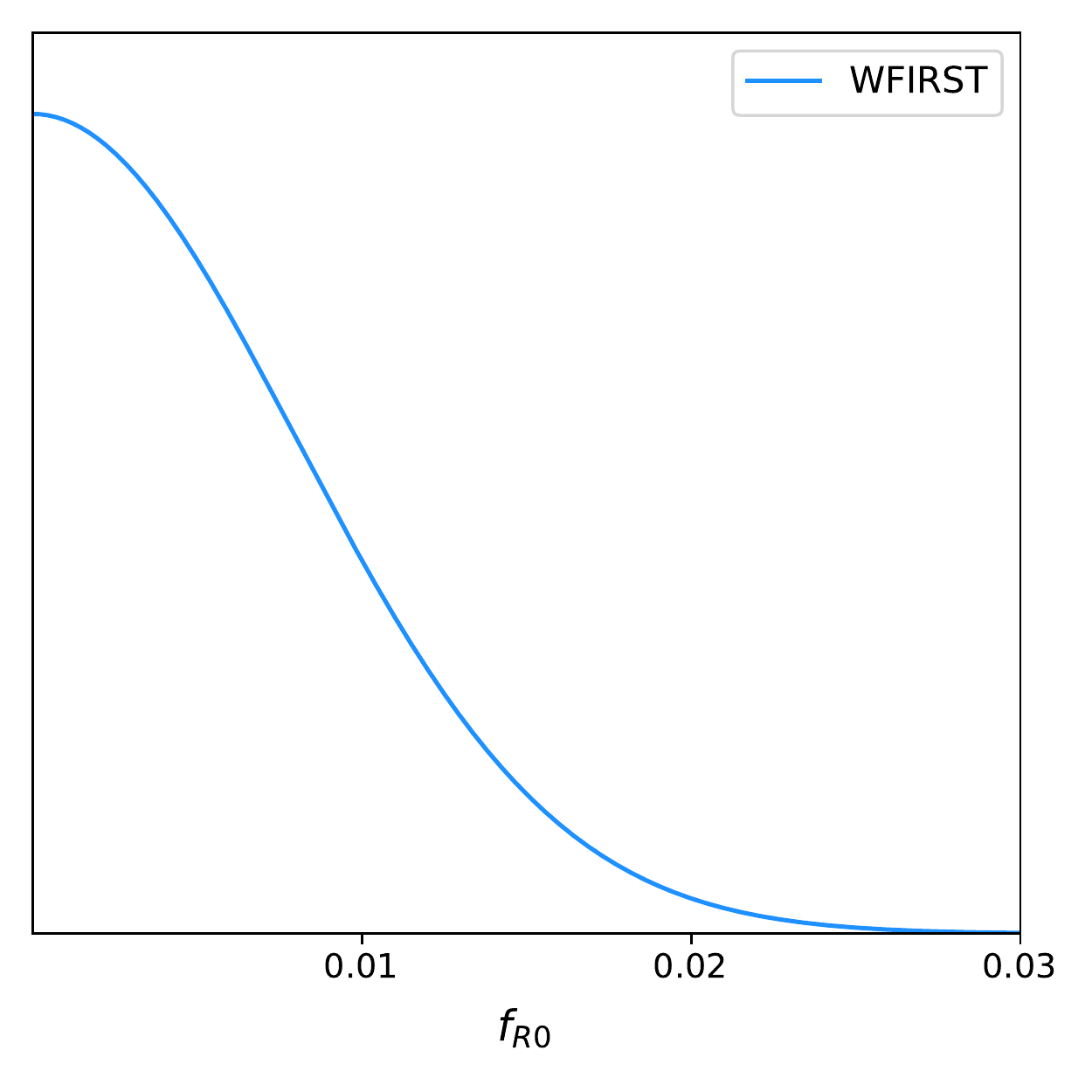}
	
	\caption{The normalized one-dimensional distributions of $f_{R0}$ in the HS $f(R)$ gravity are shown for 21 cm IM, HI galaxy redshift, CMB, optical galaxy redshift, WL, GC, SNe Ia and GW surveys, respectively. }
	\label{f10}
\end{figure}

\begin{figure}[htbp]
	\centering
	\includegraphics[scale=0.6]{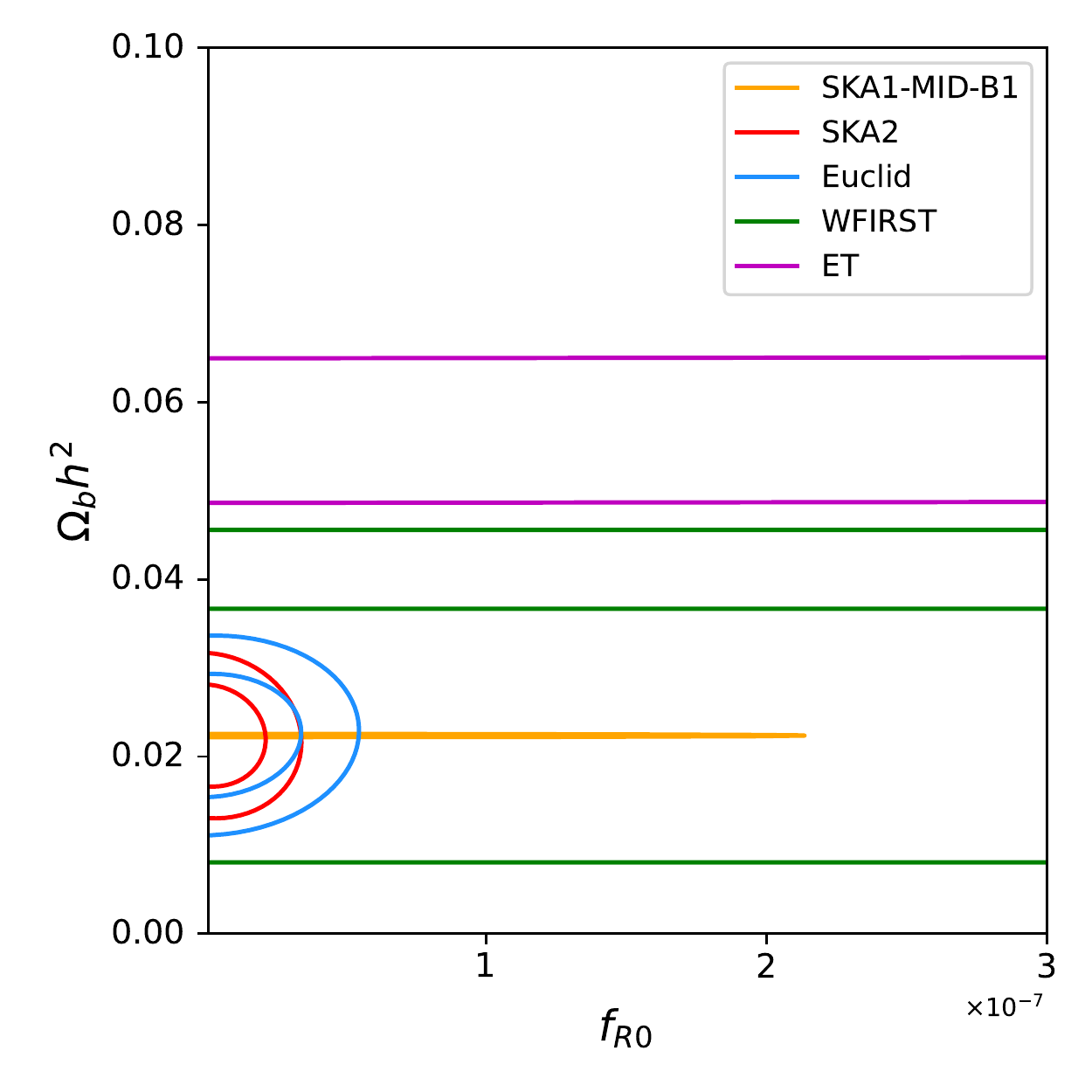}
	\includegraphics[scale=0.6]{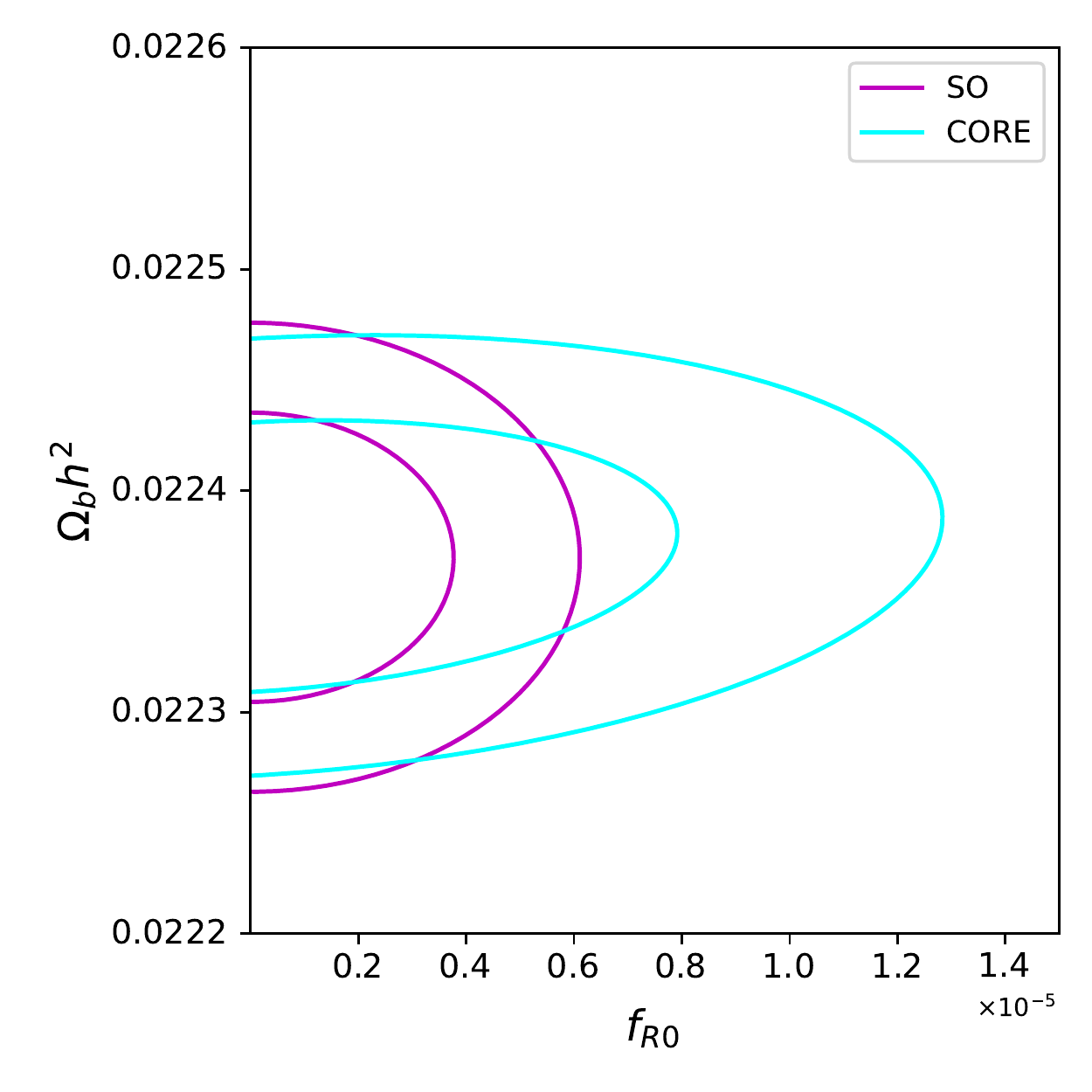}
	\includegraphics[scale=0.6]{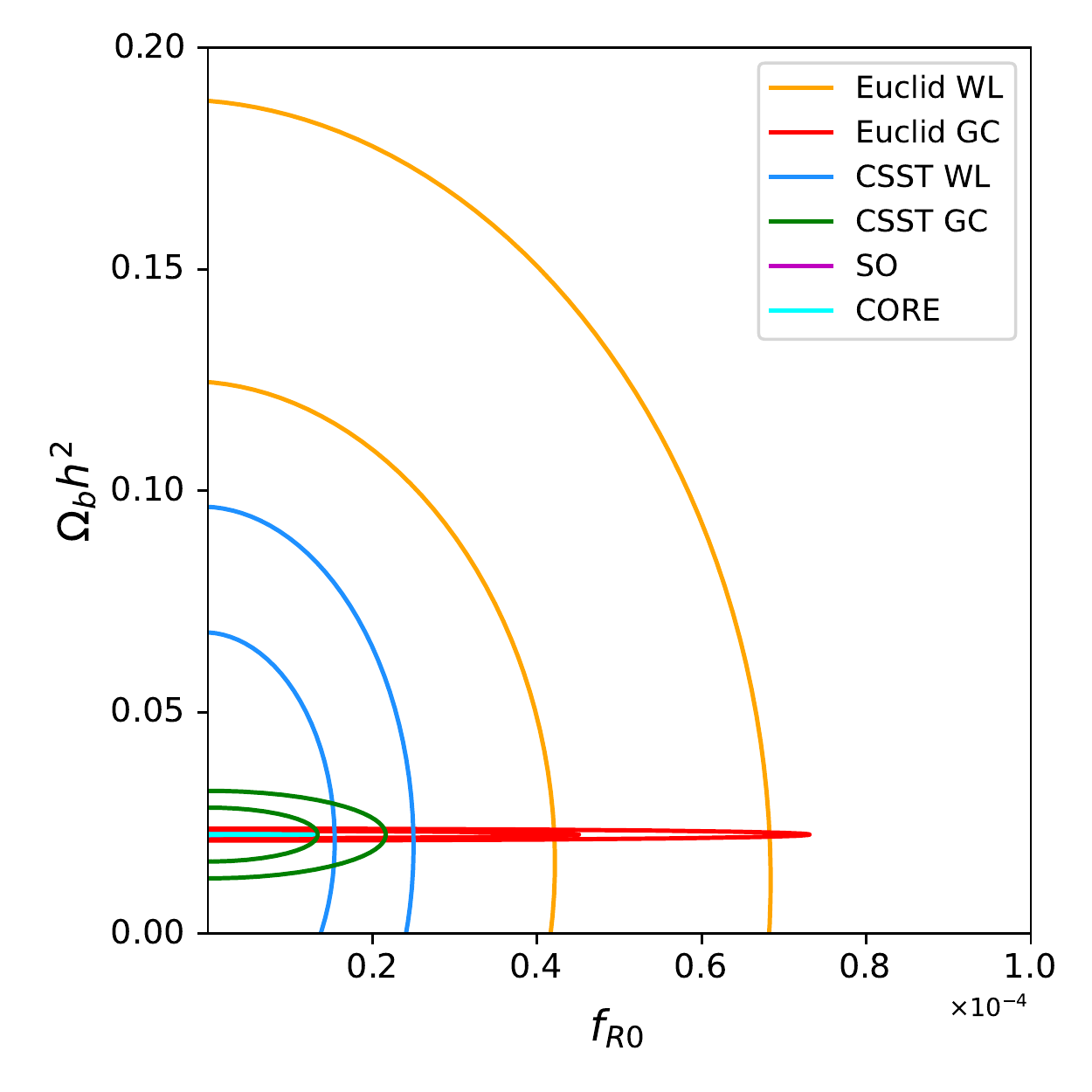}
	\includegraphics[scale=0.6]{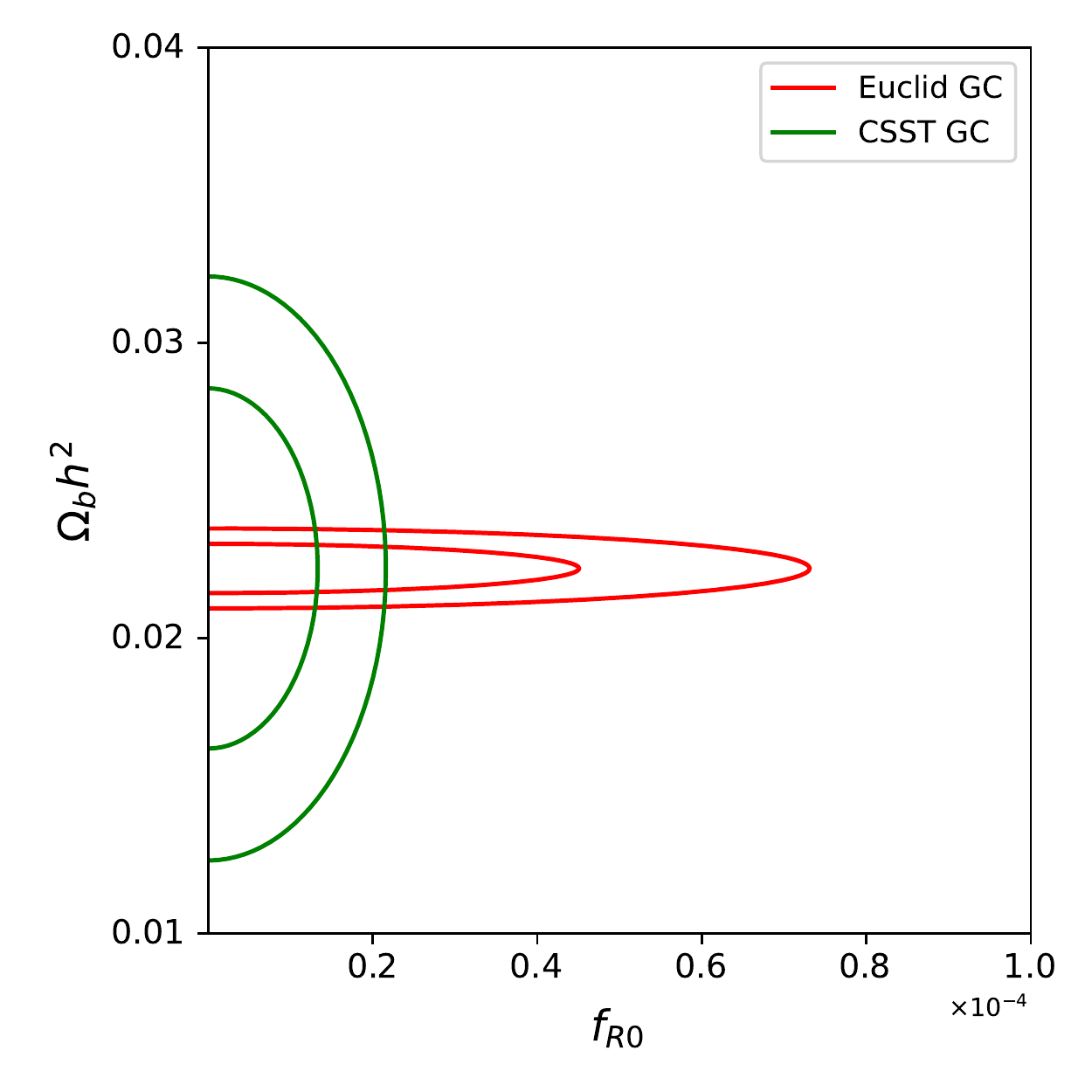}
	\caption{The predicted two-dimensional distributions of $f_{R0}$ and $\Omega_b h^2$ in the HS $f(R)$ gravity are shown for 21 cm IM, HI galaxy redshift, CMB, optical galaxy redshift, WL, GC, SNe Ia and GW surveys, respectively. }
	\label{f11}
\end{figure}

\begin{figure}[htbp]
	\centering
	\includegraphics[scale=0.6]{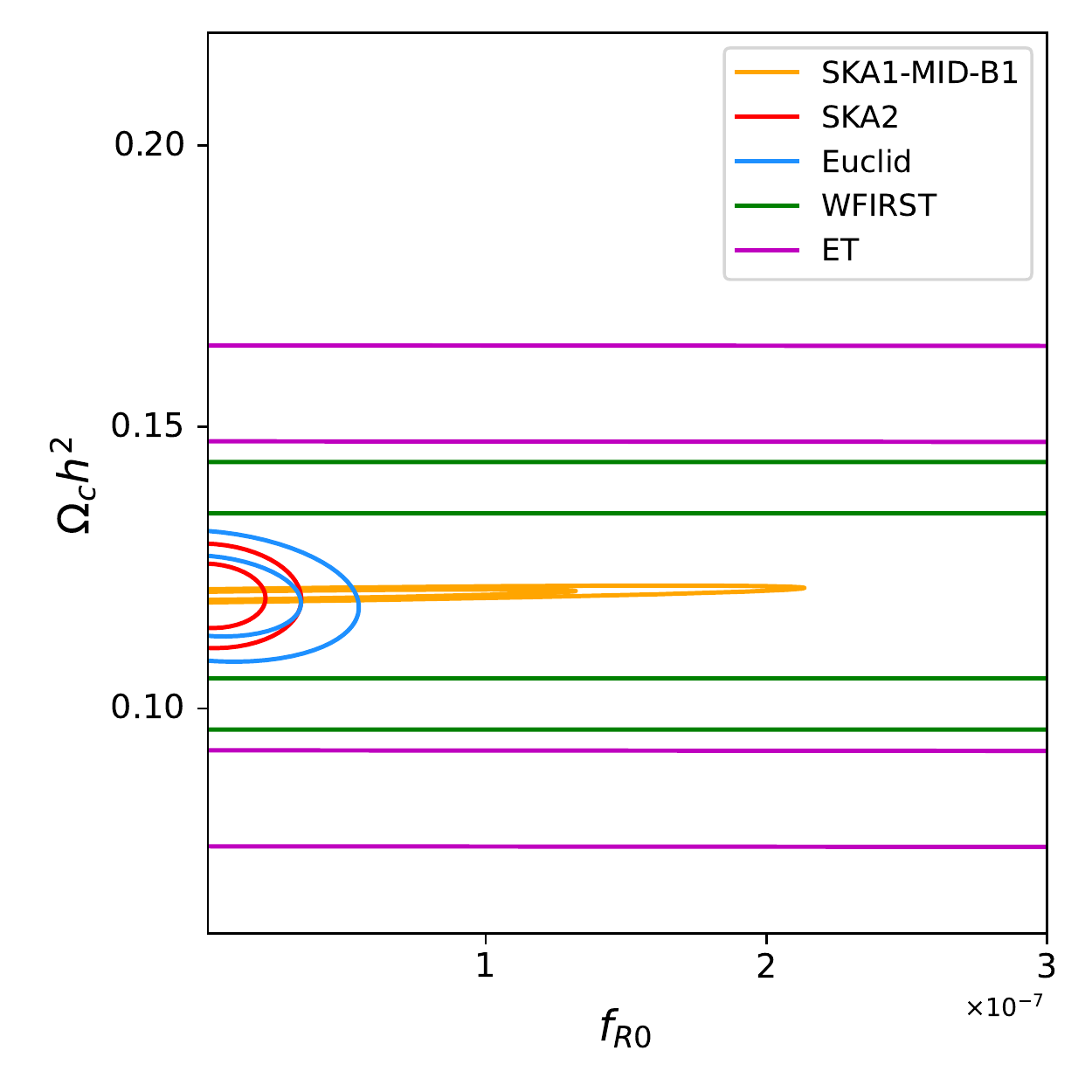}
	\includegraphics[scale=0.6]{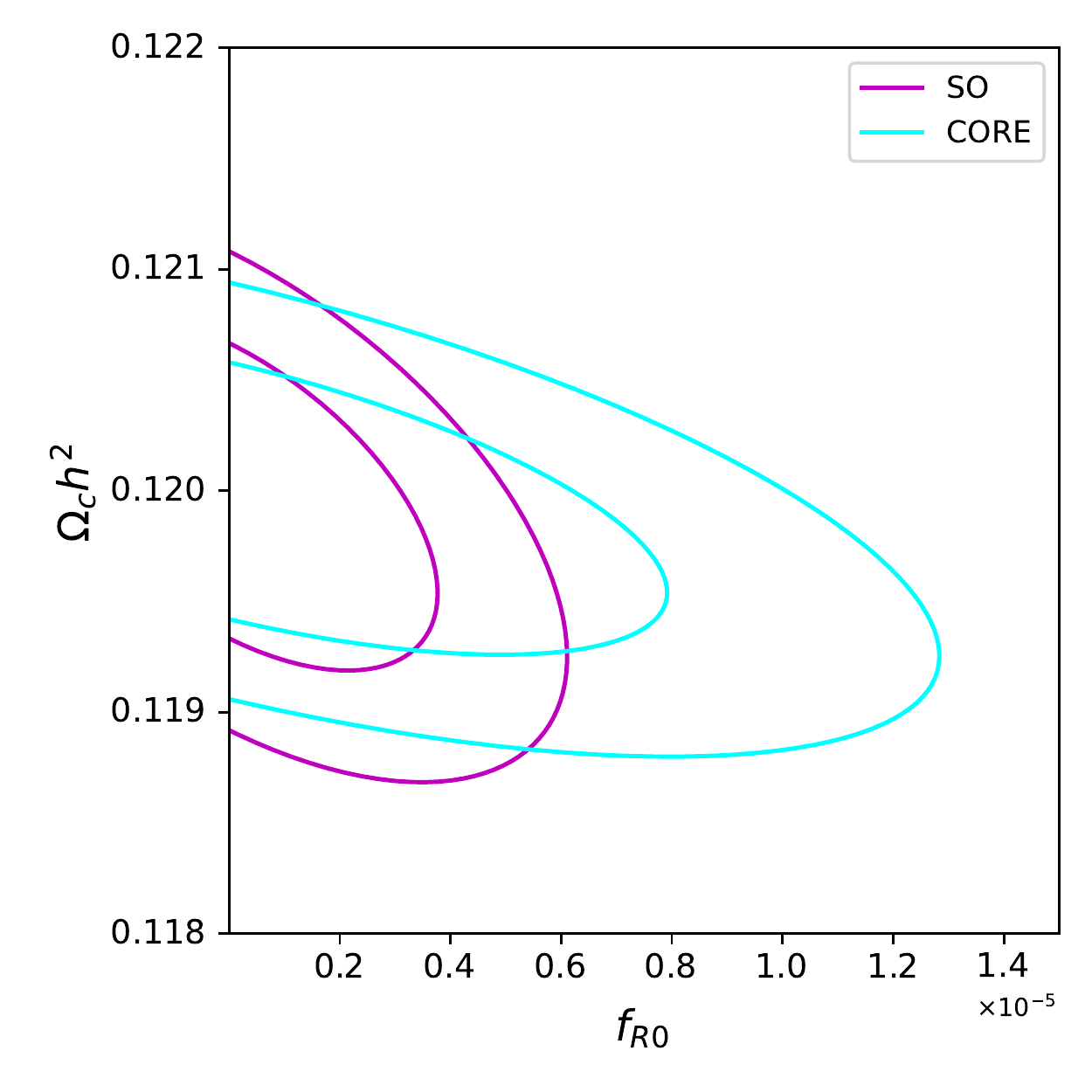}
	\includegraphics[scale=0.6]{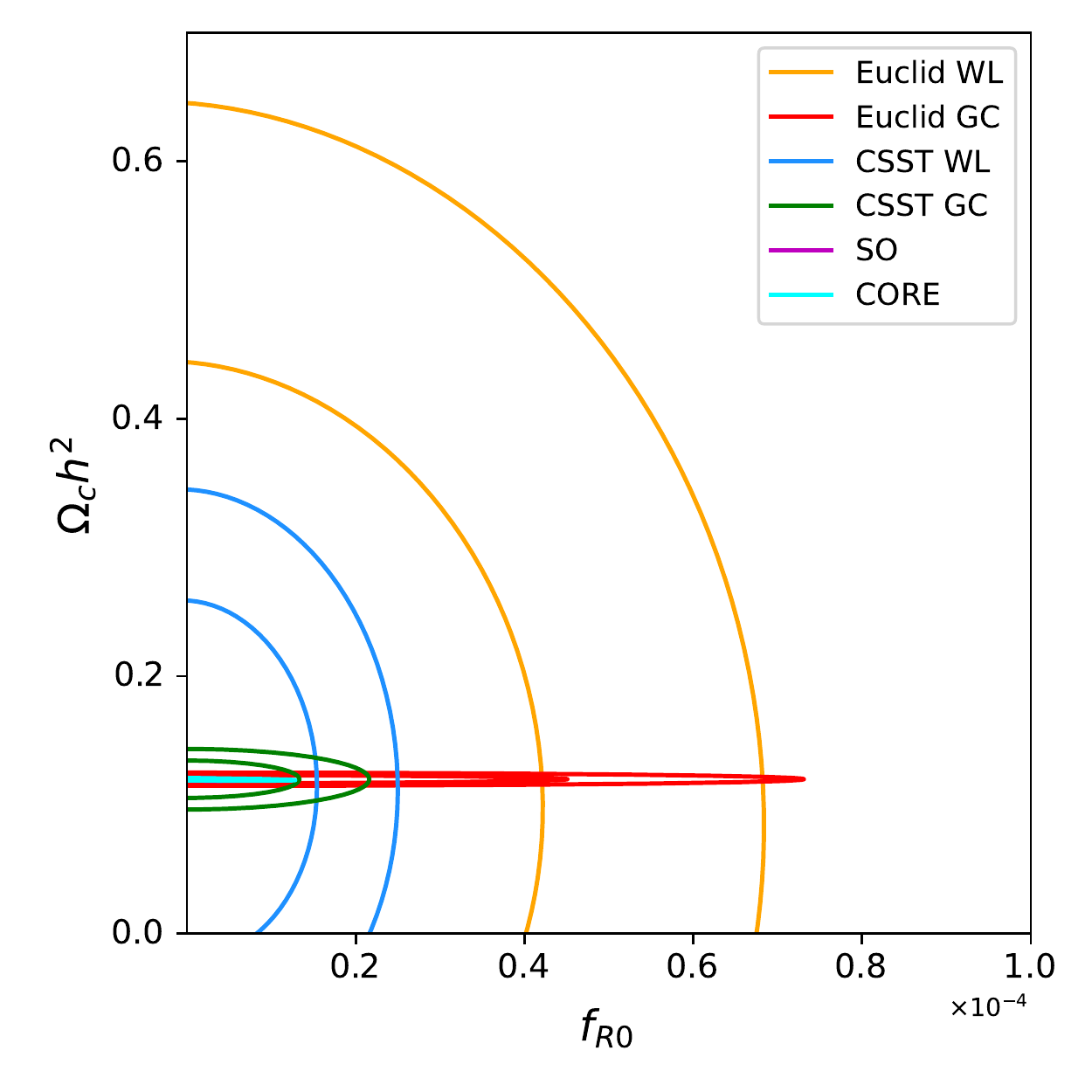}
	\includegraphics[scale=0.6]{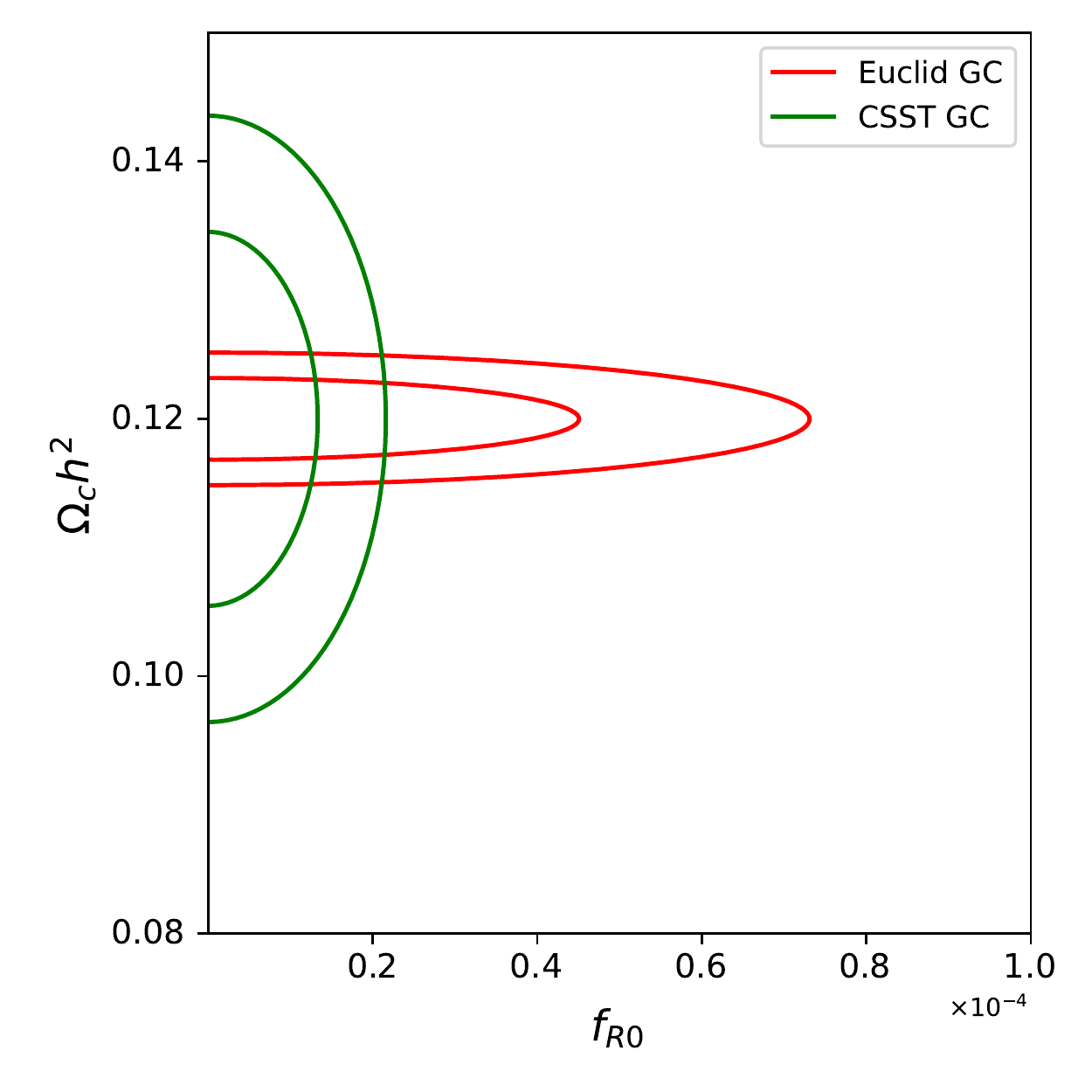}
	\caption{The predicted two-dimensional distributions of $f_{R0}$ and $\Omega_c h^2$ in the HS $f(R)$ gravity are shown for 21 cm IM, HI galaxy redshift, CMB, optical galaxy redshift, WL, GC, SNe Ia and GW surveys, respectively.}
	\label{f12}
\end{figure}

\begin{figure}[htbp]
	\centering
	\includegraphics[scale=0.6]{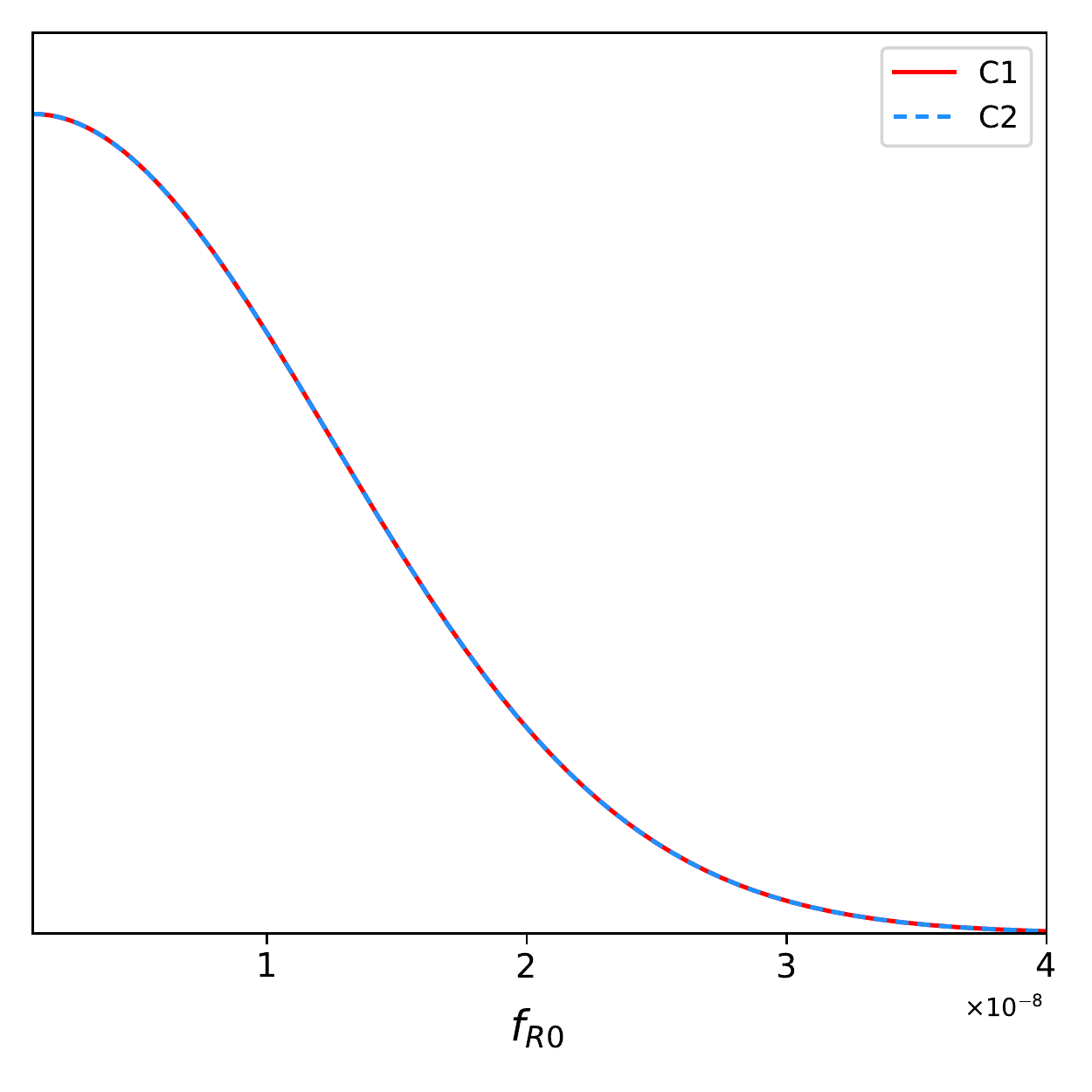}
	\includegraphics[scale=0.6]{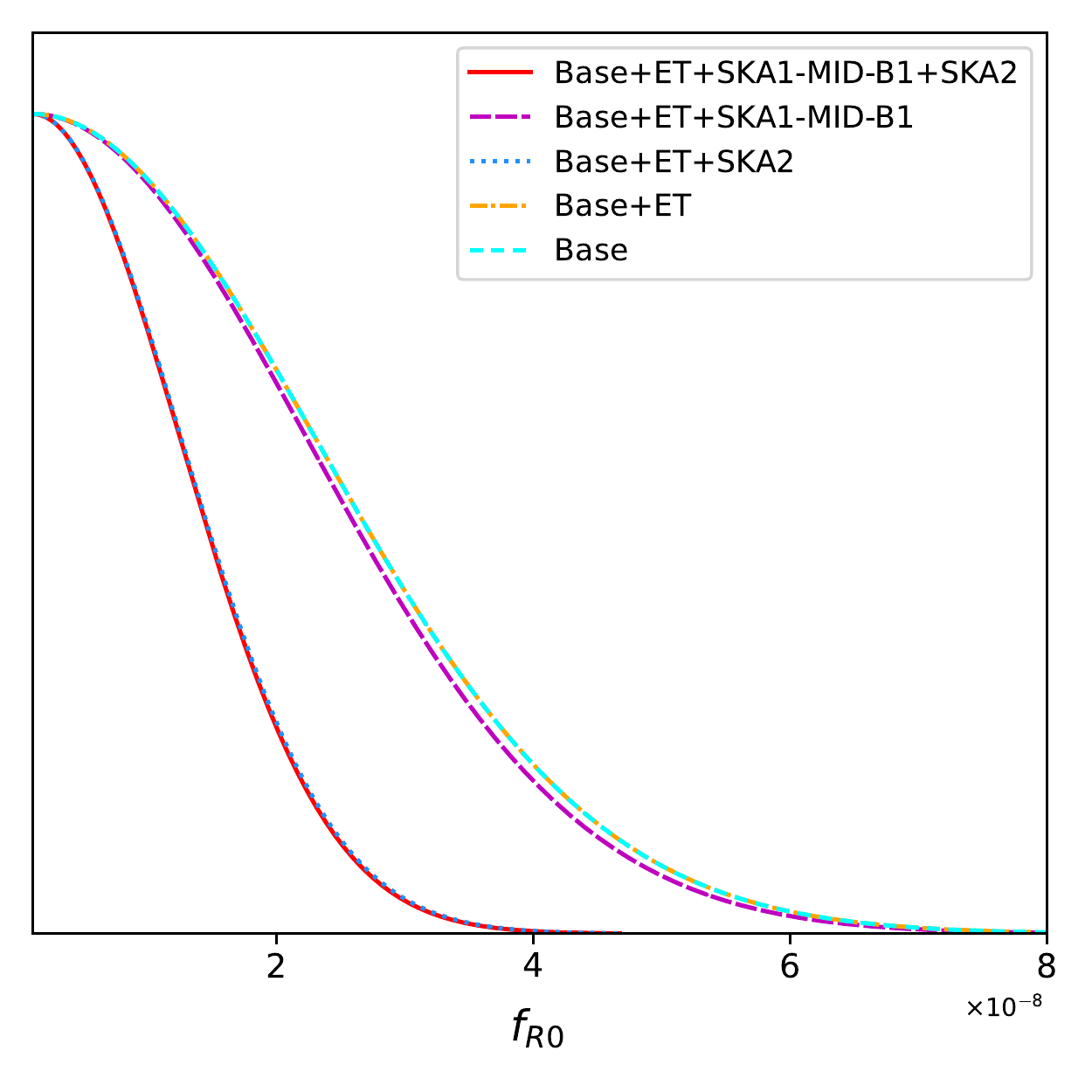}
	\caption{The normalized one-dimensional distributions of $f_{R0}$ in the HS $f(R)$ gravity are shown for total ({\it left}) and hierarchical ({\it right}) combinations, respectively.}
	\label{f13}
\end{figure}

\begin{figure}[htbp]
	\centering
	\includegraphics[scale=0.6]{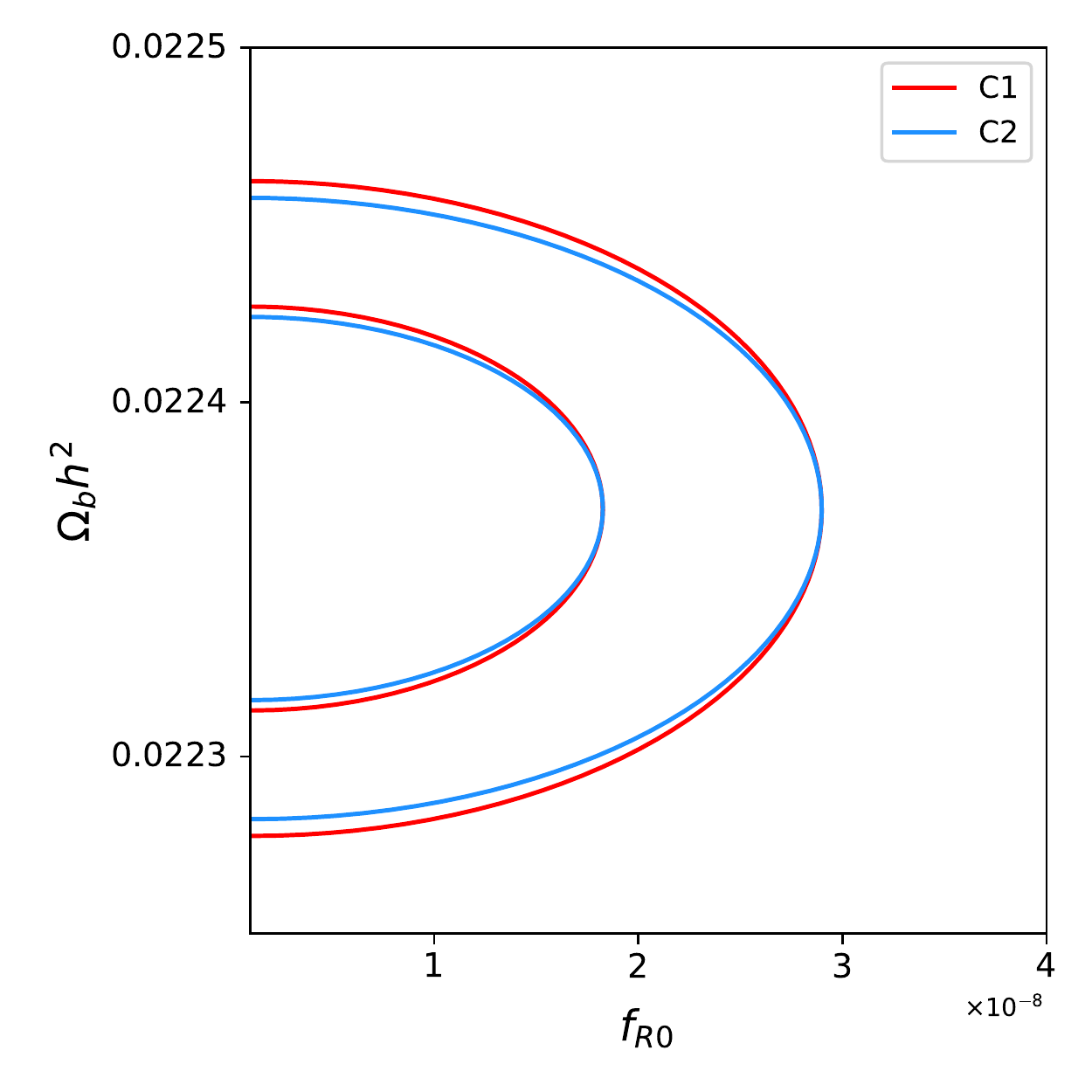}
	\includegraphics[scale=0.6]{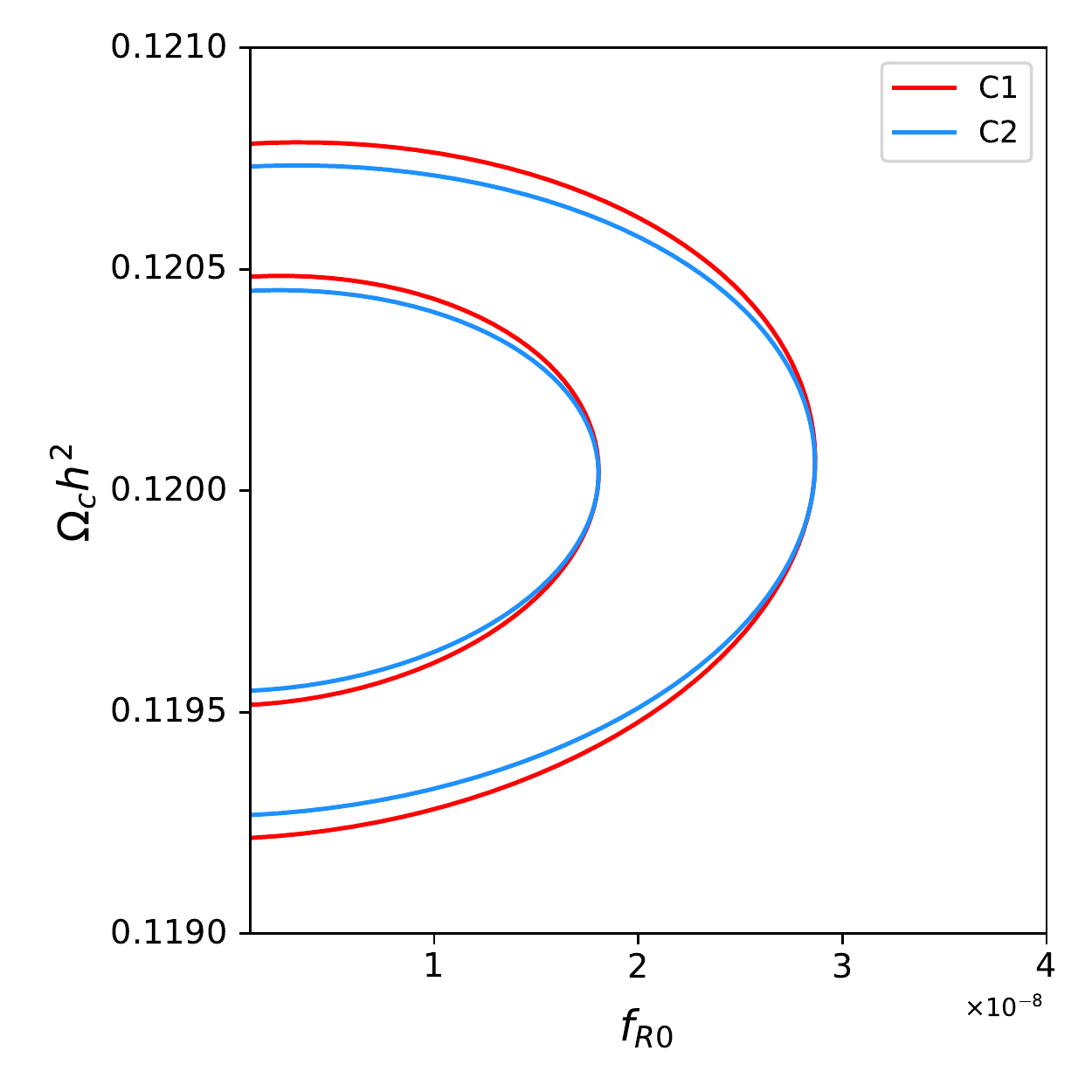}
	\includegraphics[scale=0.6]{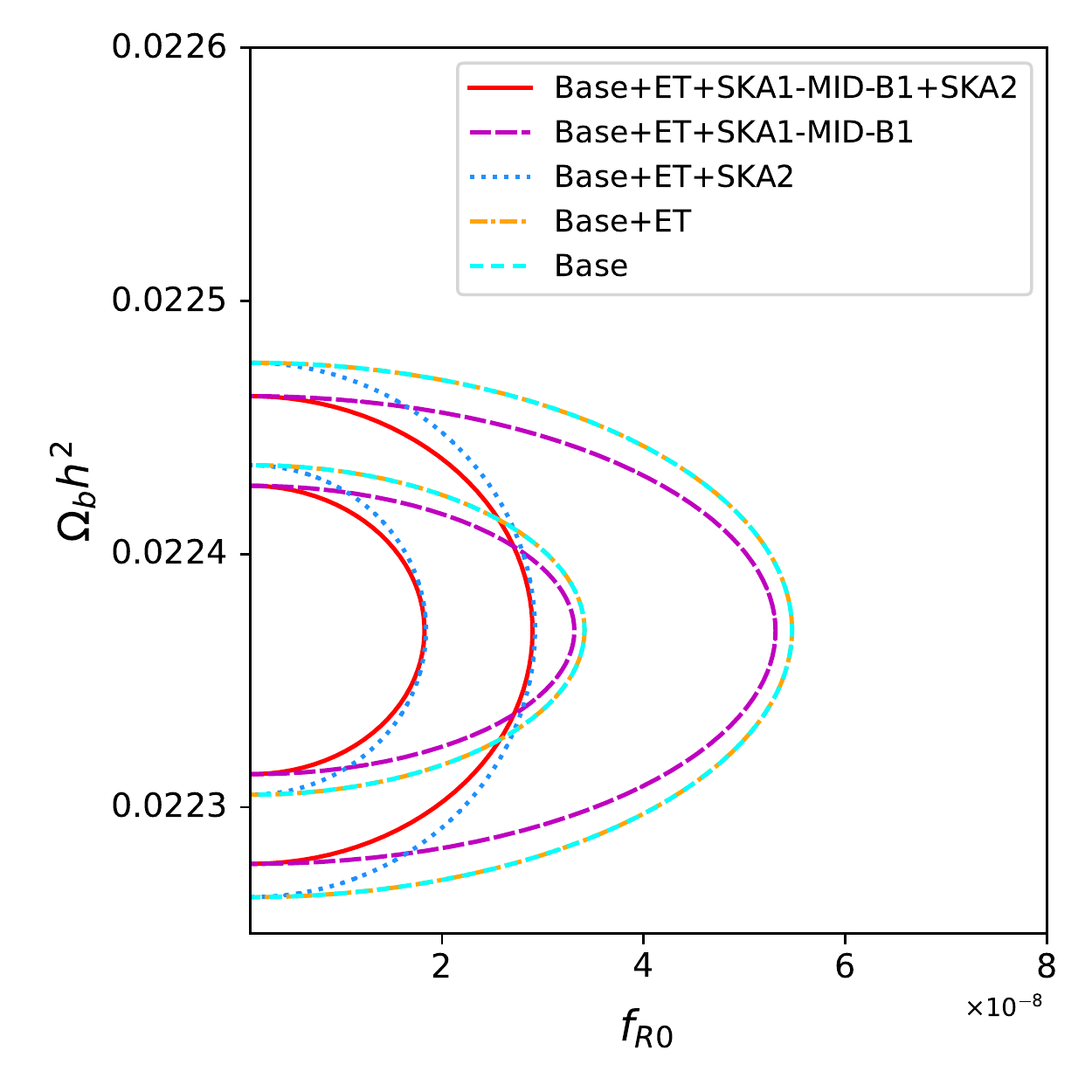}
	\includegraphics[scale=0.6]{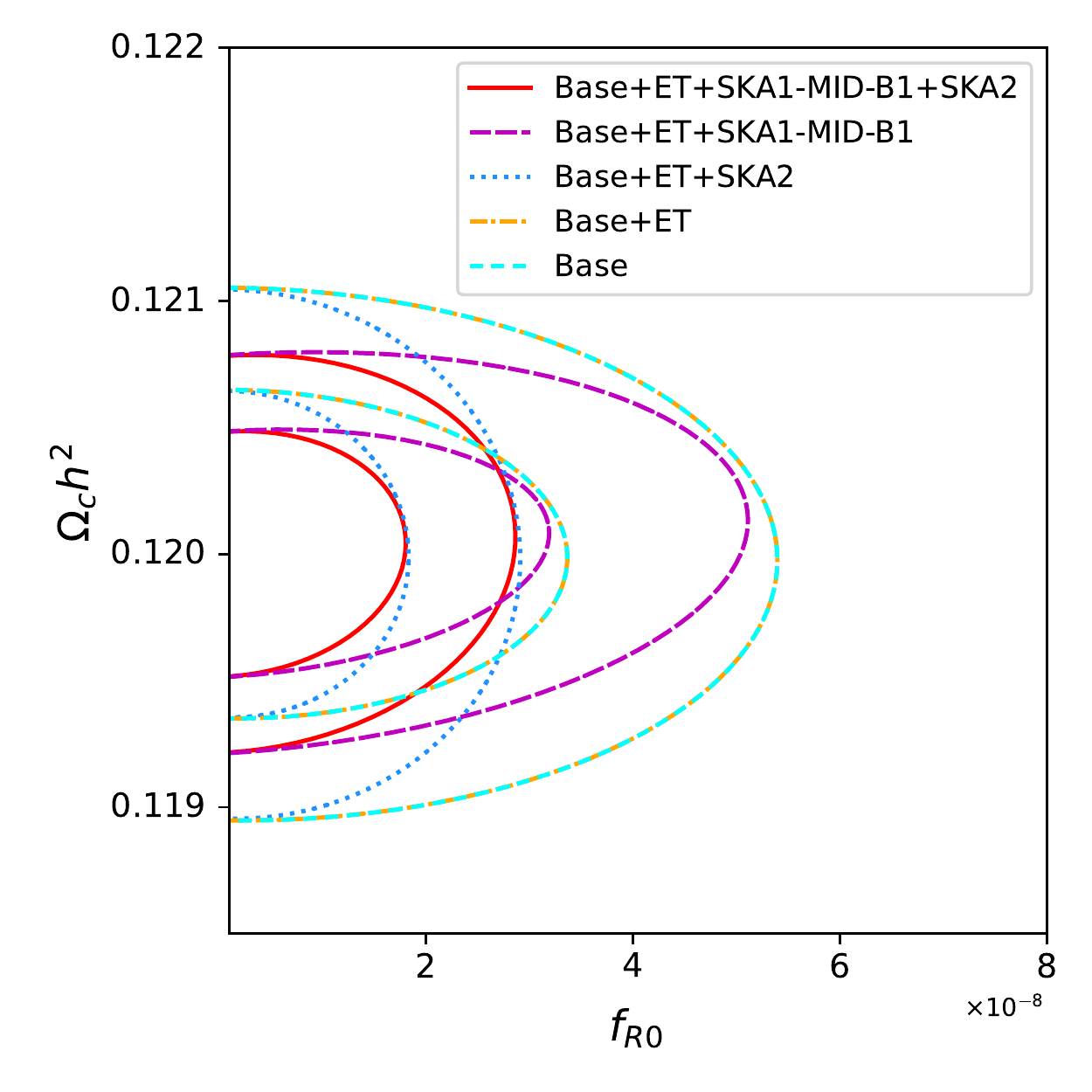}
	\caption{The predicted two-dimensional distributions of $f_{R0}$-$\Omega_b h^2$ ({\it left}) and $f_{R0}$-$\Omega_c h^2$ ({\it right}) in the HS $f(R)$ gravity are shown for total ({\it top}) and hierarchical ({\it bottom}) combinations, respectively.}
	\label{f14}
\end{figure}

\begin{figure}[htbp]
	\centering
	\includegraphics[scale=0.6]{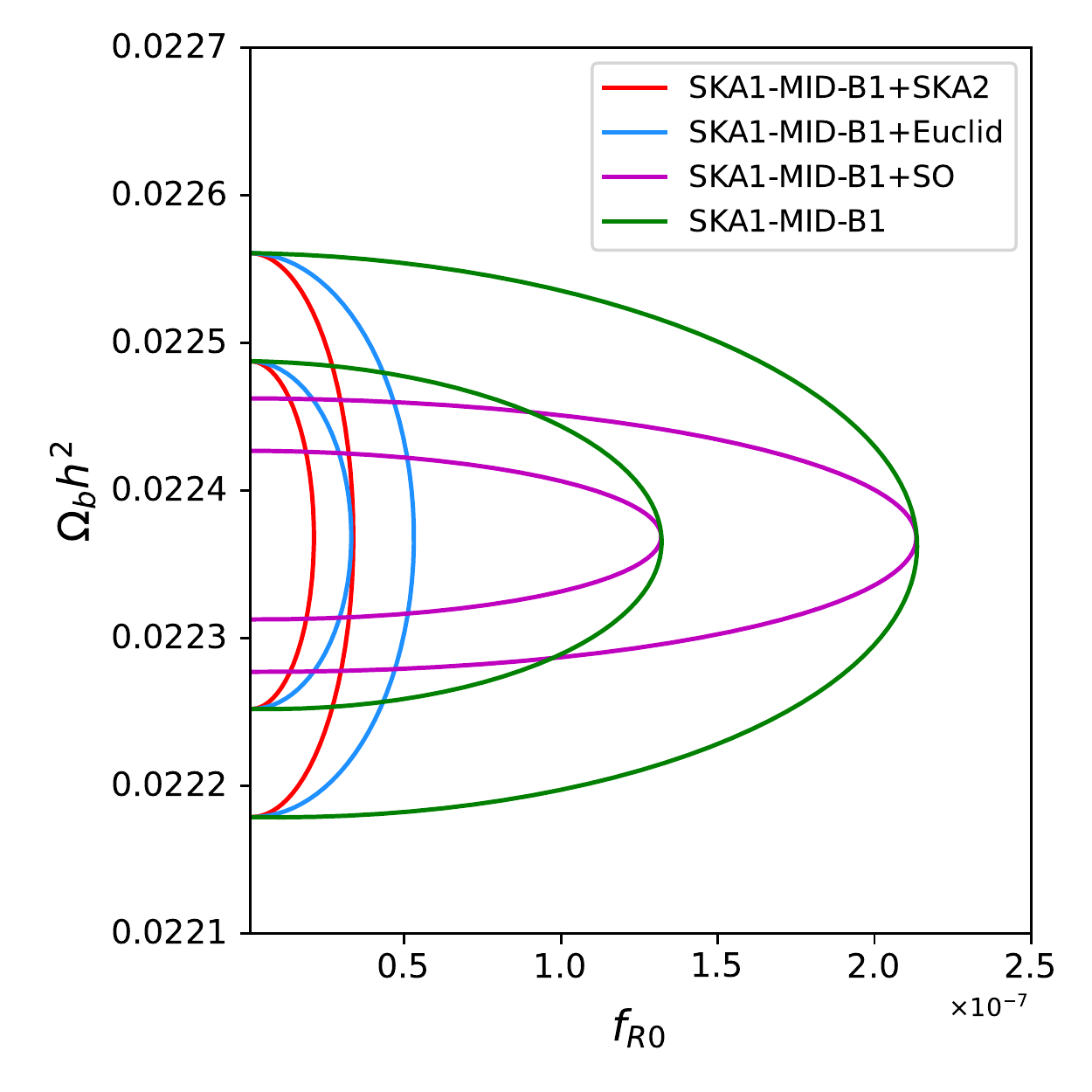}
	\includegraphics[scale=0.6]{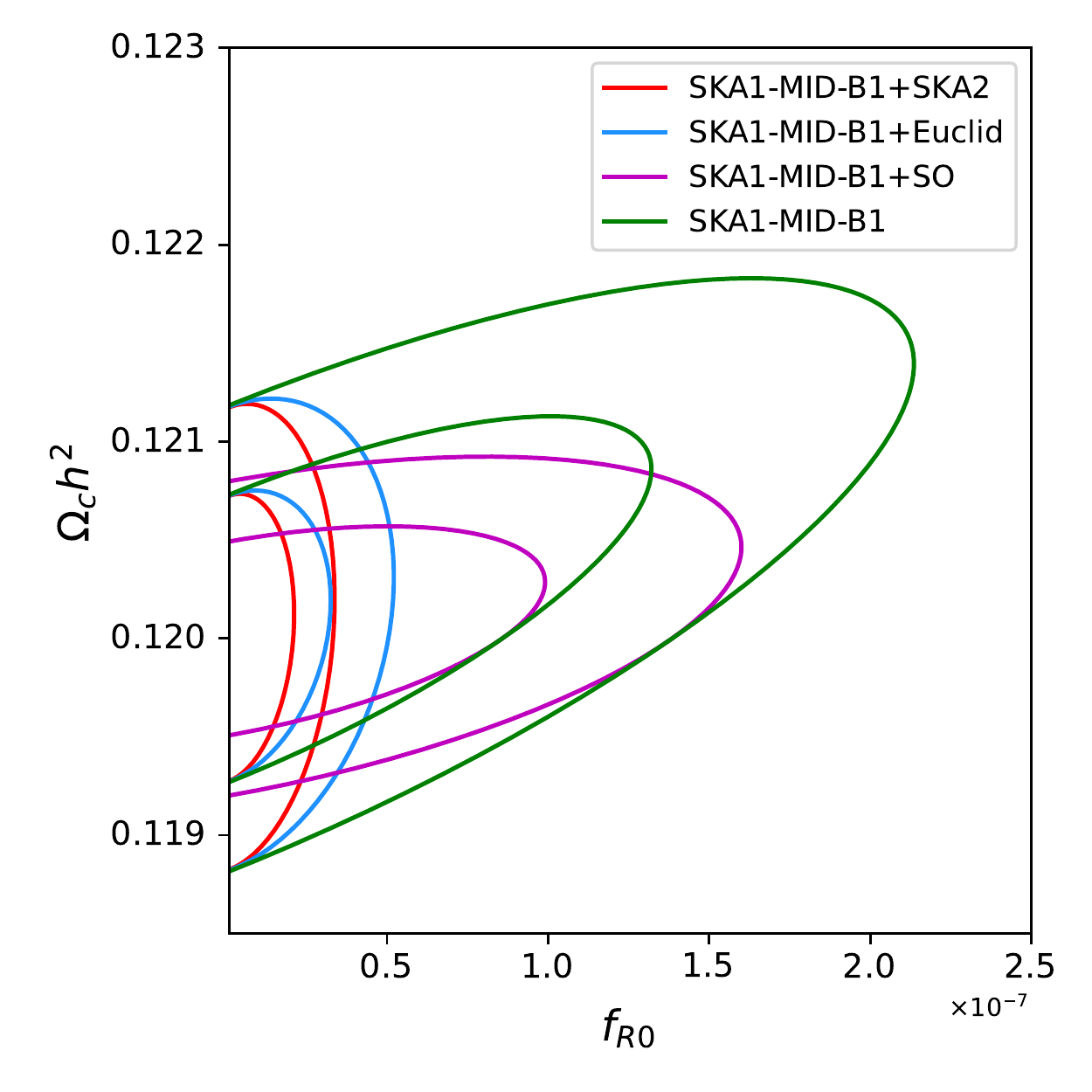}
	\caption{The predicted two-dimensional distributions of $f_{R0}$-$\Omega_b h^2$ ({\it left}) and $f_{R0}$-$\Omega_c h^2$ ({\it right}) in the HS $f(R)$ gravity are shown for 21 cm IM and its combinations with other surveys, respectively.}
	\label{f15}
\end{figure}

\begin{figure}[htbp]
	\centering
	\includegraphics[scale=0.6]{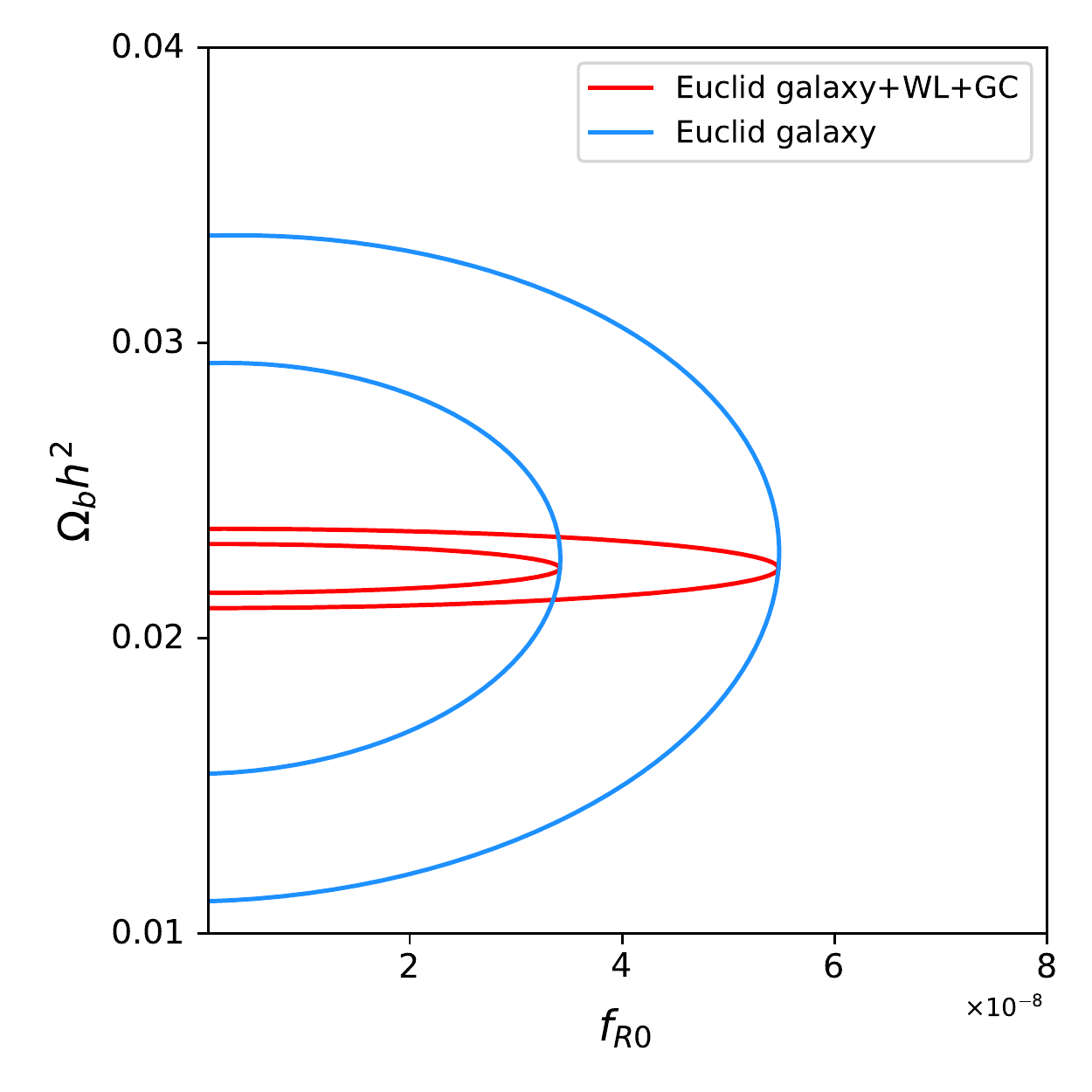}
	\includegraphics[scale=0.6]{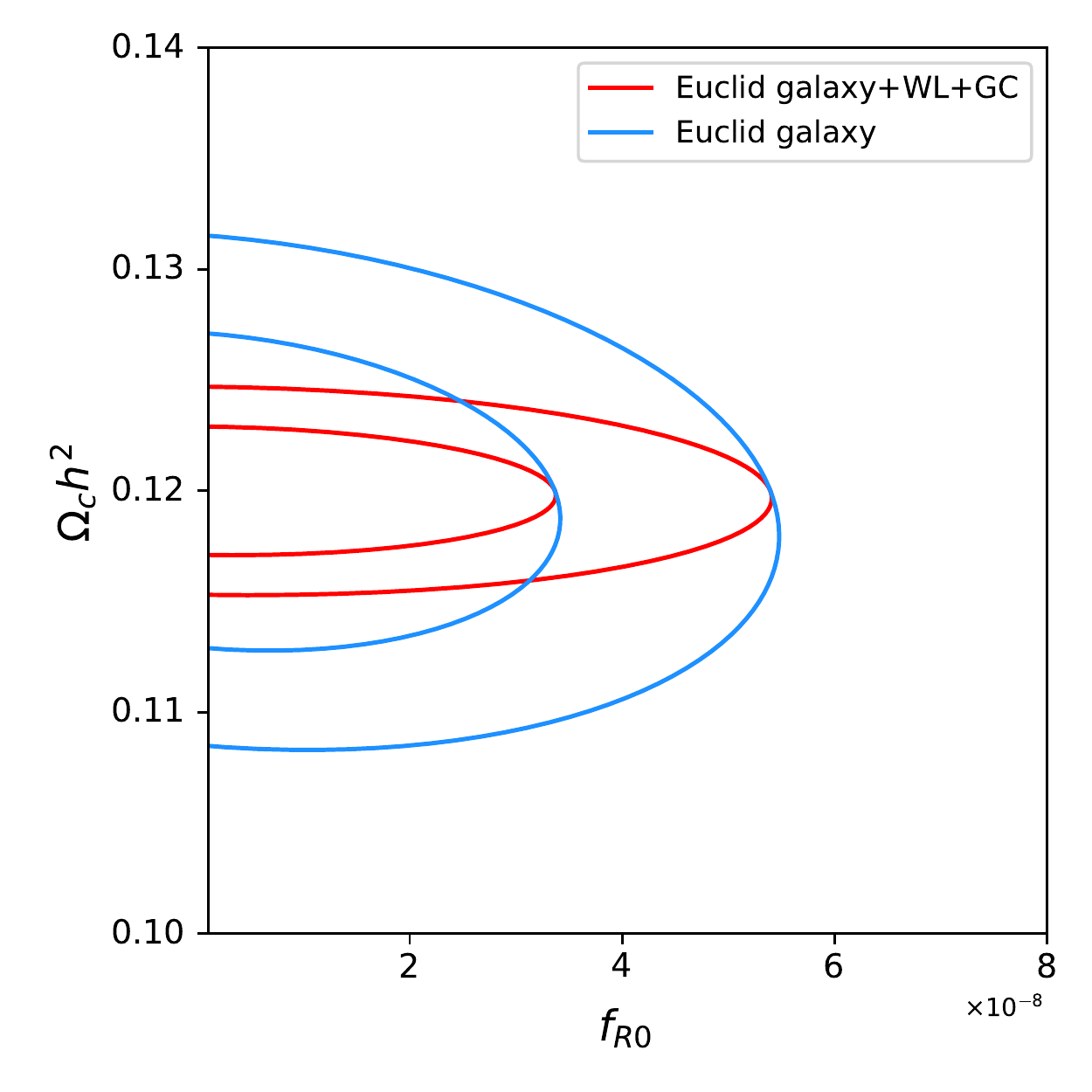}
	\caption{The predicted two-dimensional distributions of $f_{R0}$-$\Omega_b h^2$ ({\it left}) and $f_{R0}$-$\Omega_c h^2$ ({\it right}) in the HS $f(R)$ gravity are shown for Euclid galaxy redshift and its combination with Euclid WL and GC surveys, respectively.}
	\label{f16}
\end{figure}

\begin{figure}[htbp]
	\centering
	\includegraphics[scale=0.6]{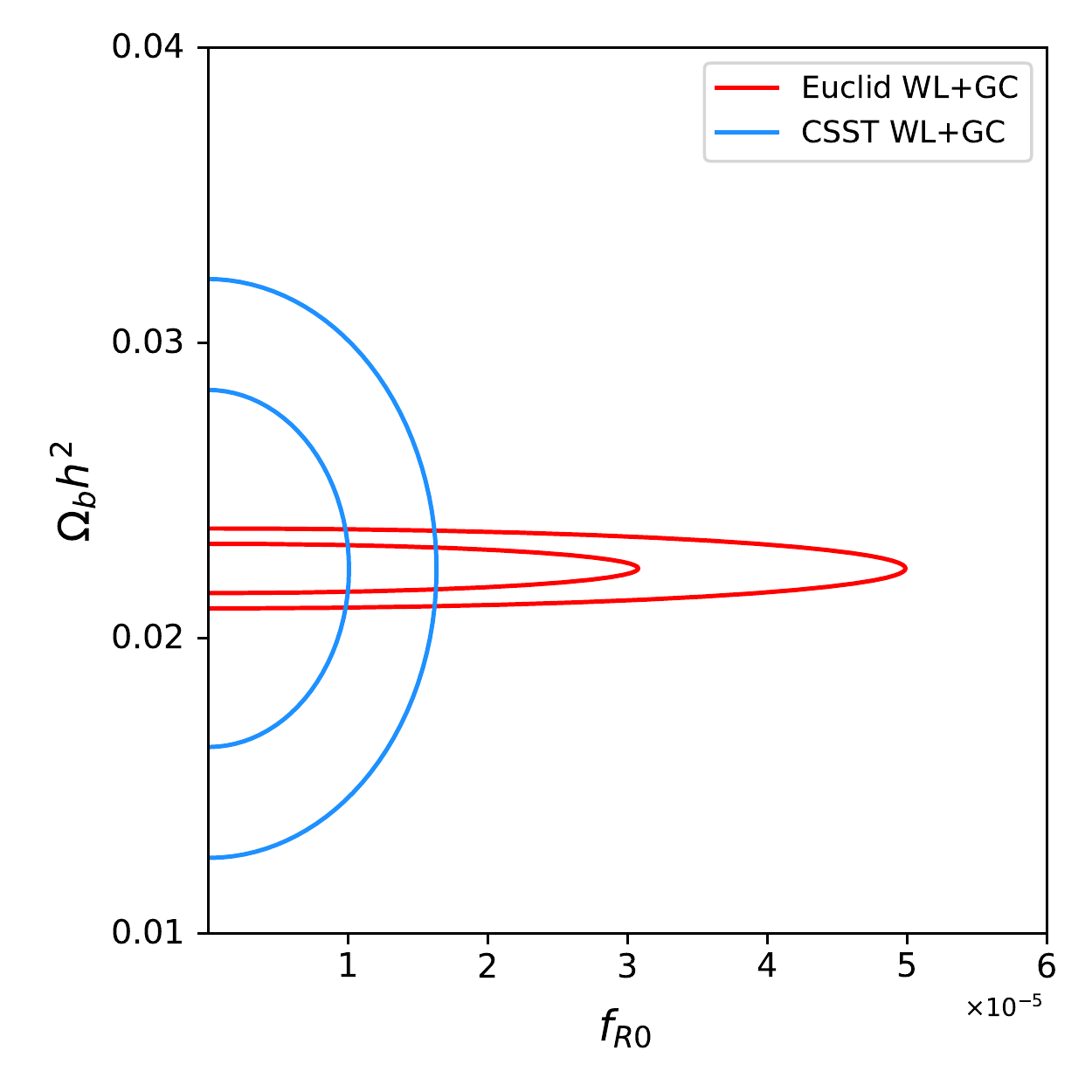}
	\includegraphics[scale=0.6]{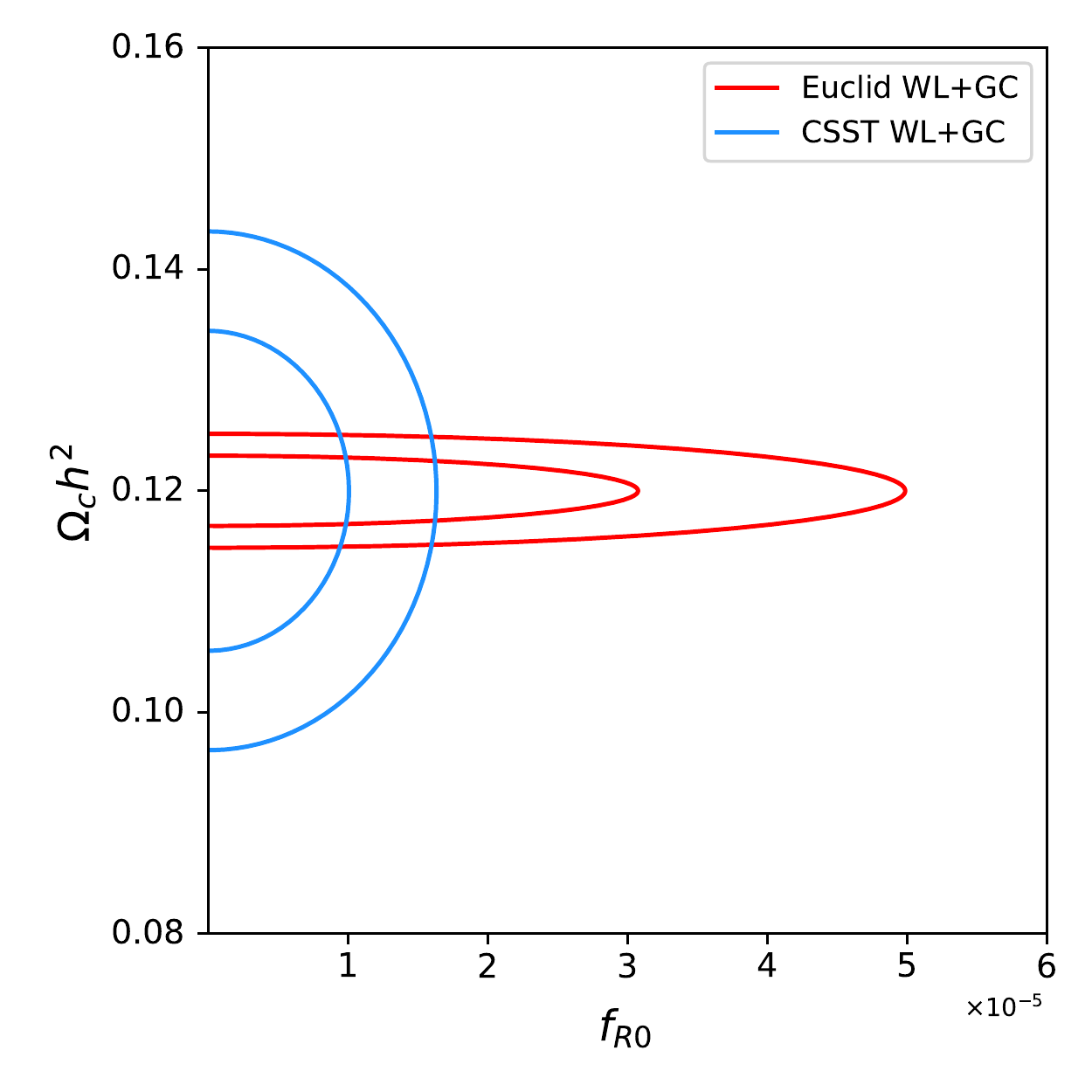}
	\caption{The predicted two-dimensional distributions of $f_{R0}$-$\Omega_b h^2$ ({\it left}) and $f_{R0}$-$\Omega_c h^2$ ({\it right}) in the HS $f(R)$ gravity are shown for combinations of WL and GC from Euclid and CSST surveys, respectively.}
	\label{f17}
\end{figure}

Assuming the likelihood function $\mathcal{L}$ of a set of parameters $\mathbf{p}$ given the data $\mathbf{d}$ to be a Gaussian distribution, it reads as
\begin{equation}
\mathcal{L}(\mathbf{p}|\mathbf{d})\propto\frac{\mathrm{exp}\left(-\frac{1}{2}\mathbf{d}^\dagger\left[\mathbf{M}(\mathbf{p})\right]^{-1}\mathbf{d}\right)}{\sqrt{|\mathbf{M}(\mathbf{p})|}}, \label{31}
\end{equation}
where $\mathbf{M}$ denotes the covariance matrix of the mock data. The specific cosmological parameters seeds in the parameter vector $\mathbf{p}$. Then at the fiducial values of parameters, the Fisher Matrix is easily constructed from the curvature of the likelihood function $\mathcal{L}$ as follows
\begin{equation}
F_{\alpha\beta}\equiv-\left<\frac{\partial^2 \mathrm{log}\mathcal{L}}{\partial p_\alpha\partial p_\beta}\right>_{\mathbf{p=p_0}}, \label{32}
\end{equation}
where $\mathbf{p_0}$ denotes the fiducial parameter vector. 

\subsection{21 cm intensity mapping}
For 21 cm IM experiments, if one uses the HI APS method to implement Fisher forecasts, the corresponding Fisher Matrix can be expressed as 
\begin{equation}
F_{\alpha\beta}=\sum\limits_\ell f_{\mathrm{sky}}\left(\frac{2\ell+1}{2}\right)\mathrm{Tr}\left[\mathbf{C}^{-1}(p)\frac{\partial \mathbf{C}}{\partial p_\alpha}\mathbf{C}^{-1}(p)\frac{\partial \mathbf{C}}{\partial p_\beta}\right], \label{33}
\end{equation} 
where $f_{\mathrm{sky}}$ denotes the sky fraction, the symbol $\mathrm{Tr}$ is the trace of a matrix and $\mathbf{C}=C^{ij}_\ell+\delta^{ij}N_\ell$ is the observed 21 cm APS between tomographic bin $i$ and $j$, which consists of true 21 cm APS $C^{ij}_\ell$ and noise PS $N_\ell$. 
The diagonal elements of matrix $\mathbf{C}$ are responsible for auto correlations of each window, when non-diagonal ones represent cross correlations of different redshift bins. Based on the fact that $\Delta_\ell$ in different windows for a given $\ell$ correspond to different comoving scales, their cross correlations are much smaller than auto correlations. Nonetheless, we will work out all the matrix elements of $\mathbf{C}$ and then calculate the Fisher matrix. 

Since the HS $f(R)$ gravity can obviously affect the 21 cm APS, we wonder the detectability of this model via the forthcoming HI IM experiments. Specifically, we take the middle frequency band 1 of SKA phase 1 (hereafter SKA1-MID-B1) survey to implement forecasts. 

The SKA is believed to be the largest radio telescope all over the world with the collecting area over one square kilometer \cite{SKA, Maartens:2015mra,SKA:2018ckk}. The SKA will be used for HI IM and measuring spectroscopic redshifts and galaxy PS, and it is made up of two phases: SKA1 being built and SKA2 being configured. SKA1 will consists of two instruments, SKA1-MID, which will operate in the frequency range $(350, 1750)$ MHz, and SKA1-LOW, which will work covering the low frequency range $(50, 350)$ MHz. 

SKA1-MID that covers a sky area of 25000 deg$^2$ is planned to contain two sub-arrays, 133 15 m SKA1 dishes and extended 64 13.5 m MeerKAT dishes \cite{SKA:2018ckk}. In this analysis, we will assume  these movable 197 dishes are all of 15 m in diameter with a dual polarization receiver \cite{Chen:2019jms}. SKA1-MID will consists of two bands, Band 1 covering the frequency range $(350, 1050)$ MHz and Band 2 covering $(950, 1750)$ MHz. Since the single-dish mode is more sensitive to the HI brightness temperature and performs better in detecting the HI signals at BAO scales than the interferometric mode, we take into account the single-dish mode for SKA1-MID-B1, the detailed experimental parameters of which are shown in Tab.\ref{t1}. 

\begin{table}[h!]
	\renewcommand\arraystretch{1.5}
	\caption{The experimental parameters of SKA1-MID-B1.}
	\setlength{\tabcolsep}{5mm}{
	{\begin{tabular}{l  c}
		
		\hline
		\hline
		Parameters         &SKA1-MID-B1      \\                                 
		\hline
		Frequency range (MHz)    & [350, 1050] \\
		Redshift range           & [0.35, 3.06] \\
        Sky area $A_{\mathrm{sky}}$ (deg$^2$) & 25000 \\
        Integration time $t_{\mathrm{obs}}$ (hours)  & 10000 \\
        System temperature $T_{\mathrm{sys}}$ (K) & 28 \\
        Number of beams $N_b$         & 1          \\
        Number of dishes $N_d$        & 197         \\
        Illuminated aperture $D_{\mathrm{d}}$ (m) & 15 \\
        Bin width (MHz)          & 20\\
		\hline
		\hline
	\end{tabular}
	\label{t1}}}
\end{table}

Because the frequencies of 21 cm observations are far lower than those of CMB surveys, the foregrounds contamination such as the above mentioned diffuse galactic synchrotron emission, bright point sources and atmospheric turbulence is more serious in 21 cm IM surveys. As a consequence, in order to extract the signal efficiently, one must adopt a foreground subtraction techniques \cite{Bigot-Sazy:2015jaa, Olivari:2015tka}. Here we consider an optimistic case by assuming a perfect foreground subtraction and instrument-only noise. The dominant component in instrumental noise is the thermal noise that reads as \cite{Bull:2014rha}
\begin{equation}
N_\ell=\frac{A_{\mathrm{sky}}T_{\mathrm{sys}}^2}{N_bN_dt_{\mathrm{obs}}\Delta\nu}, \label{34}
\end{equation}
where $A_{\mathrm{sky}}$ denotes the sky area covered by a survey, $T_{\mathrm{sys}}$ the system temperature, $N_d$ th number of dishes, $N_d$ the number of antennae, $t_{\mathrm{obs}}$ the observation time, and $\Delta\nu$ the width of frequency windows.

During the observations, the angular resolution at small scales is subject to the finite beam size. This effect will modulate the 21 cm APS as follows
\begin{equation}
C_\ell^{\mathrm{obs}}=C_\ell^{\mathrm{th}}\mathrm{exp}\left[{-\left(\frac{\ell\theta_\mathrm{FWHM}}{\sqrt{8\,\mathrm{ln}\,2}}\right)^2}\right], \label{35}
\end{equation} 
where $C_\ell^{\mathrm{obs}}$ and $C_\ell^{\mathrm{th}}$ (see Eq.(\ref{10})) are the predicted and theoretical 21 cm APS, respectively, and $\theta_{\mathrm{FWHM}}$ denotes the full width at the half maximum (FWHM) of the facilities
\begin{equation}
\theta_\mathrm{FWHM}=\frac{1.2\lambda}{D_{\mathrm{d}}}, \label{36}
\end{equation}
where $D_{\mathrm{d}}$ denotes the angular diameter of each dish and $\lambda$ is the corresponding wavelength for each frequency window.

\subsection{Galaxy redshift survey}
For galaxy redshift surveys, Fisher matrices can be shown as 
\begin{equation}
F_{\alpha\beta}=\frac{1}{2}\int\frac{\mathrm{d}^3k}{(2\pi)^3}V_{\mathrm{eff}}(\mathbf{k})\left(\frac{\partial \mathrm{ln\,S}_T}{\partial \, p_\alpha}\frac{\partial \mathrm{ln\,S}_T}{\partial \, p_\beta}\right), \label{37}
\end{equation}
where $S_T$ denotes the total covariance of measured signal, which consists of underlying signal $S_S$ and noise $S_N$, while $V_{\mathrm{eff}}(\mathbf{k})=V(S_S/S_T)^2$ represents the effective volume of the experiment covering a physical volume $V$. Note that the cosmology information seeds in $S_S$. 
The total signal encompasses two parts, GPS (see also Eq.(\ref{14})) and shot noise, i.e., $S_T=P(\mathbf{k},z)+1/n(z)$.

It is worth noting that there are two important cosmological quantities $\Omega_{\mathrm{HI}}$ and $b_{\mathrm{HI}}$. For 21 cm IM experiments, $\Omega_{\mathrm{HI}}$ can be assumed to be a constant $\Omega_{\mathrm{HI}}=0.62\times10^{-3}$ \cite{Prochaska:2008fp,Switzer:2013ewa} when $z\leqslant3$ and we take the HI bias $b_{\mathrm{HI}}=1$ in this study. 

As described above, we will include both optical and HI galaxy redshift surveys in our analysis. For the optical survey, we choose the Euclid satellite \cite{Euclid, EuclidTheoryWorkingGroup:2012gxx} as an example, which will be launched in the early 2020s and covers a sky area of 15000 deg$^{2}$. It will measure 50 million galaxy redshifts in the redshift range $z\in(0.65, \, 2.05)$ by using a infrared spectrograph. Meanwhile, it will conduct a photometric survey of 1 billion galaxies in the range $z\in(0, \, 2)$. The spectroscopic data will be used to measure BAO, RSD and GC signals, while the photometric data will be used for estimating the galaxy ellipticities and then measuring the cosmic shear signals. For the HI survey, we adopt the SKA2 \cite{Santos:2015gra} as a reference case, which will be very sensitive and achieve an root-mean-square (rms) flux sensitivity $S_{\mathrm{rms}}\approx5 \, \mathrm{\mu Jy}$ covering a sky area of 30000 deg$^2$ for 10000 hours. We expect that SKA2 will produce a catalogue of 1 billion HI galaxies in the redshift range $z\in(0.18, \, 1.84)$, which is far beyond any planned optical or near infrared experiment when $z\in(0, \, 1.4)$.
The expected galaxy number densities and bias for Euclid and SKA2 at given redshifts are shown in Tab.\ref{t2}.

\begin{table}[!t]
	\renewcommand\arraystretch{1.5}
	\caption{The expected galaxy number densities and bias for Euclid and SKA2 at given redshifts are shown. Note that the galaxy number densities are in units of Mpc$^{-3}$. }
	\setlength{\tabcolsep}{5mm}{
	\begin{tabular} { c| c | c | c | c | c }
		\hline
		\hline
		
		 \multicolumn{3}{c|}{Euclid}      &   \multicolumn{3}{c}{SKA2}                              \\
		\hline
		  $z$      & $n(z)\times10^{-3}$ & $b(z)$    & $z$  & $n_{\mathrm{HI}}(z)\times10^{-6}$  & $b_{\mathrm{HI}}(z)$  \\
		\hline
		 0.7     & 1.25    & 1.30    &  0.23    & 44300    & 0.713            \\
		 0.8     & 1.92    & 1.34    &  0.33    & 27300   &   0.772          \\
		 0.9     & 1.83    & 1.38    &  0.43    & 16500   &   0.837          \\
		 1.0     & 1.68    & 1.41    &  0.53    & 9890   &    0.907         \\
		 1.1     & 1.51    & 1.45    &  0.63    & 5880   &    0.983         \\
		 1.2     & 1.35    & 1.48    &  0.73    & 3480   &    1.066         \\
		 1.3     & 1.20    & 1.52    &  0.83    & 2050   &    1.156        \\
		 1.4     & 1.00    & 1.55    &  0.93    & 1210   &    1.254         \\
		 1.5     & 0.80    & 1.58    &  1.03    & 706   &     1.360        \\
		 1.6     & 0.58    & 1.61    &  1.13    & 411   &     1.475        \\
		 1.7     & 0.38    & 1.64    &  1.23    & 239   &     1.600        \\
		 1.8     & 0.35    & 1.67    &  1.33    & 139   &     1.735        \\
		 1.9     & 0.21    & 1.70    &  1.43    & 79.9   &    1.882         \\
		 2.0     & 0.11    & 1.73    &  1.53    & 46.0   &    2.041         \\
		 ---     & ---    & ---      &  1.63    & 26.4   &    2.214         \\
		 ---     & ---    & ---      &  1.73    & 15.1   &    2.402         \\
		 ---     & ---    & ---      &  1.81    & 9.66   &    2.566         \\

		\hline
		\hline
	\end{tabular}
	\label{t2}}
\end{table}

\subsection{Cosmic microwave background}
For CMB experiments, although Fisher matrices has the same form as Eq.(\ref{33}), the matrix $\mathbf{C}$ for CMB is different from HI IM. In our codes, we have explored the forecasting ability of CMB temperature, E-mode, B-mode and lensing APS. Hence, $\mathbf{C}=C_\ell^{TT}, \, C_\ell^{EE}, \, C_\ell^{BB}$ and $C_\ell^{\phi\phi}$. Note that the noise cross PS is zero because we just take into account the statistical noise which are uncorrelated.  

The noise PS for CMB surveys can be expressed as 
\begin{equation}
N_\ell^{AB}=u^2\mathrm{exp}\left\{{-\left[\frac{\ell(\ell+1)\theta_\mathrm{FWHM}}{\sqrt{8\,\mathrm{ln}\,2}}\right]^2}\right\}, \label{38}
\end{equation}
where $u$ denotes the total instrumental noise in a unit of $\mu K$ radian and $AB=TT, \, EE, \, BB$ and $\phi\phi$. One can easily find that this formula is very similar to Eq.(\ref{35}) but a small difference. It is worth noting that $s$ should multiply a factor of $\sqrt{2}$ in the cases of E- and B-mode polarization. To calculate the CMB lensing noise APS $N_\ell^{\phi\phi}$, we adopt a minimum variance estimator to reconstruct the lensing signals using $C_\ell^{EE}$ and $C_\ell^{BB}$. This iterative method is carefully studied in Refs.\cite{Hirata:2003ka, Okamoto:2003zw} and outlined in Ref.\cite{Smith:2010gu}.

To test the constraining power of future CMB experiments, specifically, we will take the forthcoming Simons Observatory (SO) \cite{SO, SimonsObservatory:2018koc} and Cosmic Origins Explorer (CORE) \cite{COrE:2011bfs} to carry out Fisher forecasts. SO covering 15000 deg$^2$ is made up of a 6 m Large Aperture Telescope (LAT) similar in size to the Atacama Cosmology Telescope (ACT) and three 0.5 m Small Aperture Telescopes (SATs) similar in size to the phase 3 of the low resolution Background Imaging pf Cosmic Extragalactic Polarization survey (BICEP3). It has a resolution of $1-2$ arcmin with a sensitivity of about $5\,\mu K$. The key science goals of SO are to depict the primordial fluctuations, measure the properties of neutrinos, explore the nature of DM and DE, constrain the duration of reionization and deepen our standing of galaxy evolution. We will take the LAT branch of SO to implement the Fisher forecasting.
CORE with a sky coverage 0.65 is mainly designed to detect the primordial GWs generated during the inflationary process at more than $3\,\sigma$ confidence level for tensor-to-scalar ratio $r$. Meanwhile, it will measure the CMB lensing PS with a high precision, the properties of neutrinos and the primordial non-Gaussianity with significant improvements over the Planck satellite. 
To make a comparison with SO, we take a 6-band CORE between 75 GHz and 225 GHz with an angular resolution ranging from about 5 arcmin at 225 GHz to 14 arcmin at 75 GHz. The frequency bands, FWHM and temperature sensitivities of SO and CORE are shown in Tab.\ref{t3}. More details about SO and CORE can be found in Refs.\cite{SimonsObservatory:2018koc,COrE:2011bfs}.

\begin{table}[!t]
	\renewcommand\arraystretch{1.5}
	\caption{The frequencies, angular resolution and temperature sensitivities of SO and CORE are shown. Note that frequencies, $\theta_{\mathrm{FWHM}}$ and temperature sensitivities are in units of GHz, arcmin and $\mu K$-arcmin, respectively.  }
	\setlength{\tabcolsep}{5mm}{
		\begin{tabular} { c| c | c | c | c | c }
			\hline
			\hline
			
			\multicolumn{3}{c|}{SO}      &   \multicolumn{3}{c}{CORE}                              \\
			\hline
			$\nu$      & $\theta_\mathrm{FWHM}$ & Noise    &$\nu$      &$\theta_\mathrm{FWHM}$ & Noise  \\
			\hline
			27     & 7.4   & 52    &  75    & 14.0    & 2.73            \\
			39     & 5.1    & 27    &  105    & 10.0   &  2.68         \\
			93     & 2.2    & 5.8    &  135    & 7.8   &  2.63          \\
			145     & 1.4    & 6.3    &  165    & 6.4   & 2.67         \\
			225     & 1.0    & 15    &  195    & 5.4   &  2.63         \\
			280     & 0.9    & 37    &  225    & 4.7   &  2.64         \\			
			\hline
			\hline
		\end{tabular}
		\label{t3}}
\end{table}

\subsection{Weak lensing and galaxy clustering}
For the WL and GC surveys, the expression of Fisher matrix Eq.(\ref{33}) also stands. To carry out the forecasts, we need to specify the galaxy number density per steradian in the $i$-th tomographic bin as 
\begin{equation}
n_i(z)=\frac{\int_{z_i^{\mathrm{min}}}^{z_i^{\mathrm{max}}}\frac{\mathrm{d}n}{\mathrm{d}z}\mathcal{E}(z,z_{\mathrm{pho}})\mathrm{d}z_{\mathrm{pho}}}{\int_0^\infty\frac{\mathrm{d}n}{\mathrm{d}z}\mathcal{E}(z,z_{\mathrm{pho}})\mathrm{d}z_{\mathrm{pho}}}, \label{39}
\end{equation}
where $\mathcal{E}(z,z_{\mathrm{pho}})$ denotes the so-called error function
\begin{equation}
\mathcal{E}(z,z_{\mathrm{pho}})=\frac{1}{\sqrt{2\pi}\sigma_{\mathrm{ph}}}\mathrm{exp}\left(-\frac{z-z_{\mathrm{pho}}}{2\sigma_{\mathrm{pho}}}\right). \label{40}
\end{equation}
Subsequently, we use the following galaxy surface density
\begin{equation}
\frac{\mathrm{d}n}{\mathrm{d}z}=z^{\tilde{\alpha}}\mathrm{exp}\left[-\left(\frac{z}{z_0}\right)^{\tilde{\beta}}\right], \label{41}
\end{equation}
where $z_0$ is the initial redshift, and $\tilde{\alpha}$ and $\tilde{\beta}$ are two free parameters characterizing the galaxy distribution.  

We will take the above mentioned Euclid satellite \cite{Euclid,EuclidTheoryWorkingGroup:2012gxx} and the Chinese Space Station Telescope (CSST) \cite{CSST,Gong:2019yxt} to forecast the constraining power of WL and GC on MG parameters. For Euclid, we choose the sky fraction $f_{\mathrm{sky}}=0.3636$, $z_0=0.6374$, photometric redshift error $\sigma_{\mathrm{pho}}=0.05(1+z)$, mean internal ellipticity 0.22 and observed galaxy number 30 per armin$^2$. 

The CSST covering a sky area of 15000 deg$^2$ will operate about 10 years with field of view 1.1 deg$^2$. It will conduct photometric imaging and spectroscopic surveys with high spatial resolution and multi-wavelength coverage. Its key science goals are exploring the nature of DM and DE, large scale structure and galaxy formation and evolution, etc. Its photo-z distribution has a peak around $z=0.6$ and can be divided into four bins. Its spectroscopic redshift distribution has a peak at about 0.35 and can be dissected into five bins. Furthermore, for CSST WL survey, we take $z_0=0.6$, mean internal ellipticity 0.15 and observed galaxy number 100000 per deg$^2$. 
For CSST GC survey, we choose $z_0=0.35$, mean internal ellipticity 0.15 and observed galaxy number 9765 per deg$^2$.    
For CSST, we use the same $\sigma_{\mathrm{pho}}=0.05(1+z)$, $\tilde{\alpha}=2$ and $\tilde{\beta}=1.5$ as Euclid. For both GC surveys, we also need to consider the galaxy bias model as $b(z)=\sqrt{1+z}$ (see also Eq.(\ref{28})) in our analysis. For more detailed information about CSST, we refer the readers to Ref.\cite{Gong:2019yxt}.
To implement the forecasts, we take the noise APS in the $i$-th bin for WL surveys in the following form
\begin{equation}
C_N^{ii}(\ell)=\frac{\mathcal{M}_{\mathrm{rms,i}}^2}{N_i}, \label{42}
\end{equation}  
where $\mathcal{M}_{\mathrm{rms,i}}$ and $N_i$ denotes the rms shear from intrinsic ellipticities of galaxies and galaxy number in the $i$-th bin, respectively. For GC surveys, we adopt a simple form
\begin{equation}
C_N^{ii}(\ell)=\frac{1}{N_i}. \label{43}
\end{equation}

\subsection{Type Ia supernovae}
For the SNe Ia survey, we adopt the Wide-Field Infrared Survey Telescope (WFIRST) \cite{Spergel:2015sza}, which is a planned satellite mission and will produce 2725 SNe Ia in the redshift range $z\in(0.1, 1.7)$ in order to investigate the nature of DE. To generate this mock catalogue, we use the redshift distribution and distance error model from Ref.\cite{Spergel:2015sza}. Based on the assumption that the errors of distance modulus (see Eq.(\ref{30})) of different SNe Ia are uncorrelated, the error model reads as
\begin{equation}
\sigma_{\mu}^2=\sigma_{\mathrm{ps}}^2+\sigma_{\mathrm{int}}^2+\sigma_{\mathrm{lens}}^2+\sigma_{\mathrm{sys}}^2, \label{44}
\end{equation} 
where $\sigma_{\mathrm{ps}}=0.08$ denotes the photometric redshift errror for a SNe Ia, $\sigma_{\mathrm{int}}=0.09$ is an intrinsic dispersion in SNe Ia luminosities that has taken a correction for light curve shape and spectral properties, $\sigma_{\mathrm{lens}}=0.07z$ is the contribution of gravitational lensing magnification to statistical errors, and the potential systematic errors $\sigma_{\mathrm{sys}}=0.02/(1+z)/1.8$. Subsequently, the Fisher matrix for SNe Ia can be written as 
\begin{equation}
F_{\alpha\beta}=\sum\limits_i\frac{1}{\sigma_{\mu,i}^2}\frac{\partial \mu_i}{\partial p_\alpha}\frac{\partial \mu_i}{\partial p_\beta}, \label{45}
\end{equation}
where $\mu_i$ and $\sigma_{\mu,i}$ are the distance modulus and corresponding error for $i$-th SNe Ia.

\begin{table}[h!]
	\renewcommand\arraystretch{1.5}
	\caption{The predicted $1\sigma$ errors of $f_{R0}$ are shown for various experiment combinations. SKA2 and Euclid denote the HI and optical galaxy redshift surveys, respectively.}
	\setlength{\tabcolsep}{10mm}{
		{\begin{tabular}{l   c}
				
				\hline
				\hline
				Combination         & $\sigma_{f_{R0}}$      \\    
				\hline
				Planck 2018         & $9.77\times10^{-6}$     \\               
				SKA1-MID-B1         &   $8.68\times10^{-8}$         \\
				SKA2                &   $1.36\times10^{-8}$         \\
				SO                  &   $2.49\times10^{-6}$          \\
				CORE                &   $5.24\times10^{-6}$          \\
				Euclid              &   $2.20\times10^{-8}$          \\
				Euclid WL           &   $2.79\times10^{-5}$          \\
				Euclid GC           &   $2.99\times10^{-5}$          \\
				CSST WL             &   $1.02\times10^{-5}$         \\
				CSST GC             &   $8.83\times10^{-6}$          \\
				WFIRST              &   $7.97\times10^{-3}$          \\
				ET                  &   $2.37\times10^{-5}$          \\
				SKA1-MID-B1+SKA2    &   $1.34\times10^{-8}$          \\
				SKA1-MID-B1+SO      &   $8.67\times10^{-8}$          \\
				SKA1-MID-B1+CORE    &   $8.68\times10^{-8}$          \\
				SKA1-MID-B1+Euclid  &   $2.20\times10^{-8}$          \\
				SKA2+SO             &   $1.36\times10^{-8}$          \\
				SKA2+CORE           &   $1.36\times10^{-8}$          \\
				SKA2+Euclid         &   $1.15\times10^{-8}$          \\
				Euclid+SO           &   $2.19\times10^{-8}$           \\
				Euclid+CORE         &   $2.20\times10^{-8}$           \\
				Euclid+Euclid WL+Euclid GC   & $2.20\times10^{-8}$         \\
				Euclid WL+Euclid GC          & $2.04\times10^{-5}$         \\
				CSST WL+CSST GC              & $6.68\times10^{-6}$         \\
				WFIRST+ET                    & $2.19\times10^{-5}$             \\
				SO+Euclid+Eulcid WL+Euclid GC+WFIRST    &  $2.19\times10^{-8}$       \\
				SO+Euclid+Eulcid WL+Euclid GC+WFIRST+ET    &  $2.19\times10^{-8}$       \\
				SKA2+SO+Euclid+Eulcid WL+Euclid GC+WFIRST+ET    &  $1.15\times10^{-8}$       \\
				SKA1-MID-B1+SO+Euclid+Eulcid WL+Euclid GC+WFIRST+ET    &  $2.13\times10^{-8}$       \\
				SKA1-MID-B1+SKA2+SO+Euclid+Eulcid WL+Euclid GC+WFIRST+ET    &  $1.14\times10^{-8}$       \\
				SKA1-MID-B1+SKA2+CORE+Euclid+CSST WL+CSST GC+WFIRST+ET    &  $1.14\times10^{-8}$       \\
				
				\hline
				\hline
			\end{tabular}
			\label{t4}}}
\end{table}

\subsection{Gravitational wave}
Since the key cosmological quantity of GWs is the luminosity distance $D_L(z)$ or distance modulus $\mu(z)$, which are the same as SNe Ia, we will also use Eq.(\ref{45}) to implement the Fisher forecasts of GWs in this analysis. 

Specifically, we take into account the binary mergers of a neutron star with either a neutron star (BNS) or a black hole (NSBH) in the third-generation ground-based detector, the Einstein Telescope (ET), which detects the high-frequency GW events. In our numerical calculations, the error of luminosity distances consists of two parts: an instrumental error $\sigma_i$ and an error $\sigma_l$ due to the effects of WL. As a consequence, the uncertainty of $D_L$ is $\sigma_{D_L}=\sqrt{\sigma_i^2+\sigma_l^2}$. The predicted events rate of BNS and NSBH for ET per year is of order $10^3\sim10^7$. However, only a small fraction ($\sim10^3$) can be observed in order to satisfy the constraint that GW events are accompanied with the observation of a short gamma-ray burst due to the narrow beaming angle. This indicates that, if assuming the events rate is $10^5$ per year, we will capture $10^2$ GW events with short gamma-ray bursts. Here we adopt an optimistic case, i.e., following the method taken in Ref.\cite{Zhao:2010sz}, we produce a catalogue of 1000 mock events from ET in the redshift range $z\in[0,5]$ under $\Lambda$CDM. More details about the GW simulation procedures can be found in Ref.\cite{Zhao:2010sz}.    

\section{Numerical results}
During the process of numerical analysis, we assume $n=1$ in the HS $f(R)$ gravity. The numerical results of Fisher forecasts from the above eight probes are presented in Figs.\ref{f10}-\ref{f17} and Tab.\ref{t4}. 

Since mainly concentrating on constraining MG, we just exhibit the $1\sigma$ uncertainties of $f_{R0}$ for various probe combinations in Tab.\ref{t4}.  
Specifically, in Figs.\ref{f10}-\ref{f12}, the constraining power of each probe on the parameters of HS $f(R)$ gravity are explored. We find that the HI galaxy redshift survey SKA2 gives the strongest constraint on $f_{R0}$, namely $\sigma_{f_{R0}}=1.36\times10^{-8}$. The optical galaxy redshift survey Euclid provides a relative weak constraint $\sigma_{f_{R0}}=2.20\times10^{-8}$. Although the promising 21 cm IM experiment SKA1-MID-B1 produces $\sigma_{f_{R0}}=8.68\times10^{-8}$, constraining power of these three probes are at the same order $\mathcal{O}(-8)$. The next-generation CMB survey SO gives $\sigma_{f_{R0}}=2.49\times10^{-6}$, which is tighter than $\sigma_{f_{R0}}=5.24\times10^{-6}$ from CORE. To a large extent, these two CMB surveys improve the constraining power on $f_{R0}$ relative to our previous constraint by using the Plank-2018 data, which gives $\sigma_{f_{R0}}=9.77\times10^{-6}$ \cite{Wang:2020dsc}.  
The future high precision large scale survey Euclid WL and GC provides $\sigma_{f_{R0}}=2.79\times10^{-5}$ and  $2.99\times10^{-5}$, respectively, which is a little weaker than $\sigma_{f_{R0}}=1.02\times10^{-5}$ and  $8.83\times10^{-6}$ from CSST WL and GC. One can easily find that CSST has stronger constraining power than Euclid. Except for CSST GC has the same order $\mathcal{O}(-6)$ as two CMB surveys, three left probes share the same order $\mathcal{O}(-5)$. It is worth noting that Euclid WL can give a better constraint on $f_{R0}$ than Euclid GC, but CSST does the opposite. The gravitational sirens from ET produces $\sigma_{f_{R0}}=2.37\times10^{-5}$ at the background level, which is tighter than $\sigma_{f_{R0}}=7.97\times10^{-3}$ from WFIRST SNe Ia by at least two orders of magnitude. One can easily find that CSST WL gives a close constraint on $\sigma_{f_{R0}}$ relative to the Planck-2018 result and that Euclid WL and GC and ET provide loose constraints relative to Planck-2018. Specially, CSST GC will give a tighter constraint than Planck-2018. Interestingly, we find that 21 cm IM, HI and optical galaxy experiments give better constraints than other probes such as CMB, WL, GC, SNe Ia and GW. This implies that future HI IM and galaxy redshift surveys will be very promising in compressing the parameter space.

To illustrate the ability of each probe in breaking degeneracies between parameters, we consider the correlations between $f_{R0}$ and baryons ($\Omega_b h^2$) or CDM ($\Omega_c h^2$) as examples. In Fig.\ref{f11}, we find that CORE gives the best constraint on the baryon fraction $\sigma_{\Omega_b h^2}=4.10\times10^{-5}$, which is tighter than $4.33\times10^{-5}$, $7.81\times10^{-5}$, $3.82\times10^{-3}$ and $4.61\times10^{-3}$ from SO, SKA1-MID-B1, SKA2 and Euclid galaxy redshift surveys, respectively. Although the constraining power is weak, being similar to the case of $f_{R0}$, CSST WL provides a tighter constraint $\sigma_{\Omega_b h^2}=3.03\times10^{-2}$ than $6.78\times10^{-2}$ from Euclid WL. It is interesting that Euclid GC provides a better restriction $\sigma_{\Omega_b h^2}=5.53\times10^{-4}$ than $4.04\times10^{-3}$. Two background surveys WFIRST and ET produce  $\sigma_{\Omega_b h^2}=9.50\times10^{-3}$ and $1.90\times10^{-2}$. One can find that, unlike the case of $f_{R0}$, ET shows weaker constraining power on the baryon fraction than WFIRST. In Fig.\ref{f12}, we find that the relative constraining power on the CDM fraction from each independent probe is completely consistent with the case of baryon fraction. Here we just report several important constraining results. CORE, SO and SKA1-MID-B1 gives $\sigma_{\Omega_b h^2}=4.90\times10^{-4}$, $5.38\times10^{-4}$ and $7.48\times10^{-4}$, respectively. Since the contours of baryon and CDM have similar shapes, the relative constraining power on the matter fraction $\Omega_m h^2$ will be same as $\Omega_b h^2$ and $\Omega_c h^2$. SKA1-MID-B1, CORE and SO have a great advantage in determining the matter fraction with high precision, and their predictions are more accurate than other independent probes by at least one order of magnitude. The above results indicate that future 21 cm IM and HI galaxy surveys can improve our understanding of cosmic energy budget and MG to a large extent.      

In order to reduce the statistical uncertainties of parameters as many as possible and compare with various probe combination, we define the following three combinations
\begin{equation}
\mathrm{C1}:=\rm{\mbox{SKA1-MID-B1}+SKA2+SO+Euclid+\mbox{Eulcid WL}+\mbox{Eulcid GC}+WFIRST+ET},  \label{46}
\end{equation}  
\begin{equation}
\mathrm{C2}:=\rm{\mbox{SKA1-MID-B1}+SKA2+CORE+Euclid+\mbox{CSST WL}+\mbox{CSST GC}+WFIRST+ET},  \label{47}
\end{equation} 
\begin{equation}
\mathrm{Base}:=\rm{SO+Euclid+\mbox{Eulcid WL}+\mbox{Eulcid GC}+WFIRST},  \label{48}
\end{equation} 
where C1 and C2 and $\mathrm{Base}$ denote two total data combinations from eight probes and the basic combination from five ones, respectively. An important reason to choose these three combinations is that Base represents current main probes CMB+optical BAO+WL+GC+SNe Ia and C1 and C2 represent future main probes 21 cm IM+radio BAO+CMB+optical BAO+WL+GC+SNe Ia. Hence, we can completely compare the ability of current and future probes constraining MG.   
Since we just replace SO, Euclid WL and GC with CORE, CSST WL and GC, C1 and C2 will give almost same constraint on cosmological parameters.  

In Figs.\ref{f12}-\ref{f17}, we present constraints on $f_{R0}$, $\Omega_b h^2$ and $\Omega_c h^2$ for various data combinations. From Fig.\ref{f13}, we find that both C1 and C2 give $\sigma_{f_{R0}}=1.14\times10^{-8}$ (see Tab.\ref{t4}). Due to weak constraining power, the addition of ET into Base does not change the constraint on $f_{R0}$. If we add SKA1-MID-B1 into Base+ET, the constraint becomes a little better and improves $2.7\%$. Very interestingly, when adding SKA2 into Base+ET, the constraint improves $47.5\%$. Furthermore, if we continue adding SKA1-MID-B1 into Base+ET+SKA2, the constraint will improve $48.0\%$. This also means the constraining power on $f_{R0}$ from both C1 and C2 will increase by $48\%$ relative to Base. 
In Fig.\ref{f14}, considering same data combinations for two-dimensional parameter spaces $f_{R0}$-$\Omega_b h^2$ and $f_{R0}$-$\Omega_c h^2$, we find that C2 can produce a little better constraint on the baryon and matter fractions, i.e., $\Omega_b h^2=3.58\times10^{-5}$ and $\Omega_c h^2=3.00\times10^{-4}$ than C1. This can be attributed to the fact that CORE in C2 has a better constraint on these two parameters than SO in C1. Interestingly, we find that when adding SKA1-MID-B1 into Base+ET, the improvement of constraining power in the plane of $f_{R0}$-$\Omega_c h^2$ will be larger than that in the plane of $f_{R0}$-$\Omega_b h^2$. This consequence also occurs when adding SKA1-MID-B1 into Base+ET+SKA2. The addition of SKA1-MID-B1 and SKA2 reduces obviously two parameter spaces. The former dominates the reduction of $\Omega_b h^2$ and $\Omega_c h^2$ and the latter dominates the compression of $f_{R0}$. We are interested in exploring the constraining power of the combinations of 21 cm IM and other surveys.
In Fig.\ref{f15}, we observe that combining SKA2 and SO with SKA1-MID-B1 can reduce the parameter space most along the $f_{R0}$ and $\Omega_b h^2$ ($\Omega_c h^2$) axes, respectively. From Tab.\ref{t4}, we find that the combination of two HI surveys, SKA1-MID-B1+SKA2, can give the best constraint $\sigma_{f_{R0}}=1.34\times10^{-8}$ in all kinds of two probe combinations. This implies that future HI surveys including IM and galaxy redshift can provide very strong constraint for MG parameters. If replacing SKA2 with CORE, SO and Euclid in SKA1-MID-B1+SKA2, we obtain $\sigma_{f_{R0}}=8.68\times10^{-8}$, $8.67\times10^{-8}$ and $2.20\times10^{-8}$, respectively. Subsequently, we investigate the constraining power of the combinations of HI galaxy and other surveys, and find that SKA2+Euclid gives the strongest constraint $\sigma_{f_{R0}}=1.15\times10^{-8}$, which is improved $15.4\%$ relative to $1.36\times10^{-8}$ from SKA2+SO and SKA2+CORE. It is interesting that HI and optical galaxy redshift surveys and C1 almost have the same constraining power on $f_{R0}$. Because Euclid galaxy redshift survey has a weaker constraining power than SKA2, Euclid+SO and Euclid+CORE just gives $\sigma_{f_{R0}}=2.19\times10^{-8}$ and $2.20\times10^{-8}$. 
In Fig.\ref{f16}, we show the parameter spaces of $f_{R0}$-$\Omega_b h^2$ and $f_{R0}$-$\Omega_c h^2$ from Euclid probes, and observe that Euclid+Euclid WL+Euclid GC can compress more parameter space relative to Euclid alone in the plane of $f_{R0}$-$\Omega_b h^2$ than in the plane of $f_{R0}$-$\Omega_c h^2$. This compression can be ascribed to the strong constraining power on baryon and matter fractions from Euclid GC (see Figs.\ref{f11}-\ref{f12}). Specifically, Euclid+Euclid WL+Euclid GC provides $\sigma_{f_{R0}}=2.20\times10^{-8}$, which is same as Euclid alone. This means that Euclid WL and GC have a poor constraint on $f_{R0}$ when compared with Euclid. However, this combination can clearly improve the constraint on $\Omega_b h^2$ and $\Omega_c h^2$ relative to Euclid. Moreover, we compare two future large scale WL and GC surveys and find that although Euclid WL+GC can provide better constraints on $\Omega_b h^2$ and $\Omega_c h^2$ than CSST WL+GC, CSST WL+GC can give a better constraint $\sigma_{f_{R0}}=6.68\times10^{-6}$ than $\sigma_{f_{R0}}=2.04\times10^{-5}$ from Euclid WL+GC. The combination of two background experiments WFIRST+ET gives $\sigma_{f_{R0}}=2.19\times10^{-5}$, which is a little tighter than ET and comparable with Euclid WL+GC.

\section{Discussions and conclusions}
Test the correctness of GR at cosmological scales is one of the most important topics in modern cosmology. In general, there are two approaches to explore this topic. One is studying Whether the predictions of GR is consistent with observations. Another is establishing a MG model based on some physical mechanism and investigating whether there exists a MG signal in light of observations. In theoretical cosmology, we usually take the second approach to reach this goal. However, due to the limited observational resolution and sensitivity, even the most precise CMB experiment Planck can not detect a MG signal at the current stage. Therefore, we forecast the ability of future mainstream cosmological probes encompassing 21 cm IM, HI galaxy redshift, CMB, optical galaxy redshift, WL, GC, SNe Ia, GW in constraining the HS $f(R)$ gravity.      

For independent probes, we find that the HI galaxy redshift survey SKA2 gives the strongest constraint $\sigma_{f_{R0}}=1.36\times10^{-8}$ and the geometrical probe WFIRST provides the weakest constraint $\sigma_{f_{R0}}=7.97\times10^{-3}$. Actually, constraints on $f_{R0}$ from the promising 21 cm IM experiment SKA1-MID-B1 and optical galaxy redshift survey Euclid have the same order $\mathcal{O}$(-8) as SKA2. The fourth-generation CMB experiments SO and CORE obtain the order $\mathcal{O}$(-6), while large scale structure surveys Euclid WL, Euclid GC and CSST WL give the order $\mathcal{O}$(-5). It is noteworthy that CSST GC also gives the same order $\mathcal{O}$(-6) as SO and CORE and that gravitational sirens survey ET also produces the same order $\mathcal{O}$(-5) as Euclid WL, Euclid GC and CSST WL. In addition, we find that CORE gives the best constraint on the baryon fraction $\sigma_{\Omega_b h^2}=4.10\times10^{-5}$ and the CDM fraction $\sigma_{\Omega_c h^2}=4.90\times10^{-4}$, which shows a large improvement relative to the Planck 2018 result \cite{Planck:2018vyg}.

For combined probes, two total combinations C1 and C2 gives the tightest constraint $\sigma_{f_{R0}}=1.14\times10^{-8}$, which is reduced by $15.4\%$ relative to $\sigma_{f_{R0}}=1.36\times10^{-8}$ from SKA2. Actually, they have a very small difference of order $\mathcal{O}(-14)$ that can be ignored. The basic combination Base gives a little weaker constraint $\sigma_{f_{R0}}=2.19\times10^{-8}$ than C1.  
Interestingly, we find that the combination of two HI surveys, SKA1-MID-B1+SKA2, can provide a strong constraint $\sigma_{f_{R0}}=1.34\times10^{-8}$. However, two galaxy redshift surveys SKA2+Euclid can give the strongest constraint $\sigma_{f_{R0}}=1.15\times10^{-8}$ among all kinds of two probe combinations. This indicates that the forthcoming HI IM and galaxy redshift surveys can produce very strong constraint for MG parameters. 
Based on the fact that both SKA2+Euclid and Base+ET+SKA2 give $\sigma_{f_{R0}}=1.15\times10^{-8}$, one can easily find that SKA1-MID-B1+SKA2+Euclid can give the same constraint $\sigma_{f_{R0}}=1.14\times10^{-8}$ as C1. This implies that the left five probes hardly affect the constraining power on $f_{R0}$. Moreover, we find that C2 gives best constraint on the baryon fraction $\sigma_{\Omega_b h^2}=3.58\times10^{-5}$ and the CDM fraction $\sigma_{\Omega_c h^2}=3.00\times10^{-4}$. This can be ascribed into the fact that CORE in C2 has a tighter constraint on these two parameters than SO in C1. Our results reveal that, to a large extent, future 21 cm IM and HI galaxy surveys can improve our understanding of MG and energy budget in the cosmic pie.

This study has at least two limitations. One is that we just investigate the 21 cm auto APS and various cross spectra at large scales in the multipole range $\ell\in[2,300]$ and do not consider the small scale behaviors of the HS $f(R)$ gravity. An ideal consequence is to find out a nonlinear probe that can distinguish well the HS $f(R)$ gravity from the $\Lambda$CDM model at small scales. The other is we just test MG with future cosmological surveys, which can also be used for providing high precision constraints for interesting DE, DM and inflation models.
We expect to address these issues in the forthcoming study.

\section*{Acknowledgements}
Deng Wang warmly thanks Liang Gao, Jie Wang and Qi Guo for useful discussions, and Hao-Nan Zheng, Hang Yang, Hui-Jie Hu, Kai Zhu and Ying-Jie Jing for beneficial communications. This work is supported by the Ministry of Science and Technology of China under Grant No. 2017YFB0203300 and National Nature Science Foundation of China under Grants No. 11988101 and No. 11851301.

\end{document}